\documentclass[preprint]{aastex}
\usepackage{graphicx}
\def\ha{\hbox{H$\alpha$}}

\begin{document}

\title{Pre-main sequence stars in the Cepheus flare region}
\author{M\'aria Kun}
\affil{Konkoly Observatory, H-1121 Budapest, Konkoly Thege \'ut 15--17, Hungary}
\email{kun@konkoly.hu}
\author{Zolt\'an Balog\altaffilmark{*\dag}}
\affil{Steward Observatory 933 N. Cherry Av. Tucson, AZ, USA}
\author{Scott J. Kenyon}
\affil{Harvard--Smithsonian CfA, 60 Garden St. Cambridge, MA, USA}
\author{Eric E. Mamajek}
\affil{Department of Physics \& Astronomy, University of Rochester, Rochester, NY, 14627-0171 USA}
\author{Robert A. Gutermuth}
\affil{Harvard--Smithsonian CfA, 60 Garden St. Cambridge, MA, USA}

\altaffiltext{*}{On leave from Dept. of Optics and Quantum Electronics,
University of Szeged, D\'om t\'er 9, Szeged, Hungary}
\altaffiltext{\dag}{Present address: Max-Planck-Institut f\"ur Astronomie,
K\"onigstuhl 17, D-69117 Heidelberg, Germany}


\begin{abstract} We present results of optical spectroscopic and
$BVR_\mathrm{C}I_\mathrm{C}$ photometric observations of 77
pre-main sequence (PMS) stars in the Cepheus flare region. A total of
64 of these are newly confirmed PMS stars, originally selected from
various published candidate lists. We estimate effective temperatures
and luminosities for the PMS stars, and comparing the results with
pre-main sequence evolutionary models we estimate stellar masses of
0.2--2.4\,M$_{\sun}$ and stellar ages of 0.1--15~Myr. Among the PMS
stars, we identify 15 visual binaries with separations of
2--10~arcsec. From archival IRAS, 2MASS, and Spitzer data, we
construct their spectral energy distributions and classify 5\% of the
stars as Class~I, 10\% as Flat SED, 60\% as Class~II, and 3\% as
Class~III young stellar objects (YSOs). We identify 12 CTTS and 2 WTTS
as members of NGC~7023, with mean age of 1.6~Myr. 
The 13 PMS stars associated with L\,1228 belong to three
small aggregates: RNO~129, L\,1228\,A, and L\,1228\,S. The age
distribution of the 17 PMS stars associated with L\,1251 suggests that
star formation has propagated with the expansion of the Cepheus flare
shell. We detect sparse aggregates of $\sim$6--7~Myr-old PMS stars
around the dark clouds L\,1177 and L\,1219, at a distance of $\sim
400$\,pc. Three T~Tauri stars appear to be associated with the Herbig
Ae star SV~Cep at a distance of 600~pc. Our results confirm that the
molecular complex in the Cepheus flare region contains clouds  
of various distances and star forming histories.
\end{abstract}


\section{Introduction} 
\label{Sect_1}

Star forming regions in our Galactic environment offer a unique
opportunity for understanding the birth of planetary systems. There is
growing evidence that not only the stellar content and structure of
young stellar groups, but also the evolution of protostellar envelopes
and accretion disks depend on the star forming environment
\citep[e.g.][]{Hester05,Dullemond06}. Cross-comparing various star forming 
environments is important for understanding the origins and early
evolution of stars and planets.  In this paper we present our results
of spectroscopic and photometric studies of a population of pre-main
sequence stars associated with the molecular clouds of the Cepheus
flare, located at 200--400~pc from the Sun.

A detailed review on the ISM and young stellar populations in the
Cepheus flare region can be found in \citet*{Kun08}. Here we briefly
review the highlights that are relevant for this study. The Cepheus
flare \citep{Hubble} molecular cloud complex consists of a large
amount of dense ISM between $100\degr \leq l \leq 120\degr$ and above
$b \geq 10\degr$ \citep*{LyndsD,TDS, CB88,DUK}.  Though the CO maps of
the region presented by \citet{Lebrun} and \citet{GLADT} indicate a
coherent cloud complex, there is evidence for multiple cloud layers
over the Cepheus flare region.  For instance, stars illuminating
reflection nebulae or displaying 100\,\micron \ excess due to their
dusty environment, can be found at various distances
\citep*[e.g.][]{Racine,KVSz}, suggesting the presence of more 
than one dust layer between about 200 and 800~pc. Wolf diagrams of the
region indicate extinction layers at 200, 300, and about 450~pc
\citep{Kun98}.  Radial velocity measurements of the interstellar
matter suggest the presence of several gas layers
\citep[e.g.][]{Heiles,YDMOF}.  \citet{Olano} modeled the space
distribution and kinematics of the interstellar neutral and molecular
gas by an expanding shell, centered on $(l,b)\approx
(120\degr,+17\degr)$, at a distance of about 300~pc. The presence of
the {\em Cepheus flare Shell\/} (CFS), an old supernova remnant,
readily explains the divergent cloud distances and the wide range of
radial velocities. The giant radio continuum feature {\em Loop~III}
\citep{Berk73, Spoelstra}, conspicuous in the {\it WMAP\/} K-band
polarization map \citep{Page07}, is nearly concentric with the
molecular shell. Several giant far-infrared loops, possibly old
supernova shells, have been identified in the IRAS images of the
Cepheus flare region \citep*{KMT}.

Signposts of low-mass star formation in the Cepheus flare region have
been detected in the molecular clouds L\,1082 (B\,150, GF\,9),
L\,1148/L\,1158, L\,1172/L1174 (associated with the young cluster
NGC\,7023), L\,1228, L\,1177 (CB 230), L\,1219 (B\,175), L\,1221,
L\,1251, and L\,1261 (CB~244) \citep[see review of ][]{Kun08}.
Recently, \citet{Kirk09} presented a list of 133 young stellar objects
and candidates, based on Spitzer observations of L\,1148/L\,1158,
L\,1172/L1174, L\,1228, L\,1221, and L\,1251. Independent distance
estimations of individual clouds suggest various distances between 150
and 500~pc, but are clustered around 300\,pc. The young open cluster
NGC\,7129, associated with the dark cloud L\,1181, is projected on the
Cepheus flare region at ($l$,$b$)=(105.4,+9.9), but is more distant
than the bulk of clouds (800--1250~pc, see \citet{Kun08}. The central region of
the complex is lacking signatures of active star formation. The
physical connection between the star-forming clouds remains rather
uncertain.

Scarce information is available on the pre-main-sequence (PMS) stellar
population of the region. \citet{HBC} list only 14 young objects in
the Cepheus flare region: RNO\,124 and PV Cep (associated with the
cloud complex L\,1148/L\,1158), HD\,200775, EH Cep, FU Cep, FV Cep
(members of NGC\,7023), Lk\ha~234, BD\,+65\degr1637, V350~Cep,
[SVS76]~NGC\,7129~6 (members of NGC\,7129), AS~507
(associated with L\,1261), and the apparently isolated Herbig Ae/Be
stars BH Cep, BO Cep, and SV Cep. Six further stars in NGC\,7129 
(MEG\,1--MEG\,3, [HL85]\,14: \citealt{MEG93}; GGD~33a: 
\citealt{MEFG94}, and V391~Cep: \citealt{Semkov93}) show T Tauri-like 
emission spectra, but no spectral types have been given for them in the
literature. \citet{Tachihara05} identified 16 weak-line T~Tauri stars
in the region, by spectroscopic and photometric follow-up studies of
X-ray emitting stars found in the ROSAT database. These stars are not
obviously associated with any dark cloud. Understanding star formation
in each dark cloud and relating star formation rates across the dark
clouds requires a comprehensive deep survey.

Here, we use optical and infrared observations to develop a better
picture of star formation in this region. To constrain the nature of
the candidate young stars, we combine new optical spectroscopy and
BVRI photometry with archival infrared data from IRAS, 2MASS, and
Spitzer.  We describe our observations and data reduction in
Sect.~\ref{Sect_obs}. We use these data to estimate ages,
luminosities, masses, and disk properties of individual young stars
and to analyze the distributions of these physical parameters among
each cloud. Section~\ref{Sect_res} contains the main results, namely
the HR diagrams and spectral energy distributions of the target stars.
We discuss the implications of our results for the star formation
history of the region in Sect.~\ref{Sect_disc}.

\section{Observations and data reduction}
\label{Sect_obs}

\subsection{Target selection}

Several lists of potential PMS stars, such as \ha\ emission stars
detected by objective prism surveys, irregular variables, far- and
mid-infrared sources can be found in the literature.  The targets of
our studies are the stars brighter than about $V=16$~mag from the
lists as follows: irregular variables associated with NGC\,7023
\citep{Rosino} and NGC\,7129 \citep{Semkov}, \ha\ emission stars
associated with L\,1228 \citep{OS90}, L\,1251 \citep{KP93}, NGC\,7129
\citep{MMN04}, and distributed over the whole area of the Cepheus
flare \citep{Kun98}, nebulous stars associated with L\,1251
\citep{ETM94}; IRAS point sources of T~Tauri-like flux ratios,
distributed over the whole region
\citep{Kun98}, near-infrared sources associated with L\,1251 \citep{RD95},
and Spitzer sources associated with L\,1228\,S \citep{Padgett}.
In addition we found four new $K_s$ band-excess stars in the 2MASS
\citep{Cutri} database, not coinciding with previously identified PMS
star candidates (2MASS\,21223461+6921142, 21225427+6921345,
21544662+6708255, and 22283821+7119023). Only one of them (2MASS
21225427+6921345) was sufficiently bright in the optical wavelengths
for our spectroscopic observations. In all, we observed spectra of 
170 stars, including 13 confirmed PMS stars of the region. 
Spectral types for some of these latter stars were
presented by \citet{CK79} and \citet{Mora}, however no previous
studies have presented spectroscopy of the red part of the optical
spectrum nor $BVR_\mathrm{C}I_\mathrm{C}$ photometry for these stars.

The accuracy of the target coordinates given in the source papers and
in the SIMBAD database is typically several arcseconds. Therefore we
used the published finding charts of the objects with the DSS to
refine the coordinates, and then used these improved coordinates to
find the associated 2MASS source (usually with positional accuracy of
$<$0\farcs2).

\subsection{Spectroscopy}

\subsubsection{CAFOS spectra}

Spectra of 155 stars were observed using the CAFOS instrument on the
2.2-m telescope of Calar Alto Observatory between 1999 August
7--11, 2004 August 10--15, 2005 September 11--14, and 2008 August
26--31. At some target positions two stars can be found within a few
arcseconds. We obtained spectra of both stars provided both were
sufficiently bright.  During the observing run in 1999 we used the
G--100 grism, covering the wavelength range
5000--7000\,\AA. Observations in 2004--2008 were performed through the
R--100 grism, covering the wavelength interval 5800--9000\,\AA. The
spectral resolution of CAFOS observation, using a 1.5-arcsec slit, was
$\lambda / \Delta \lambda
\approx 3500$ at $\lambda=6600$\,\AA. We estimated the exposure time
required for obtaining $S / N \approx 50$ at 6600\,\AA\ based on the
available optical magnitudes of the target stars. The spectrum of a
He--Ne--Rb lamp was regularly observed for wavelength calibration.  We
reduced and analysed the spectra using standard IRAF\footnote{IRAF is
distributed by the National Optical Astronomy Observatories, which are
operated by the Association of Universities for the Research in
Astronomy, Inc., under cooperative agreement with the National Science
Foundation. http://iraf.noao.edu/} routines.

The wavelength range of spectra taken with the R--100 grism 
was suitable for determining several
flux ratios defined for spectral classification by
\citet{Kirkpatrick} ({\sl A, B, C, B\,/\,A, B\,/\,C\/}), \citet{MK}
({\sl I$_2$, I$_3$\/}), and \citet*{PGZ} ({\sl T1, T2\/}). We measured
these spectral features on the spectra of our stars, and calibrated
them against the spectral type and luminosity class by measuring them
in a series of standard stars published by \citet{LeBorgne}. 
Each flux ratio resulted in an estimate of spectral type, and we
adopted as final spectral type the average of those inferred from the
individual flux ratios. The accuracy of our spectral classification,
estimated from the range of spectral types obtained from different
flux ratios, is $\pm1$ subclass. We have not tried to correct the
classification for the veiling.  The above spectral indices can be
used for classification of stars later than about K5. The brighter,
and thus possibly earlier type, stars of the region were therefore
observed with other instruments, covering shorter wavelengths.  To
classify K-type stars in the CAFOS+R--100 sample we used the
\ion{Ca}{1} lines at 6122 and 6161~\AA. Spectral classification
criteria for the spectra obtained with G--100 grism included
the \ion{Mg}{1}, \ion{Na}{1}, and \ion{Ca}{1} lines. Preliminary results
of the classification of these stars were presented in
\citet{EK03}.

\subsubsection{FLWO spectra}

With the help of the FLWO staff (P. Berlind, M. Calkins) we acquired
low-resolution optical spectra of 46 target stars with FAST, a high 
throughput slit spectrograph mounted on the Fred L. Whipple Observatory 1.5-m
telescope on Mount Hopkins, Arizona \citep{Fabricant}.  In addition to
the candidate PMS stars, we observed the spectrum of VDB\,141, a star
illuminating a reflection nebula near L\,1177, whose spectral type is
not included in \citeauthor{Racine}'s \citeyearpar{Racine} list of
reflection nebulae. We used a 300~g~mm$^{-1}$ grating blazed at
4750~\AA, a 3\arcsec\ slit, and a thinned Loral 512 $\times$ 2688
CCD. These spectra cover 3800--7500~\AA\ at a resolution of $\sim$
6~\AA. We reduced and analysed the spectra with IRAF.  After trimming
the CCD frames at each end of the slit, we corrected for the bias
level and flat-field for each frame, applied an illumination
correction, and derived a full-wavelength solution from calibration
lamps acquired immediately after each exposure. The wavelength
solution for each frame has a probable error of $\pm$0.5--1.0\,\AA. To
construct final 1-D spectra, we extracted object and sky spectra using
the optimal extraction algorithm within APEXTRACT. Most spectra have
moderate signal-to-noise per pixel ($S/N \sim 30$).
Using the method described by \citet{BK02}, we detected \ha\ emission 
in the spectra of 22 stars. The spectral classification was performed
by visually comparing the spectra of MK standard stars with our
candidates following the procedure described by \citet{Jaschek87}.

\subsubsection{The final list of PMS stars}

\begin{figure*} \centerline{\includegraphics[width=16cm]{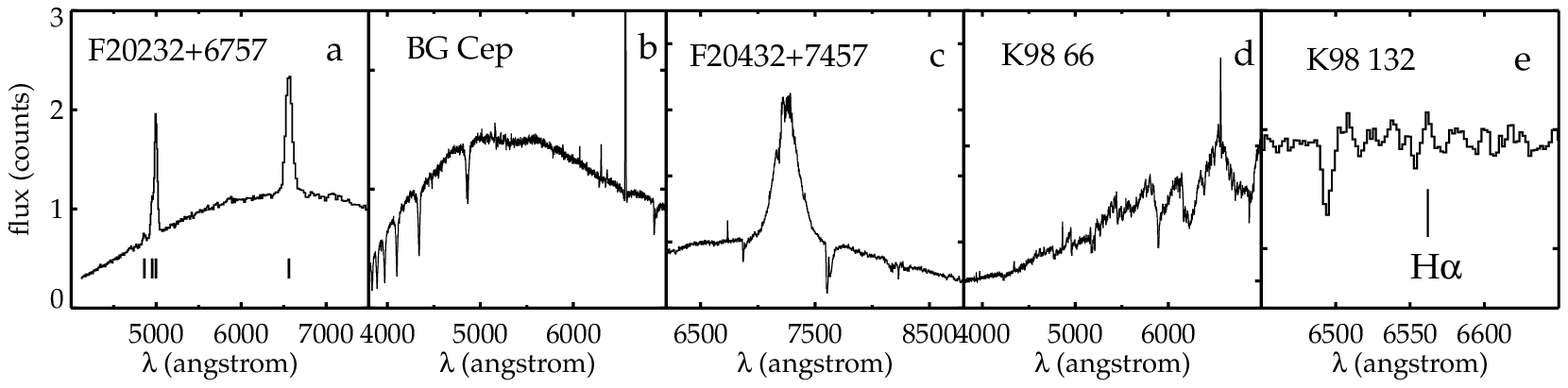}}
\caption{Examples of \ha\ emission stars not related to star
formation.
(a) Objective prism spectrum of the star [K98c]\,Em*\,4, associated
    with IRAS F20232+6757,
taken without red filter.  The strong \ion{O}{3} emission lines at
4953 and 4996\,\AA, whose positions are marked along with the \ha\ and
H$\beta$, show that this object is a planetary nebula. (b) FAST
spectrum of BG~Cep, a post-AGB star;
(c) CAFOS spectrum of IRAS~F20432+7457, which is proved to be a quasar 
(NED); (d) FAST spectrum of [K98c]\,Em*\,66, an M-type star with
Balmer and \ion{Ca}{2} H and K line emission but without detectable
lithium absorption; (e) part of the spectrum  around the \ha \ line of 
[K98c]\,Em*\,132, which proved to be a non-emission star.}
\label{Fig_other}
\end{figure*}

We accepted an object as a classical T~Tauri star if it fulfilled the
criterion established by \citet{Barrado}, i.e. EW(H$\alpha$) exceeded
the saturation limit of the chromospheric activity, depending on the
spectral type. The \ha\ emission of such stars is thought to originate
from the gas accreted by the star from the disk. Many of
our target stars claimed to be \ha\ emission stars based on objective prism
observations \citep{OS90,Kun98} proved to be M-type dwarfs with EW(\ha)
$< 10$\,\AA, evolved stars, or stars without \ha \ emission. Part of these 
latter stars may be PMS objects with variable \ha \ emission. Possibly such 
an object is K98c\,Em*\,132, a K-type star whose  \ha \ absorption 
line is filled in by emission (see Fig.~\ref{Fig_other}). 
We excluded these stars from the further study, unless
the LiI\,$\lambda$\,6708 line was conspicuous in their spectrum or
some evidence of youth (strong X-ray emission, infrared excess) could
be found in the literature. Strong \ion{Li}{1} absorption in low
resolution spectra alone is not sufficient criterion of the PMS
nature, given that blending with several neighboring lines may result
in a large apparent equivalent width. Therefore we regard the M-type
stars whose EW(\ha) $<$ 10\,\AA, and EW(\ion{Li}{1})$<$ 0.4\,\AA, {\em
candidate} weak-line T~Tauri stars (WTTS)
\citep[cf.][]{Tachihara05}. Earlier (G and K) spectral types
need lower EW(\ha) for the CTTS classification. Our input list
contains a few A- and F-type stars, whose candidate PMS status is based
on their IRAS associations and which display \ha\ in absorption, but
EW(\ha) is lower than the value characteristic of the spectral
type, suggesting that the photospheric absorption is partly filled by 
emission. We regard these stars candidate Herbig Ae stars. 
According to the above criteria, 77 of the observed 170 stars have 
shown spectra characteristic of pre-main sequence stars. These stars 
are listed in Table~\ref{TabList0}. The observed stars rejected as PMS 
objects are listed in Table~\ref{Tab_rej}.  We show examples of non-PMS
stars from the candidate list by \citet{Kun98} in Fig.~\ref{Fig_other}. 

\placefigure{Fig_other}

\placetable{TabList0}

\placetable{Tab_rej}

\begin{figure*} \centerline{\includegraphics[width=15cm]{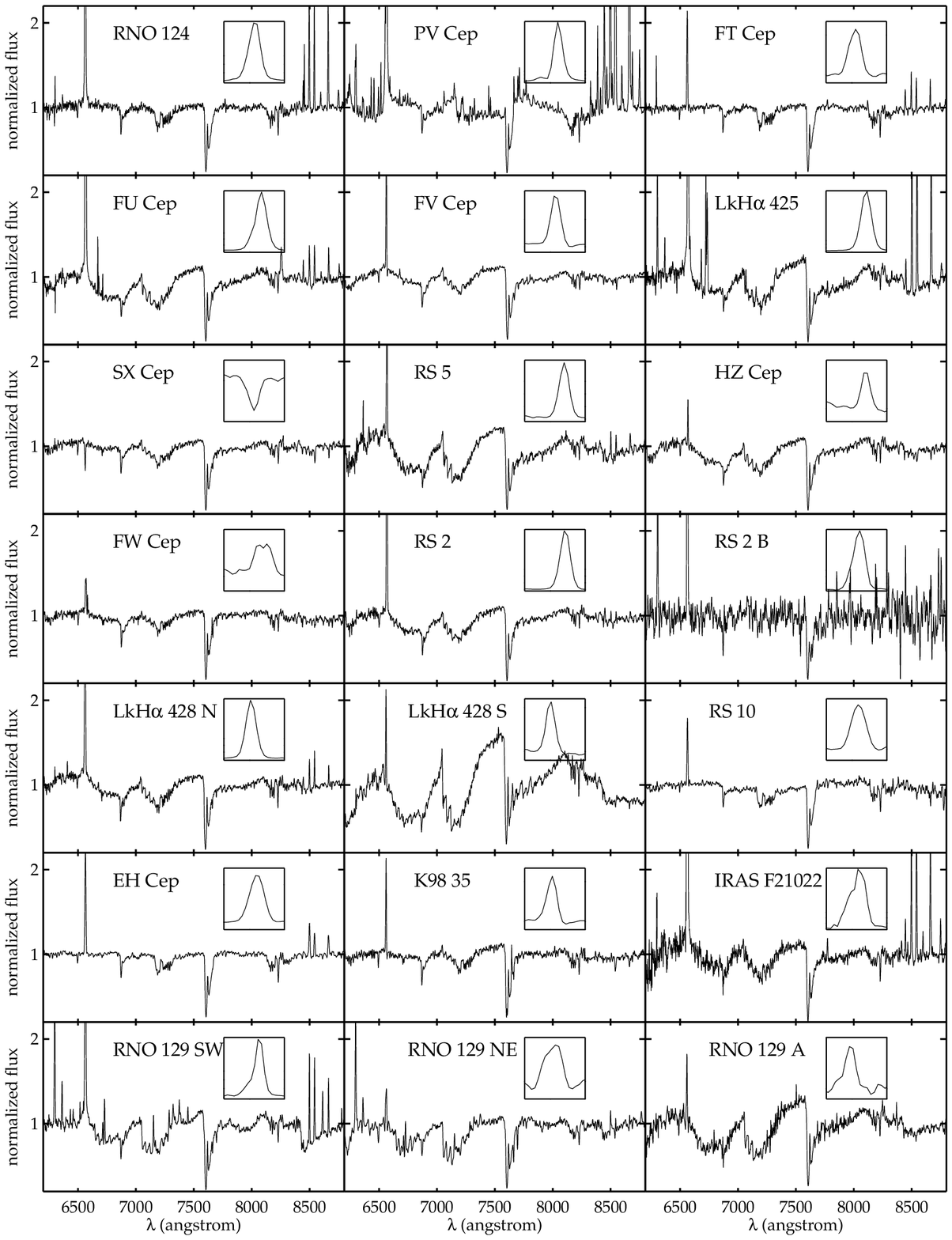}}
\caption{Spectra of pre-main sequence stars observed with CAFOS and
R--100 grism.}  \label{Fig_cafosr} \end{figure*}

\placefigure{Fig_cafosr}

\placefigure{Fig_cafosb}

\placefigure{Fig_cafosc}

\placefigure{Fig_cafosg}

\placefigure{Fig_FAST}

The spectra of the stars listed in Table~\ref{TabList0} are shown in
Figs.~\ref{Fig_cafosr}--\ref{Fig_FAST}. Figure~\ref{Fig_cafosr} shows
the spectra of the PMS stars, normalized to the continua, observed
with CAFOS+R--100. Fig.~\ref{Fig_cafosg} shows those observed with CAFOS+G--100. The
FAST spectra of stars classified as PMS objects, normalized to the
value of the continuum at 6560\,\AA, are shown in
Fig.~\ref{Fig_FAST}. The insets show the shape of the \ha\ line in
more detail, i.e. the section of the spectrum between 6545 and
6580\,\AA. The vertical size of the enlarged section is scaled to the
peak intensity of the line.

Results of the spectroscopic observations are presented in
Tables~\ref{Tab_sp0} and \ref{Tab_sp1}. The instruments of
observation, the derived spectral types, the type of the \ha\ emission
line shape, according to the scheme introduced by \citet*{RPL}, 
and a list of further emission lines identified in the
spectra are listed in Table~\ref{Tab_sp0}.  
We show in Table~\ref{Tab_sp1} the equivalent widths of the \ha,
\ion{Li}{1}, and  \ion{Ca}{2} infrared triplet lines in \AA, 
as well as $\Delta$v(10\%), the width of the \ha\  line 10\% above the continuum,
deconvolved by the response of the instrument, in km\,s$^{-1}$. 
Whereas \ion{Li}{1} is an important indicator of youth, EW(\ha), EW(\ion{Ca}{2}\,8542), and
$\Delta$v(10\%) are used as accretion indicators 
\citep{Muzerolle98,Natta04,Dahm08}. The equivalent widths were 
determined by fitting a Gaussian or multiple Gaussians to the lines using the
IRAF task `splot'.  The uncertainties, derived from repeated 
measurements, are about 10\%. The real uncertainties of the 
\ion{Li}{1} equivalent widths may be higher due to the blending of the line with 
neighboring \ion{Fe}{1} lines (at 6703.5, 6705.1, 6707.4, and 6710.3\,\AA). 
We used the `deblend' command of
the `splot' task of IRAF to separate the \ion{Li}{1} line
from the \ion{Ca}{1}\,$\lambda$\,6718 and [\ion{S}{2}]\,$\lambda$\,6717
lines. The last column of Table~\ref{Tab_sp1} shows the PMS type of the object,
based on our data. 

\placetable{Tab_sp0}

\placetable{Tab_sp1}

\subsection{$BVR_\mathrm{C}I_\mathrm{C}$ photometry}

Photometric observations in the $BVR_\mathrm{C}I_\mathrm{C}$ bands of
the young stars identified by our spectroscopic observations were
undertaken during one night in 2004 December, two nights in 2005 June,
and eight nights in 2006 September and October using the 1-m
RCC-telescope of the Konkoly Observatory. We used a Princeton
Instruments VersArray:1300B camera, that utilizes a back-illuminated,
1300$\times$1340 pixel Roper Scientific CCD. The pixel size is
20\,$\mu$m, corresponding to 0.31~arcsec on the sky.  Integration
times were between 30\,s and 600\,s. To calibrate the photometry we
observed the open cluster NGC\,7790 several times each night, at
various airmasses, as well as NGC\,188 once each night. We reduced the
images in IRAF. After bias subtraction and flatfield correction
PSF-photometry was performed using the `daophot' package. The field of
view, 6.7$\times$6.9~arcmin, contained several stars suitable for
establishing the point spread function of the images. The
transformation formulae between the instrumental and standard
magnitudes and color indices as a function of the airmass were
established each night by measuring the instrumental magnitudes of
some 80 photometric standard stars published by \citet{Stetson} for
NGC\,7790 and 40 stars in NGC\,188. Each target was observed at least
twice, on two different nights, and several stars on three or more
nights. The photometric errors, obtained as quadratic sums of the
formal errors of the instrumental magnitudes and those of the
coefficients of the transformation equations are about $\pm0.02$~mag
for stars brighter than about 16th~mag, and about $\pm0.05$~mag for the
fainter stars.

The results of the photometry are presented in
Table~\ref{Tab_phot}. The magnitudes listed are averages of 2--4
independent measurements. In several cases magnitude differences well
above the formal errors, indicative of variability, were detected. We
show in parentheses the mean absolute deviation of the magnitudes, and
indicate the variability in Col.~6.  During the psf photometry faint
visual companions were revealed near some objects. We designate these
stars in Table~\ref{Tab_phot} by adding a letter to the name of the
primary, and give the angular distance from the primary and the
position angle in Col.~6 of
Table~\ref{Tab_phot}. Figure~\ref{Fig_doubles} shows the $I_C$ images
of the visual double stars observed.
 
\placetable{Tab_phot}

\subsection{Infrared archive data}

We supplemented our observational data with the near infrared data of
the 2MASS All Sky Catalog \citep{Cutri} and the far-infrared data of
the IRAS PSC \citep{PSC} and FSC \citep{FSC}. One of the target stars,
Lk\ha\ 425, is not included in the 2MASS point source catalog. A
nonstellar object (artifact?) appears in the 2MASS images next to this
star. We took JHK photometric data for this star from \citet{Sellgren}. 

Spitzer IRAC and MIPS fluxes are available for parts of L\,1228 and
L\,1251, as well as for RNO\,124 and RNO\,129 in the c2d data archive
\citep{Evans}\footnote{http://data.spitzer.caltech.edu/popular/c2d/20071101\_enhanced\_v1/}.
NGC\,7129 was observed by IRAC and MIPS as part of the Spitzer young
stellar cluster survey \citep{Guter09}. Four cluster
members included in the Spitzer survey are common with our
targets. The magnitudes of these stars are listed in Table~4 of
\citet{Guter09}. IRAC and MIPS photometric data for NGC\,7023 are
presented in \citet{Kirk09}. All of our program stars in NGC\,7023 but
SX~Cep are included in their Table~6. The identifiers of our
stars in both papers are shown in Col.~4 of Table~\ref{TabList0}.

To obtain IRAC and MIPS 24 $\mu$m photometric data for SX Cep we
downloaded the basic calibrated data (BCD) frames processed with the
SSC IRAC Pipeline v14.0 for Prog ID 30574 and created mosaics from the
BCD frames using a custom IDL program. For details see
\citet{Guter08}. Aperture photometry was carried out using PhotVis
version 1.10 which is an IDL-GUI based photometry visualization
tool. See \citet{Guter04} for further details on PhotVis. The radii of
the source aperture, and of the inner and outer boundaries of the sky
annulus were 2.4, 2.4 and 7.2 arc-second respectively. For MIPS the
frames were processed using the MIPS Data Analysis Tool
\citep{Gordon05}. PSF fitting in the IRAF/DAOPHOT package was used to
obtain the following photometry for SX Cep: $[3.6]=8.91\pm0.01$,
$[4.5]=8.65\pm0.02$, $[5.8]=8.24\pm0.06$, $[8.0]=7.66\pm0.04$,
$[24]=5.85\pm0.07$.

\section{Results}
\label{Sect_res}

\subsection{Visual binaries, spectroscopic and photometric variability}
\label{Sect_bin}

Figure~\ref{Fig_doubles} shows the $I_\mathrm{C}$ images of the visual
double stars detected during our observations. Photometry was
performed separately for the components, whereas we observed only the
spectrum of the brighter component of FT~Cep, NGC~7023~RS\,5, OSHA~50,
[K98c]\,Em*~46, NGC~7129~S\,V3, [K98c]\,Em*~95, and AS\,507. The
uncertainty ellipse of IRAS~22152+6947 contained two nearly equally
bright stars separated by 4\farcs4.  The eastern component is a G2
dwarf, while the western component is an F5 star with weak \ha\
emission. We include the F5 star in our final list of PMS stars. Our
observations find RNO~129, OSHA~48, Lk\ha\ 428, and [K98c]\,Em*~119 to
be new binary T Tauri stars. 
 
\placefigure{Fig_doubles}

Variability of several stars has already been established
\citep[e.g.][]{Rosino,SY89,Grankin}. We observed strong night-to-night
variation of FT~Cep and most of the stars in L\,1251. Spectra of some
stars were observed at more than one epoch, and significant variations
were found in the equivalent width of the \ha\ line in some
cases. Figure~\ref{Fig_havar} shows the three most conspicuous cases.

\placefigure{Fig_havar}

\subsection{Surface distribution of the PMS stars}
\label{Sect_surfmap}

Figure~\ref{Fig_map} shows the distribution of the PMS stars
overplotted on the extinction map of the region published by
\citet{DUK}. In addition to our target stars, known Class~I objects,
Herbig~Ae/Be stars and WTTSs are also indicated.
Dotted lines indicate the nominal boundaries of the Cepheus flare
Shell \citep{Olano}, a section of Loop~III, and the far-infrared loop
G\,109+11 \citep{KMT}. The dashed line shows a section of the Cepheus
Bubble, located at some 800~pc, and possibly associated with NGC\,7129
\citep{ABK00}. For better visibility, parts of the map centered on
NGC\,7023, L\,1228, L\,1251, and NGC\,7129 are shown enlarged in
Figs.~\ref{Fig_l1174map}--\ref{Fig_n7129map}.

The clusters of young stars are concentrated near the edges of the
cloud complex, and avoid the central part of the region. All known
star forming clouds but L\,1221 are associated with optically visible
young stars. L\,1172/L\,1174 displays head--tail structure, with the
optically visible cluster in the head, indicative of star formation
propagating from higher towards lower Galactic longitudes. The 13
stars associated with L\,1228 (Fig.~\ref{Fig_l1228map}) apparently
belong to three small aggregates: L\,1228\,S \citep{Padgett},
L\,1228\,A \citep{Bally95}, and RNO\,129 \citep{MM04}. Most of the
young stars of L\,1251 form two small clusters, associated with two
dense C$^{18}$O cores, L\,1251~C and L\,1251~E \citep{SMN94}. On-going
star formation has been observed in both cores
\citep[e.g.][]{Beltran01,Reipurth04,Lee06,Lee07,Kirk09}. The strong
X-ray source [KP93]~2-43 \citep{Simon} is projected on a lower density
part of the cloud between the two cores. The cometary shape of
L\,1251, and the presence of three PMS stars on its high longitude
side suggest star formation propagating from the higher towards the
lower Galactic longitudes. Two CTTSs and a WTTS are projected near
L\,1261.  The extinction map of the NGC\,7129 region, displayed in
Fig.~\ref{Fig_n7129map} shows that its associated cloud, L\,1181 contains
two star-forming clumps. The cluster is embedded in the low-latitude
clump, whereas V391~Cep and four newly confirmed CTTSs
(NGC\,7129~S\,V1--V3, and [K98c]\,Em*~72) are associated with another
clump at the high-latitude part of the cloud.

In addition to the clusters associated with known star-forming clouds, PMS 
stars can be found widely scattered at the low-latitude side of the region, between  
L\,1177 and L\,1219, apparently along the periphery and inside of the shell G\,109+11.
A small group can be seen close to L\,1235 and SV~Cep. The A1 type \ha\ emission star 
coinciding with IRAS~22219+7901 (TYC 4608-2063-1), and a group of WTTSs are projected
on the cleared central region of the CFS.

\placefigure{Fig_map}

\placefigure{Fig_l1174map}

\placefigure{Fig_l1228map}

\placefigure{Fig_l1251map}

\placefigure{Fig_n7129map}

\subsection{Line-of-sight distribution of the PMS stars}
\label{Sect_dist}

Various distance values can be found in the literature for the star
forming clouds of the Cepheus flare. A comprehensive table of
published distance estimates is presented in Table 3 of \citet{Kun08}.
We adopt the distances for the clouds and the associated PMS stars
listed in Table~\ref{Tab_dist}.

\placetable{Tab_dist}

The distances of PMS stars far from the known star forming clouds require further
consideration. A small aggregate can be found around $l \approx 106\degr, b
\approx+13.6\degr$, close to the star forming cloud L\,1177.
The members of this group are [K98c]\,Em*\,53, [K98c]\,Em*\,58,
2MASS 21225427+6921345, the HAe star BD\,+68\degr1118, and a candidate 
PMS star, 2MASS 21223461+6921142.
We estimated the distance of this aggregate by the assumption that
BD\,+68\degr1118 is a member and lies on the ZAMS. This assumption
results in a distance of 390~pc. The star BD\,+67\degr1300 (VDB 141)
illuminates part of the L\,1177 dark cloud (see
Fig.~\ref{Fig_VDB141}), and provides us another distance estimate.
Our FAST spectrum for BD\,+67\degr1300 indicates a spectral type of
K0III. The absolute magnitudes of K0 giants, tabulated by
\citet{Wainscoat}, together with the photometric data of
BD\,+67\degr1300 and its associated 2MASS source, found in the
NOMAD\footnote{http://www.nofs.navy.mil/nomad} catalogue, result in a
distance $D \approx 370$\,pc.  The large dispersion of the absolute
magnitudes of K~giants, $\sigma_\mathrm{M}=0.7$ \citep{Wainscoat},
corresponds to a 1$\sigma$ range of distances of $270\,\mathrm{pc}
\la D \la 515\,\mathrm{pc}$. 
We derive the luminosities of the stars near L\,1177 using a 
distance of 390~pc. This value is close to the distance of L\,1219/B\,175,
surrounded by a group of PMS stars (BH~Cep, BO~Cep, IRAS~22129+6949, 
IRAS~F22144+6923, IRAS~22152+6947, [K98c]\,Em*~119, [TNK2005]~37 \citep{Tachihara05}).  
We assume that these stars are associated with L\,1219 at $\sim$400\,pc.
The candidate PMS star 2MASS 22283821+7119023 
is probably a further member of this group.
Both L\,1177 and L\,1219 are bordered by a high density
rim on the low-latitude side \citep[cf.][see also Fig.~\ref{Fig_VDB141}]{BR01}, 
indicative of shocks propagating from the direction of low Galactic latitudes.
Their similar morphologies and distances suggest that both clouds and the associated 
star formation may be influenced by the large Galactic shell G\,109+11 \citep{KMT}.
Therefore we refer to this group of PMS stars as the {\em G\,109+11 association}.

Another small aggregate ([K98c]\,Em*~95, 108, and 109) can be seen
projected near the HAe star SV~Cep, and at the same time close to the
head of the starless dark cloud L\,1235. Whereas L\,1235 is situated at 300\,pc
\citep{Kun08}, SV~Cep is much more distant  (715~pc: \citealt{Friede92};
$700\pm70$~pc: \citealt{KVSz}; $596^{+106}_{-43}$~pc: \citealt{Montes09}).  
Plotting these stars in the HRD with a distance of 300\,pc,
we obtain ages between 10 and 20 million years, inconsistent with
typical CTTSs. Assuming that the stars are as distant 
as SV~Cep (600\,pc), we estimate ages of 1--5 Myr, similar to
that of SV~Cep \citep{Montes09}. Therefore we assume that these stars 
formed in the same cloud as SV~Cep.

We assume that the apparently isolated stars IRAS 20535+7439 and RNO~135
belong to the cloud complex at 300~pc, and [K98c]\,Em*~73 is a member
of the G\,109+11 association at 400-pc. We have no guideline for finding the distance
and luminosity of IRAS~22219+7901. If we assume that it is a 
main sequence star, we estimate a distance of 1.3\,kpc. 

\subsection{Interstellar extinction and luminosities}

We derived the interstellar extinction suffered by the program stars
from their $E_{R_\mathrm{C}-I_\mathrm{C}}$ and $E_{I_\mathrm{C}-J}$
color excesses. The unreddened color indices of the spectral types
were adopted from \citet{KH95}. We determined $A_\mathrm{V}$ from both
color excesses independently, using \citeauthor{Cardelli}'s
\citeyearpar{Cardelli} extinction law with $R_V=3.1$, and then
accepted the average of both values as the visual extinction of the
star. For visual binaries that are resolved in our optical images, 
but unresolved in 2MASS, we used only $E_{R_\mathrm{C}-I_\mathrm{C}}$ 
to determine $A_\mathrm{V}$. We corrected the observed magnitudes 
using the $A(\lambda)/A_\mathrm{V}$ relations according to the same 
extinction law. The extinction law with $R_V=3.1$ proved inadequate 
in the case of the most heavily reddened stars, above 
$A_\mathrm{V} \approx 4$~mag: the unreddened
SEDs of these stars over the {\it V\/}--{\it J\/} region poorly fitted
the shape of the photospheric SED. Therefore we rederived the
extinctions of these stars using $R_V=5.5$, using the relations
$A(\lambda)/A_\mathrm{V}$ given by
\citet{Cardelli}. $R_V=3.1$ gave definitely better results for most of
the lightly reddened stars: for instance, in L\,1251 we used $R_V=5.5$
for [KP93]~2-1, 46, [KP93]~3-10, [RD95]~A, [RD95]~B, and
[ETM94]~Star~3, and $R_V=3.1$ for the other members. Exceptions are 
the apparently lightly reddened stars associated with small 
reflection nebulae or exhibiting Class~I SEDs, namely 
NGC~7023~RS~10, [K98c]\,Em*58, and [K98c]\,Em*119. The apparently low 
reddening of these stars results from the contribution of scattered light to their
color indices. We corrected their fluxes for the extinction using $R_V=5.5$.
The difference between the two $A_\mathrm{V}$ values, derived from 
$E_{R_\mathrm{C}-I_\mathrm{C}}$ and $E_{I_\mathrm{C}-J}$, allowed us
to estimate the uncertainty as $\pm0.5$~mag.
Bolometric magnitudes were derived by applying the bolometric
corrections, $BC_{I_\mathrm{C}}$, tabulated by
\citet{Hartigan}. Effective temperatures of the spectral types were
adopted from \citet{KH95}. Luminosities were derived using the
distances listed in Table~\ref{Tab_dist}. The $T_\mathrm{eff}$ and
$L_\mathrm{bol}$ values are listed in Table~\ref{Table_res}.

\subsection{Hertzsprung--Russell diagrams}
\label{Sect_hrd}

The Hertzsprung--Russell diagrams of the pre-main sequence stars are
plotted in Figs.~\ref{Fig_hrd1}--\ref{Fig_hrd5}, separately for
L\,1228, NGC\,7023, L\,1251, and NGC\,7129. The G\,109+11 association, the stars
associated with L\,1155/L\,1157, SV~Cep, and L\,1261, as well as the
isolated stars IRAS~20535+7439 and RNO~135 are plotted with
varied symbols in Fig.~\ref{Fig_hrd5}. Since we could not find absorption 
features suitable for spectral classification in the spectrum of PV~Cep, we accepted a
spectral type of K0, reported by \citet{MM01}.  Evolutionary tracks
and isochrones, as well as the position of the birthline and zero-age
main-sequence \citep{PS99} are also shown. Masses and ages, derived
from these diagrams, are listed in Table~\ref{Table_res}.

\placefigure{Fig_hrd1}

\placefigure{Fig_hrd2}

\placefigure{Fig_hrd3}

\placefigure{Fig_hrd4}

\placefigure{Fig_hrd5}

\subsection{Accretion signatures}
\label{Sect_acc}

Both the emission spectra and the shape of the SED carry information
on the mass accretion rate from the disk onto the star.
The accretion rates $\dot{M}_\mathrm{acc}$ of our PMS stars were
estimated from the \ha \ line width $\Delta v(\ha)$ using the formula
of \citet{Natta04}, and from the \ha \ and \ion{Ca}{2} luminosities 
using the formulae of \citet{Dahm08}.  Since our spectra are not flux
calibrated, we derived the emission line luminosities from the
equivalent widths, using the extinction-corrected $R_\mathrm{C}$ and
$I_\mathrm{C}$ magnitudes for determining the underlying stellar flux
for the \ha\ and \ion{Ca}{2} lines, respectively. In order to convert
the luminosity to accretion rate according to the relationship

\begin{equation}
L_\mathrm{acc} \approx \frac{G\dot{M}M_{*}}{R_{*}}\left(1-\frac{R_{*}}{R_0}\right)
\end{equation}

\noindent we took the stellar mass, $M_{*}$, and radius, $R_{*}$ from
the HRD, and adopted $R_{0} \approx 5\,R_{*}$ for the inner radius of
the gaseous disk \citep{Gullbring}.  The accretion rates,
$\dot{M}_\mathrm{acc}$, obtained from the \ha\ and 
\ion{Ca}{2}\,8542\,\AA \ lines are listed in
Table~\ref{Table_res}.

\placefigure{Fig_mdot1}

Figure~\ref{Fig_mdot1} shows the relationship between the accretion
rates, $\dot{M}_\mathrm{acc}$, resulted from $\Delta v(\ha)$ and
$L(\ha)$, respectively.  Given the empirical nature of both
relationships the large scatter is not surprising.  An additional
source of uncertainty of the results is that our spectroscopic and
photometric observations were obtained at different epochs.
 We notice that for stars exhibiting double-peaked \ha\ emission 
line accretion rates derived from $\Delta v(\ha)$ are systematically 
larger than those derived from $L(\ha)$.
Double-peaked \ha\ lines (Types II and III in
Table~\ref{Tab_sp0}) have among the strongest emission lines in the
sample. All but one ([K98c]\,Em*~109) belong to the types IIB or IIIB,
i.e. the absorption dip is on the blue side of the main peak,
indicating the contribution of wind to the line shape. A
reason for the discrepancy may be that, due to the absorption by the
wind, the peak of the \ha\ emission is lower than it would be without
the contribution of wind. Thus the effects of the wind are a smaller
equivalent width and, since the 10\% level of the peak is also lower,
a larger line width at this level. We also note that the \ha\ lines of
these stars are wider than those used by
\citet{Natta04} for calibrating the $\dot{M}_\mathrm{acc}~vs.~\Delta
v(H\alpha)$ relationship.  Accretion rates derived from $L(\ha)$  and 
 $L(8542)$ are consistent with each other
according to Fig.~\ref{Fig_haca}. We note, however, that there is a
large scatter in $\dot{M}$ from \ion{Ca}{2} at fixed $\dot{M}$ from
\ha, thus we cannot state that there is a 1:1 relation between
them.

\placefigure{Fig_haca}

We examined the dependence of accretion rate on stellar mass.
Figure~\ref{Fig_mdotm} shows the result. The accretion rates plotted
are averages of both methods derived from the \ha\ emission line.  The
line fitted to the data is $\log \dot{M}_\mathrm{acc} \propto 1.74
\times \log M_{*}$, compatible with similar relationships found for
other nearby star forming regions
\citep[$\log \dot{M}_\mathrm{acc} \propto 1.8 \times \log M_{*}$;][]{Natta06,Herczeg08}. 
The correlation, however, is rather weak: the Pearson correlation 
coefficient is 0.48.
The large scatter of this plot (exceeding three orders of magnitude at
a fixed $M_{*}$) partly originates from plotting together stars of
several regions, having different ages and accretion histories. For
instance, most of the (older) dispersed stars appear below the fitted
line, whereas the stars of L\,1251 show higher accretion rates than
the average. However, no clear dependence of $\dot{M}_\mathrm{acc}$ on
the stellar age can be found in our data set. As \citet{Tilling} 
pointed out, biases are hard to remove in quantifying the
$\dot{M}_\mathrm{acc}$ vs. $M_{*}$ relation.

\placetable{Table_res}

\placefigure{Fig_mdotm}

\subsection{Spectral energy distributions}
\label{Sect_sed}

The optical, near- and mid-infrared parts of the spectral energy
distributions (SEDs) of the programme stars were constructed using all
available photometric data, each corrected for the interstellar
extinction. The fluxes corresponding to zero magnitude were obtained
from \citet{Glass} for the $V R_\mathrm{C} I_\mathrm{C}$ bands, and
from the 2MASS All Sky Data release web
document\footnote{$\mathrm{http://www.ipac.caltech.edu/2mass/releases/allsky/doc/}\mathrm{sec6\_4a.html}$}
for the $J H K_\mathrm{s}$ bands. IRAS fluxes are also plotted when
available.  IRAS fluxes of 2MASS 21225427+6921345 were obtained by
Scanpi\footnote{http://scanpiops.ipac.caltech.edu:9000/applications/Scanpi/index.html}. The
resulting SEDs are shown in Figs.~\ref{Fig_sed1}--\ref{Fig_sed5},
separately for the clusters/aggregates of the region. Photospheric
SEDs have been drawn by dashed lines. Their values were determined
from the dereddened $I_{C}$ magnitudes, and from the color indices
corresponding to the spectral types. We also plotted for comparison
the median SED of the T~Tauri stars of the Taurus star forming region
\citep{DAlessio}, the best-studied sample population of 1--2 Myr~old low-mass
stars.

The shape of the SED allows us a qualitative classification of the
circumstellar environment of the young stars. According to the slope between 
2 and 24\,\micron, $\alpha_{24-K} = d\log(\lambda F_{\lambda)}/ d\log(\lambda)$, 
Class~I ($\alpha > 0.3$),
Flat ($0.3 > \alpha >  -0.3$), Class~II ($-0.3 > \alpha >  -1.6$),
and Class~III ($\alpha < -1.6$) YSOs are distinguished \citep{Lada,Evans09}. 
Class~I infrared sources are
probably among the youngest PMS stars, close to the end of the main 
accretion phase of their evolution. Their circumstellar environment is 
composed of a disk and an envelope. Class~II sources are CTTSs surrounded
by optically thick accretion disks, and the low infrared excesses of
Class~III stars originate from optically thin disks. 

Four stars, namely PV~Cep,
NGC~7023\,RS\,10, [K98c]\,Em*~58, and GGD\,33a exhibit SEDs
characteristic of Class~I objects. The luminosities of these stars,
derived from the optical observations, are very low, indicating that
we observe the light of these stars scattered from edge-on disk or
envelope. The underestimated luminosity is also reflected in the
discrepancy between the accretion rates obtained from $\Delta v(\ha)$
and $L(\ha)$.  PV~Cep and GGD\,33a are associated with Herbig--Haro
objects \citep*{Gomez97,EGM92}, suggesting that they are probably real
Class~I objects, embedded in massive envelopes. The small reflection 
nebula associated with NGC~7023\,RS\,10 may
also be an indicator of the envelope. The nature of
[K98c]\,Em*~58 is uncertain. It may either be an edge-on disk or an
unresolved binary system. The histogram of $\alpha_{24-K}$ for our 
stars is shown in the left panel of Fig.~\ref{Fig_dhist}.
Five percent of the studied stars belong to the Class~I, 13\% exhibit Flat SEDs,
60\% are Class~II, 3\%  Class~III, and 18\% (14 stars) remain unclassified 
due to the lack of both 24 and 25\,\micron\ fluxes. 

\citet{Cieza} introduced a two-dimensional way of classification of
SEDs, based on  $\lambda_\mathrm{turnoff}$, the longest wavelength at which 
the observed SED is photospheric, and $\alpha_\mathrm{excess}$, the slope of
the SED for $\lambda > \lambda_\mathrm{turnoff}$. 
$\lambda_\mathrm{turnoff} \lesssim 2$\,\micron\ may indicate accreting stars,
with the inner edge of the disk at the dust sublimation radius. SEDs 
with $\lambda_\mathrm{turnoff}$ in the IRAC wavelength range
may result from more evolved, purely irradiated disks with inner holes. 
The right panel of Fig.~\ref{Fig_dhist}
shows the histogram of $\lambda_\mathrm{turnoff}$ for our target stars. 
We found $\lambda_\mathrm{turnoff} \le 2\micron$ for some 45\% of the studied 
objects, while $\lambda_\mathrm{turnoff}$ cannot be determined for 22\%  due to 
the lack of IRAC data. These stars obviously  
have $\lambda_\mathrm{turnoff} > 2$\,\micron. Thus slightly less than 
half of our studied stars exhibit near-infrared excesses.
Remarkably SEDs of various shape \citep[cf.][]{Evans09} occur in each small cluster 
of the region. For instance, we find both Flat SEDs (KP93~2-2) and transitional/annular
disks (KP93 2-3) in L\,1251, and both Class~I (NGC 7023~RS\,10)   
and purely photospheric (HZ Cep) SEDs in NGC~7023.
On the contrary, the seven observed SEDs of L\,1228\,S are less diverse: 
none of them show NIR excess, indicating either evolved disks, or star forming 
conditions  different from those in the other parts of the region.

\placefigure{Fig_sed1}

\placefigure{Fig_sed2}

\placefigure{Fig_sed3}

\placefigure{Fig_sed4}

\placefigure{Fig_sed5}

\placefigure{Fig_dhist} 

\section{Discussion}
\label{Sect_disc}

\subsection{Distribution of stellar and disk properties}
\label{Sect_disc1}

Figure~\ref{Fig_hist} shows the histogram of the
effective temperatures and masses, derived from the HRD, and mass
accretion rates, derived from the \ha\ line (average from both
methods). It is apparent that due to the magnitude limit of our
observations spectral types later than about M2, and masses lower than
about 0.2\,M$_{\sun}$ are largely missing from our
sample. Comparison of the mass distribution with the shape of the
Salpeter IMF shows that the incompleteness of our sample starts at
around 0.5\,M$_{\sun}$. The median mass of the observed stars is 0.8\,M$_{\sun}$.
The median accretion rate for our sample, derived
from the \ha\ emission line, is $\dot{M}_\mathrm{acc} = 3.6 \times
10^{-8}$\,M$_{\sun}$/yr.  
\citet{Gullbring} measured a median accretion rate of $9.6\times
10^{-9}$\,M$_{\sun}$/yr for 17 CTTSs in Taurus, and \citet{Natta06}
measured a median accretion rate of $4.4\times 10^{-9}$\,M$_{\sun}$/yr
for 43 members of $\rho$~Oph (upper limits excluded). These figures
show that our target selection resulted in a limited interval of
effective temperatures and masses, and in the detection of the most
actively accreting members of the population. 
Median stellar parameters, as well as median $\alpha_{24-K}$ 
and $\lambda_\mathrm{turnoff}$ for various parts of the
Cepheus flare are summarized in Table~\ref{Tab_med}.

\placefigure{Fig_hist}

\placetable{Tab_med}
 
\subsection{A possible origin of star formation in the Cepheus flare}

The CFS/Loop~III system may represent a link between the apparently
distinct star forming clouds \citep{Olano,Kun07}. L\,1147/L\,1158 and
NGC\,7023 are located outside the CFS, but close to the periphery of
Loop~III. The giant, nearly concentric shells are suggestive that
multiple supernova explosions occurred inside the Cepheus void, that
is, the presence of OB stars there in the past few million
years. Looking for the remnants of such an association we find the
small R-association consisting of the classical Cepheid SU~Cas and a
small group of A and B type stars at a distance of 258\,pc
\citep{Turner84}. We have no sufficient data for deciding whether this 
small group is a remnant of a large OB association. The G\,109+11 association 
at 400-pc may result from a star formation event due
to the impact of the shell G\,109+11 6--7 million years ago. The
expansion of the shell towards higher latitudes might have triggered
star formation in L\,1177. The 50-Myr old association Cep~OB6 at a
distance of 300~pc and towards low-latitudes from the Cepheus flare
may be the source of the shell \citep{BR01}.

\section{Conclusions}

From spectroscopic and photometric observations, we estimated the
effective temperatures and luminosities of 78 PMS stars in the Cepheus
flare region, 65 of which are newly confirmed PMS stars. Comparing the
results with pre-main sequence evolutionary models we estimate stellar
masses of 0.2--2.4\,M$_{\sun}$ and stellar ages of 0.1--15~Myr. From
archival IRAS, 2MASS, and Spitzer data, we constructed the spectral
energy distributions; 5\% of the stars are Class~I, 5\% Flat, 81\%
Class~II, and 9\% Class~III.

Star forming clouds and PMS stars of the region belong to the complexes as follows. 
\begin{itemize}
\item Star formation associated with the large expanding system of Cepheus flare Shell/Loop~III 
occurs in the clouds L\,1148/L\,1158, L\,1172/L\,1174, L\,1251,
L\,1228 and L\,1261/L\,1262. The latter two clouds are located on the
approaching side of the expanding shell.
\item A sparse association consisting of 6--7~Myr old PMS stars at the low-latitude part 
of the region ($105\degr \lesssim l \lesssim 112\degr$, $10\degr\lesssim b \lesssim 14\degr$) can be found at a distance of 380--400\,pc. 
The  shapes of the clouds in this part of the region suggest shocks from the
direction of the Galactic plane. 
\item Low-mass PMS stars are associated with the Herbig~Ae star SV~Cep, located at $\sim$600\,pc. 
\item NGC\,7129, at $\sim$800\,pc from the Sun, is probably associated with the Cepheus Bubble,
created by the older subgroup of the association Cep~OB2. Stars have been formed
in at least two clumps of its natal cloud L\,1181.
\end{itemize}
The A1-type \ha\ emission star associated with IRAS~22219+7901 is probably more distant 
than the Cepheus flare clouds. Its nature remains uncertain.

\acknowledgements

Our results are partly based on observations obtained at the Centro
Astron\'omico Hispano Alem\'an (CAHA) at Calar Alto, operated jointly
by the Max-Planck-Institut f\"ur Astronomie and the Instituto de
Astrof\'{\i}sica de Andaluc\'{\i}a (CSIC). We thank Calar Alto
Observatory for allocation of director's discretionary time to this
programme. This publication makes use of data products from the Two
Micron All Sky Survey, which is a joint project of the University of
Massachusetts and the Infrared Processing and Analysis
Center/California Institute of Technology, funded by the National
Aeronautics and Space Administration and the National Science
Foundation.  This work makes use of observations made with the
\textit{Spitzer Space Telescope}, which is operated by the Jet
Propulsion Laboratory, California Institute of Technology under a
contract with NASA. Our observations were supported by the
OPTICON. OPTICON has received research funding from the European
Community's Sixth Framework Programme under contract number
RII3-CT-001566. This research was supported by the Hungarian OTKA
grant T49082. Support for Z. Balog was provided by NASA through
contract 1255094, issued by JPL/Caltech. We are grateful to L\'aszl\'o
Szabados for careful reading of the manuscript.

\clearpage

\clearpage

\clearpage

\addtocounter{figure}{-1}
\begin{figure*}
\centerline{\includegraphics[width=15cm]{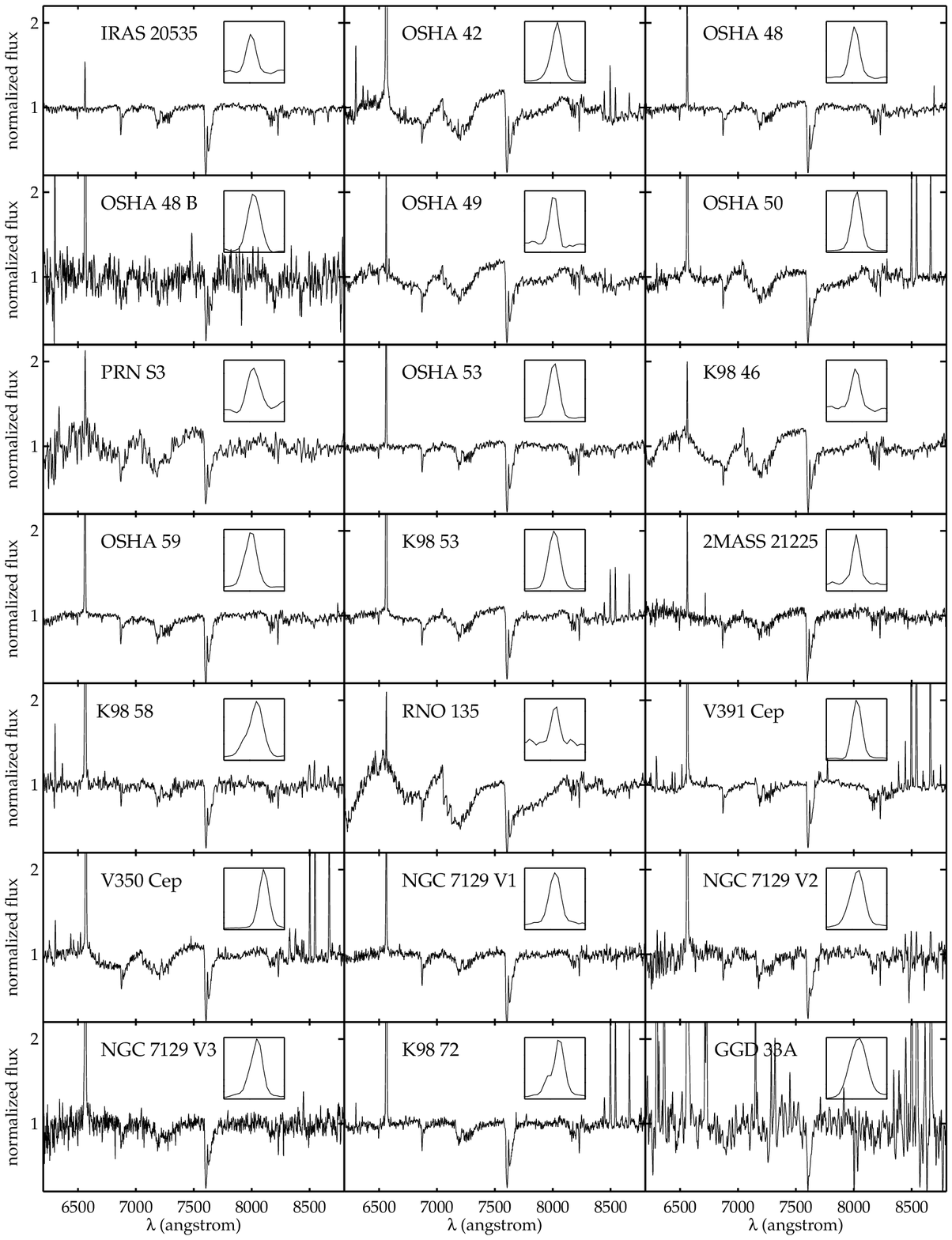}}
\caption{Spectra of pre-main sequence stars observed with CAFOS and R--100 grism
(continued).}
\label{Fig_cafosb}
\end{figure*}

\clearpage

\addtocounter{figure}{-1} \begin{figure*}
\centerline{\includegraphics[width=15cm]{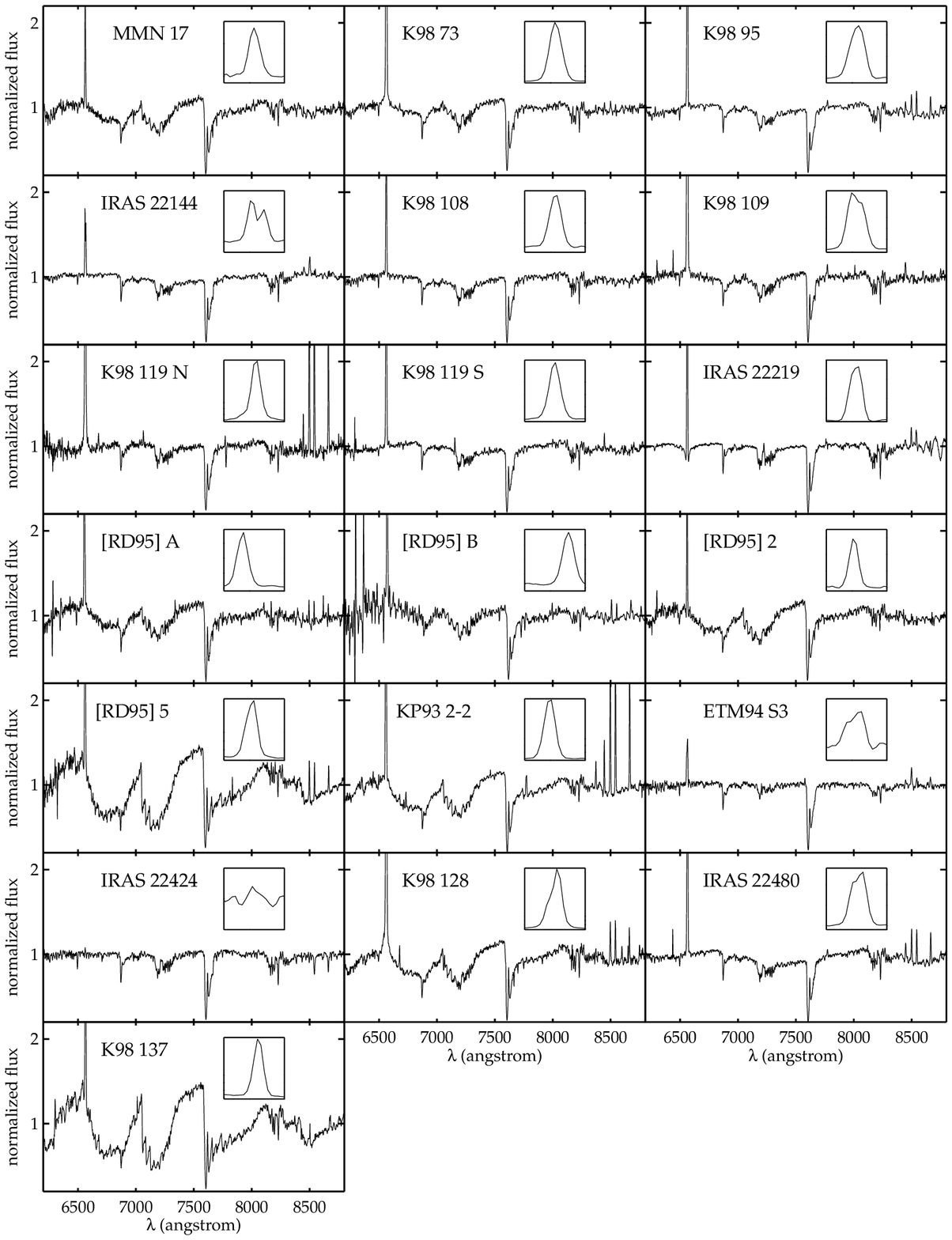}} \caption{Spectra
of pre-main sequence stars observed with CAFOS and R--100 grism
(continued).}  \label{Fig_cafosc} \end{figure*}

\clearpage

\begin{figure*}[!ht]
\centerline{\includegraphics[width=15cm]{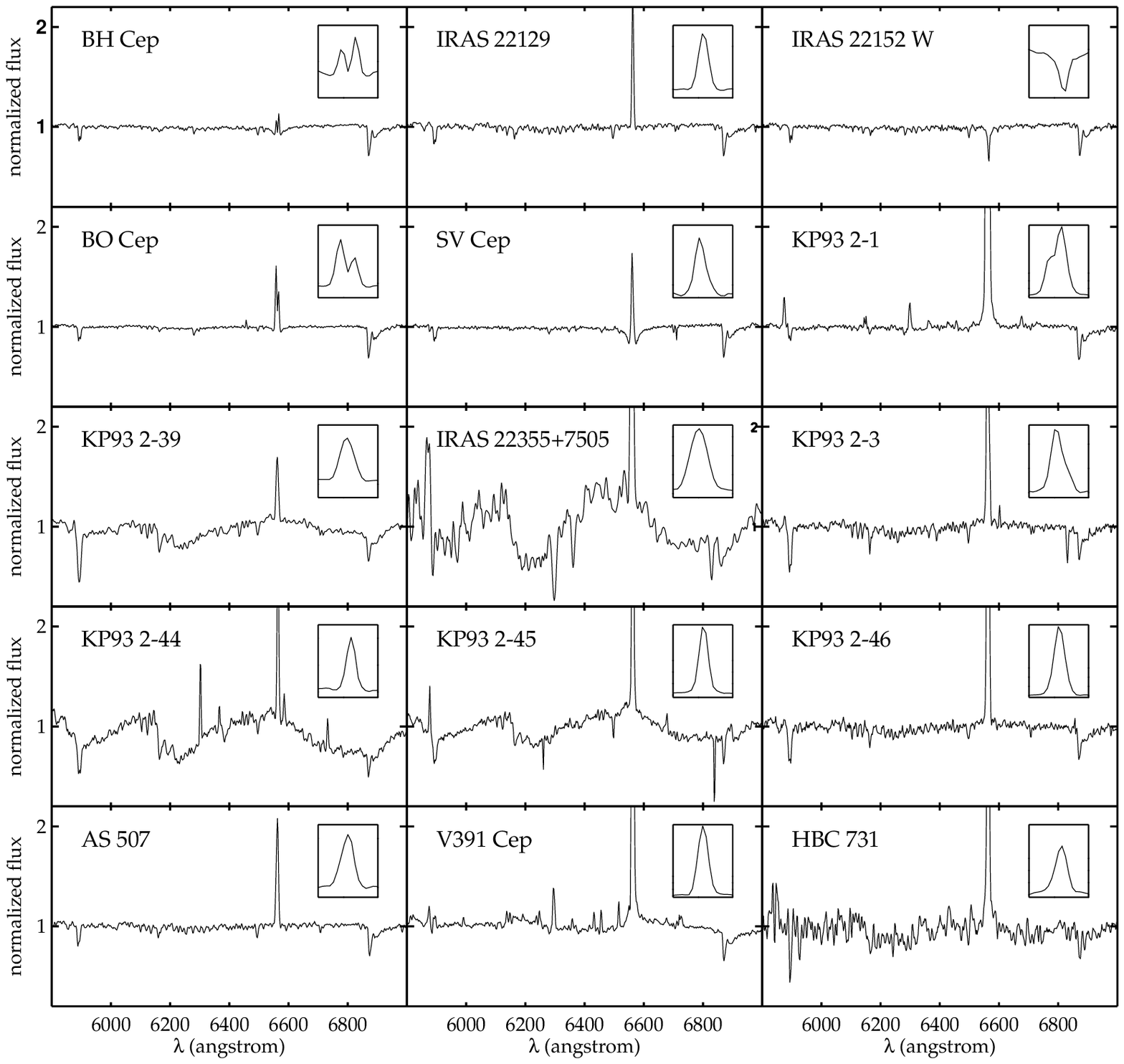}} \caption{Spectra of
pre-main sequence stars observed with CAFOS and G--100 grism.}
\label{Fig_cafosg} \end{figure*}

\clearpage

\begin{figure*}
\centerline{\includegraphics[width=15cm]{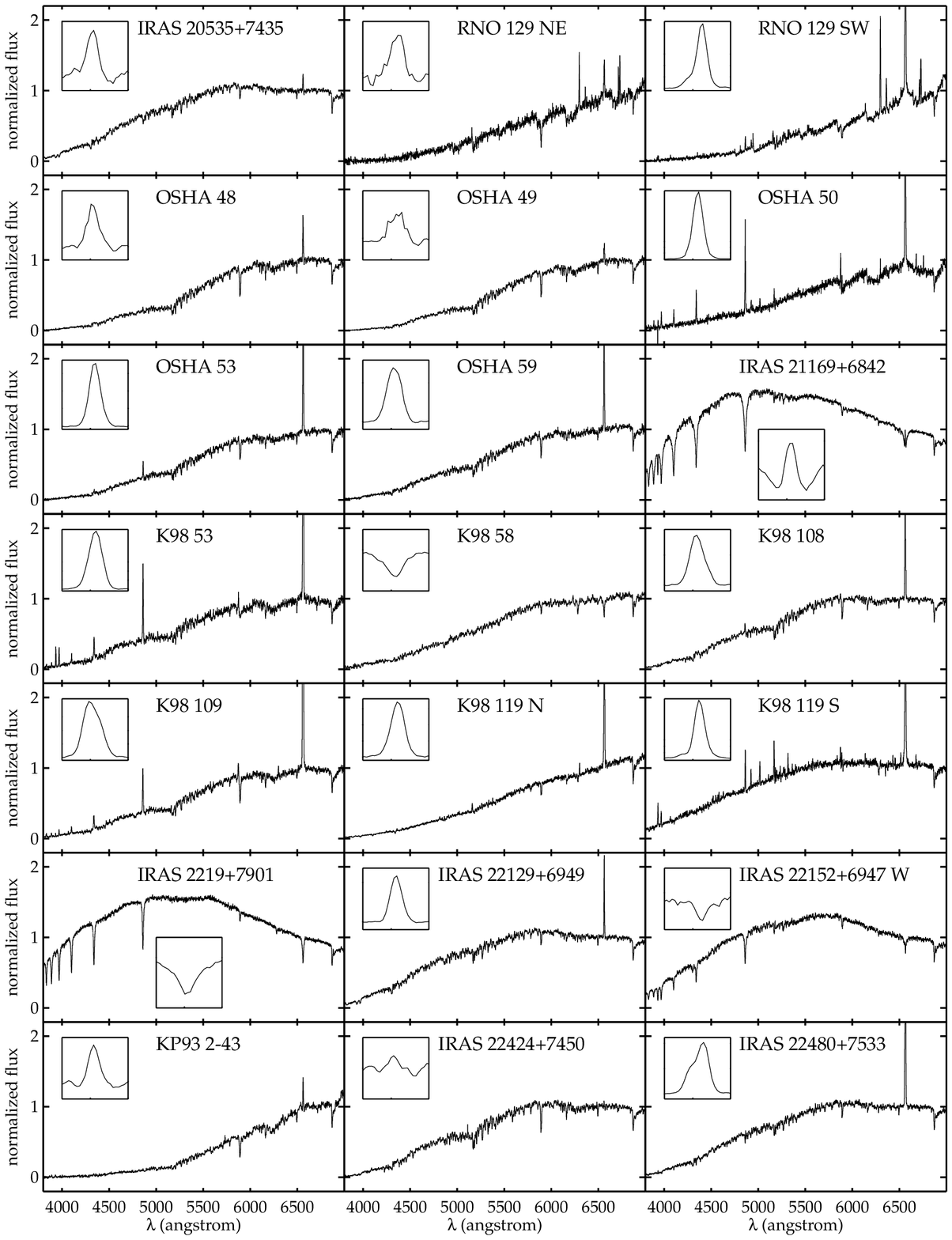}}
\vskip-1cm
\caption{Spectra of pre-main sequence stars observed with FAST.}
\label{Fig_FAST}
\end{figure*}

\clearpage

\begin{figure*}
\centerline{\includegraphics[width=15cm]{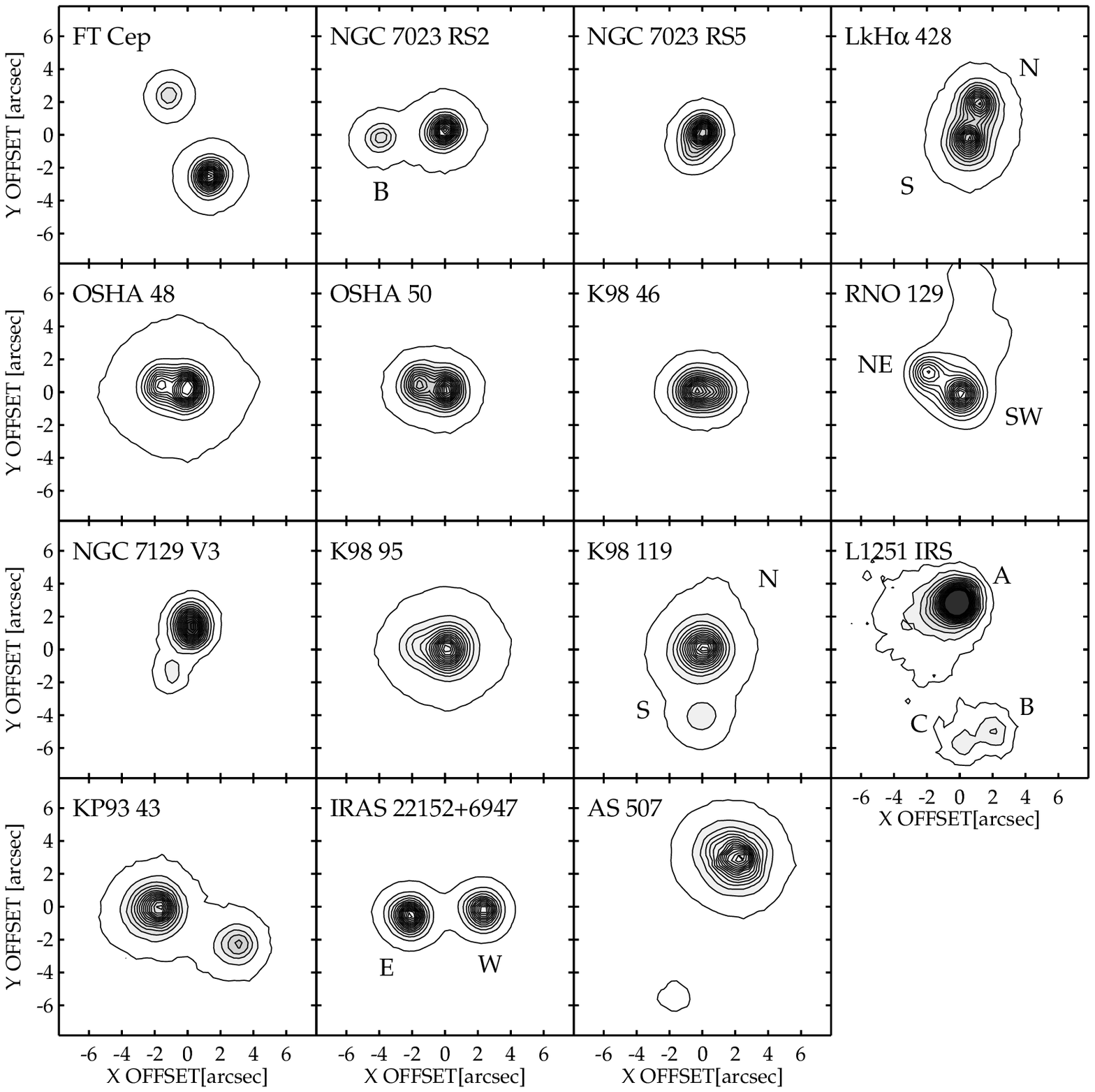}}
\caption{Visual double stars of our sample, consisting of components
separated by $\sim 1-10\arcsec$. $15.6\arcsec \times15.6\arcsec$
($51\times51$ pixels) parts of $I_\mathrm{C}$ images are plotted.}
\label{Fig_doubles}
\end{figure*}

\clearpage

\begin{figure}[!ht]
\centerline{\includegraphics[width=12cm]{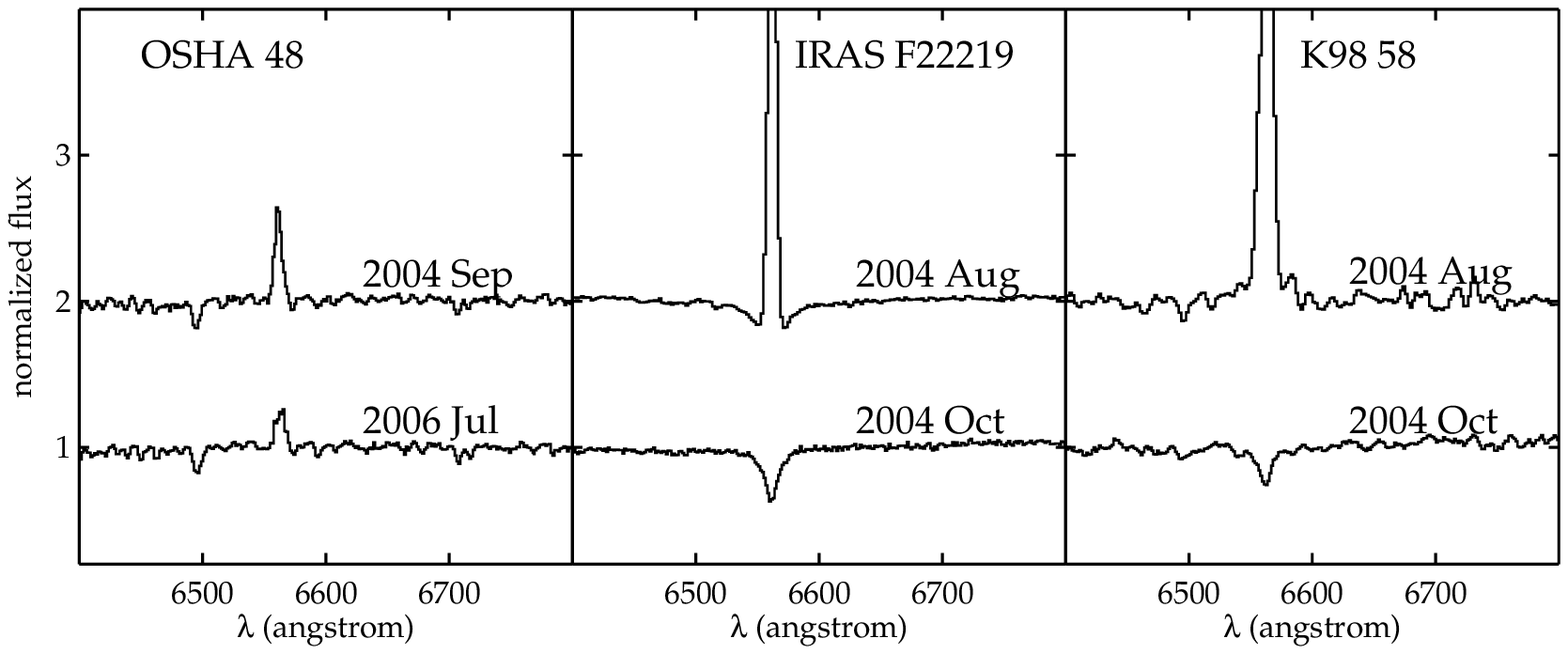}}
\caption{Spectra displaying variable H$\alpha$ line.}
\label{Fig_havar}
\end{figure}

\clearpage

\begin{figure*}[!ht]
\centerline{\includegraphics{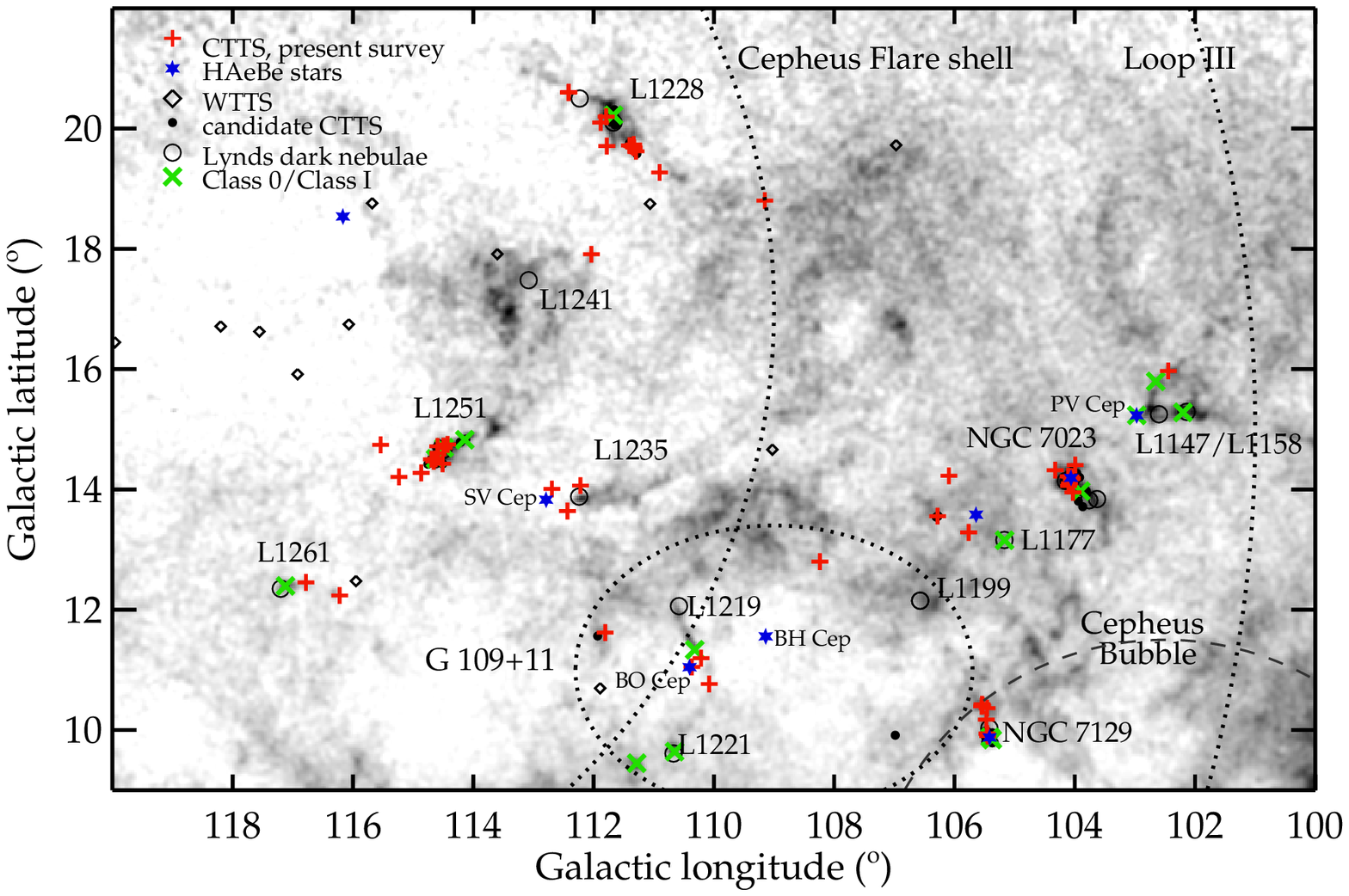}}
\caption{Distribution of the PMS stars overplotted on the 
extinction map of the region \citep{DUK}. Plusses (red in the
electronic version) indicate the CTTSs studied in this paper, and diamonds
are WTTSs identified by \citet{Tachihara05}. Star symbols show Herbig
Ae/Be stars, and black dots are infrared-excess stars based on their
near (2MASS) or mid-infrared (Spitzer IRAC) color indices.  Crosses
(green in the electronic version) show the known Class~0/Class~I type
objects of the region, indicative of active low-mass star
formation. Dotted lines indicate the nearby giant loop structures
Loop~III, Cepheus flare Shell, and G\,109+11, and dashed line shows the
more distant Cepheus Bubble.}
\label{Fig_map}
\end{figure*}

\clearpage

\begin{figure} 
\centerline{\includegraphics[width=8cm]{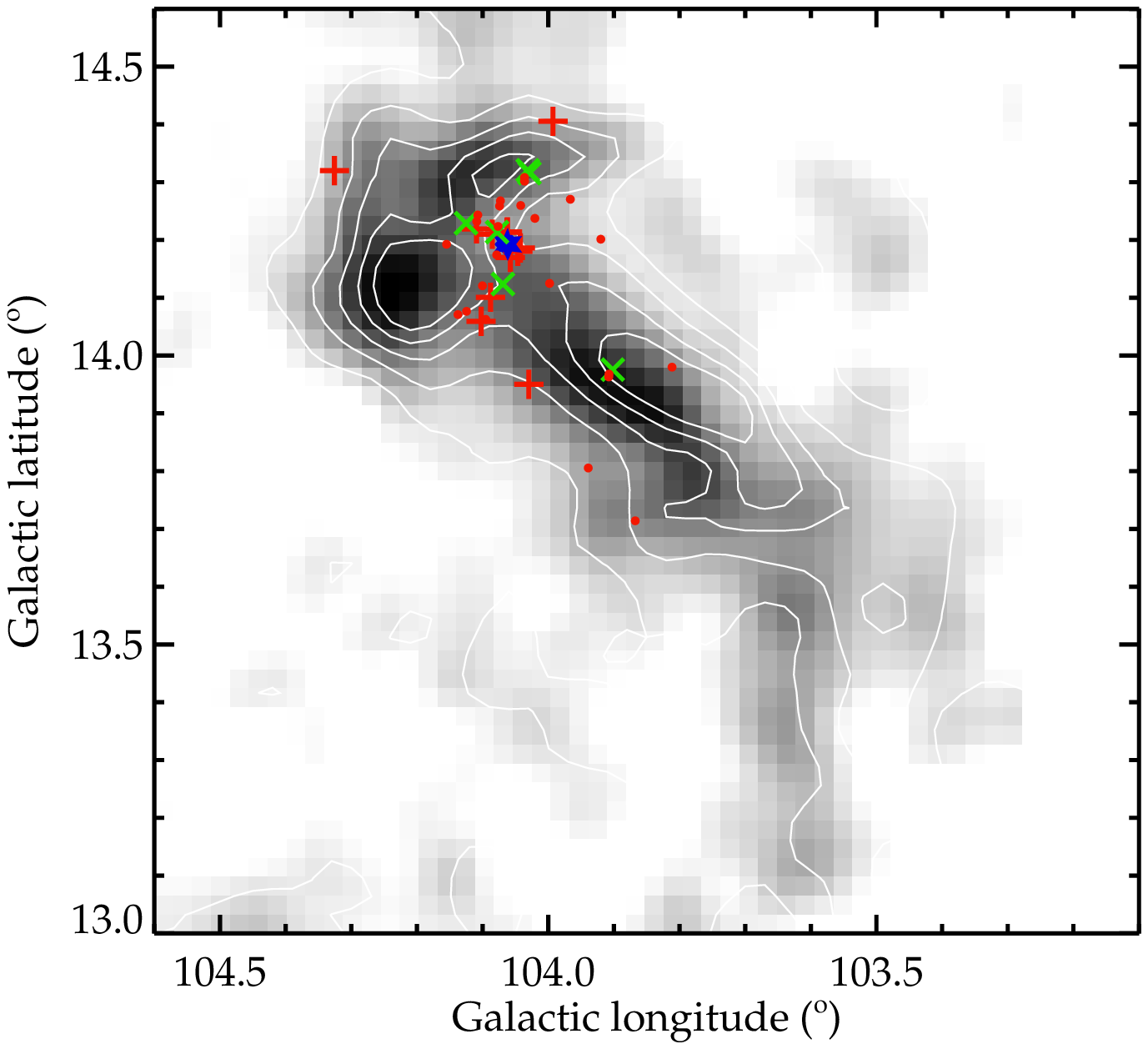}}
\caption{Enlarged part of the extinction map shown in
Fig.~\ref{Fig_map}, centered on L\,1174. The lowest contour is at
$A_\mathrm{V}=0.7$~mag, and contour interval is 0.5~mag. Plusses (red
in the electronic version) show our program stars, crosses (green in
the electronic version) are protostellar (Class~I) objects. The star
symbol (blue in electronic version) indicates HD\,200775, the most
massive member of NGC\,7023. Small dots (red in the electronic version)
are Class~II YSOs detected by Spitzer \citep{Kirk09}.}
\label{Fig_l1174map}
\end{figure}


\begin{figure} 
\centerline{\includegraphics[width=8cm]{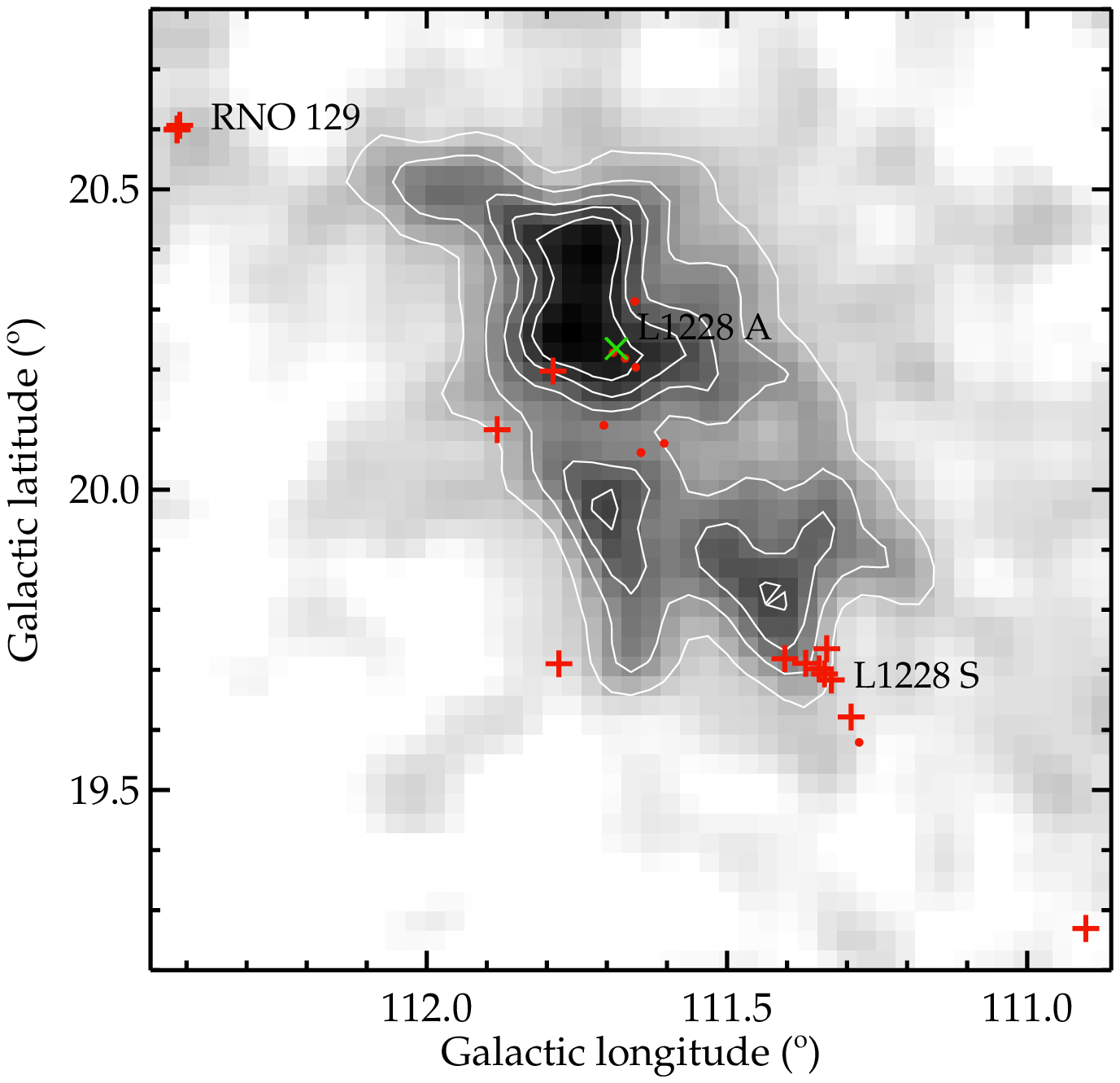}}
\caption{Enlarged part of Fig.~\ref{Fig_map}, centered on  L\,1228. Extinction 
contours and symbols are same as in Fig.~\ref{Fig_l1174map}.}
\label{Fig_l1228map} 
\end{figure}

\clearpage

\begin{figure} 
\centerline{\includegraphics[width=8cm]{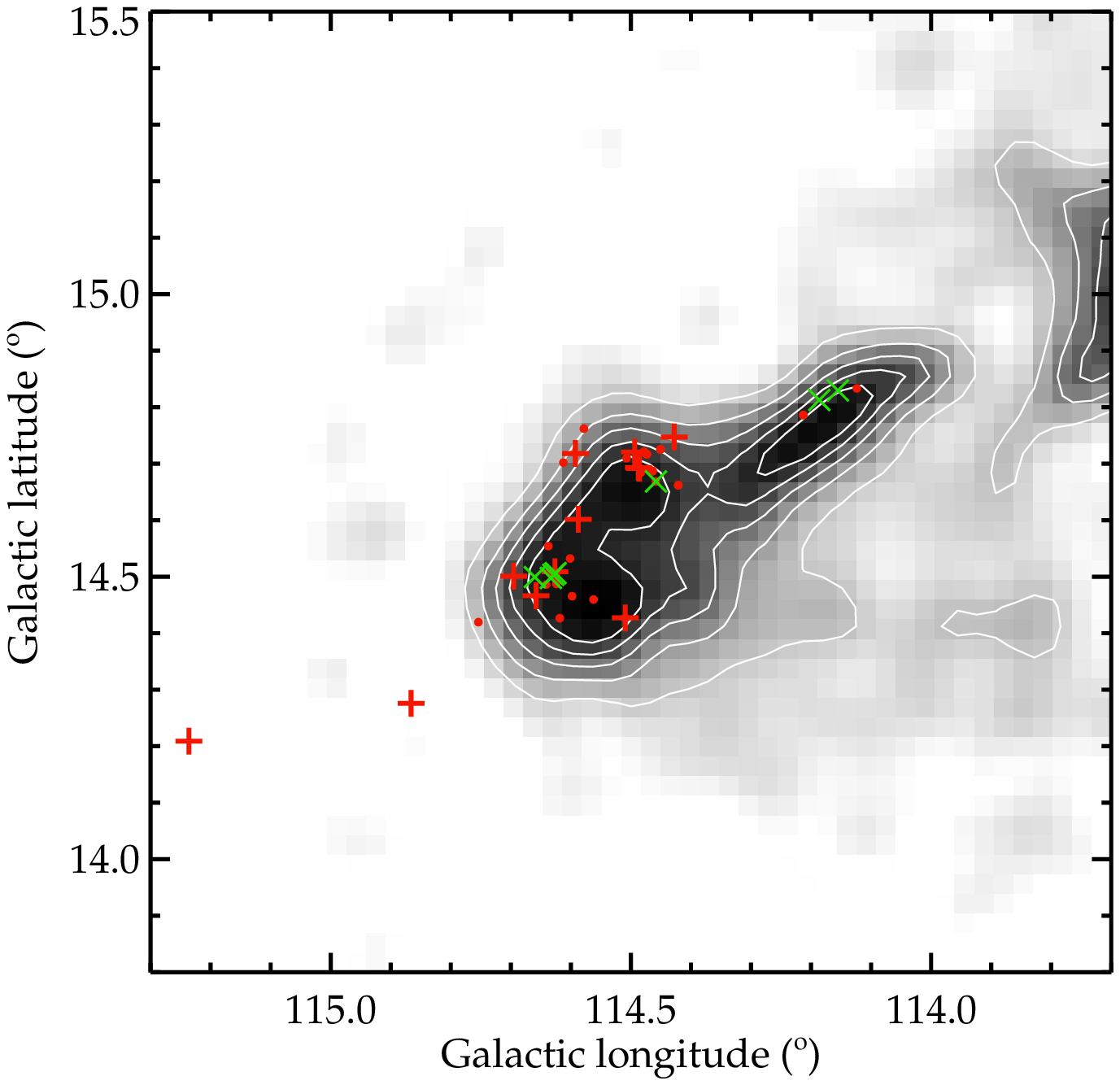}}
\caption{Same as Fig.~\ref{Fig_l1174map}, but centered on
L\,1251. Symbols are same as in Fig.~\ref{Fig_l1174map}.}
\label{Fig_l1251map}
\end{figure}


\begin{figure} 
\centerline{\includegraphics[width=8cm]{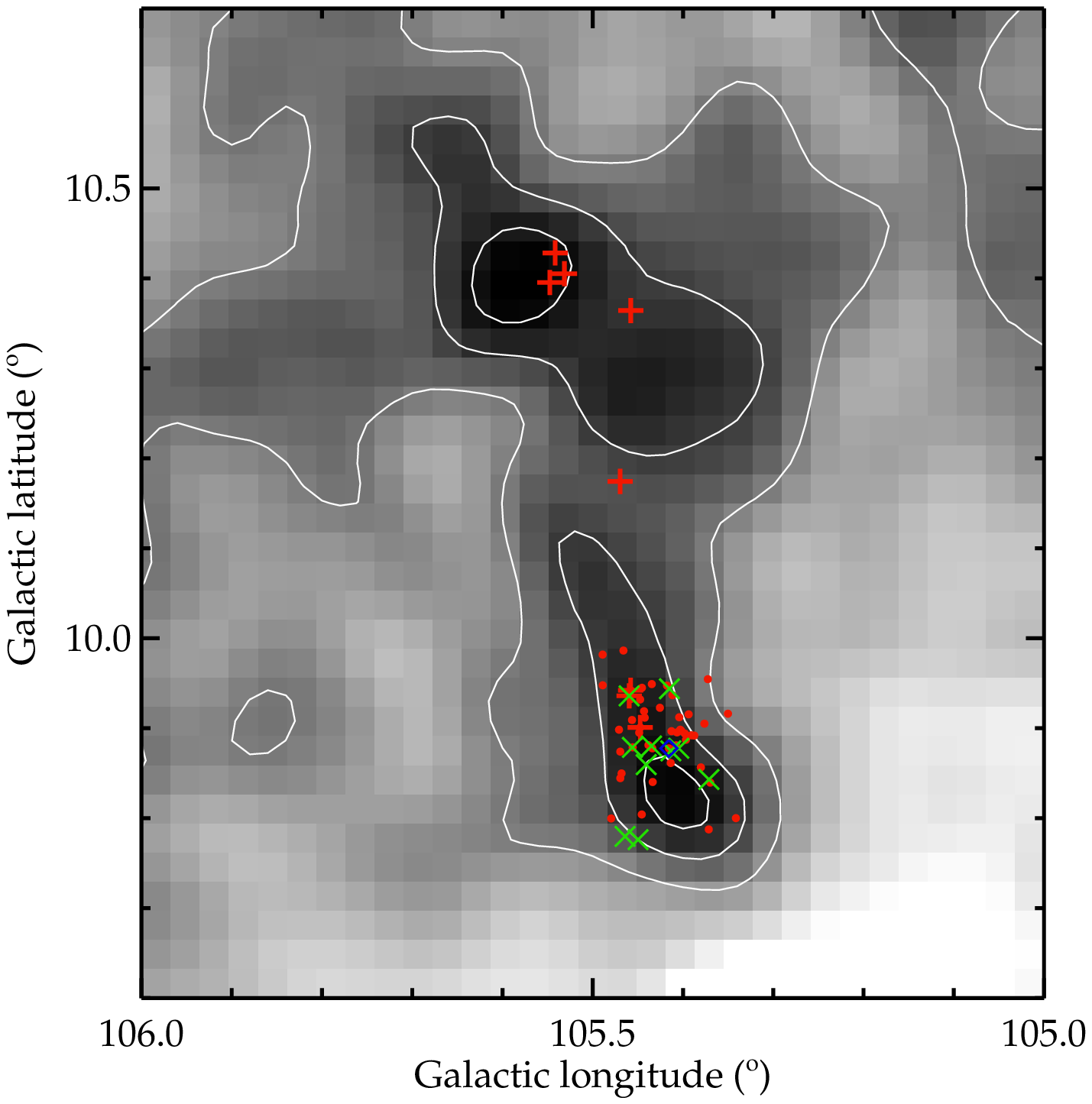}}
\caption{Enlarged part of Fig.~\ref{Fig_map}, centered on
NGC\,7129. Symbols are same as in Fig.~\ref{Fig_l1174map}.}
\label{Fig_n7129map}
\end{figure}

\begin{figure}
\centering
\includegraphics[width=7cm]{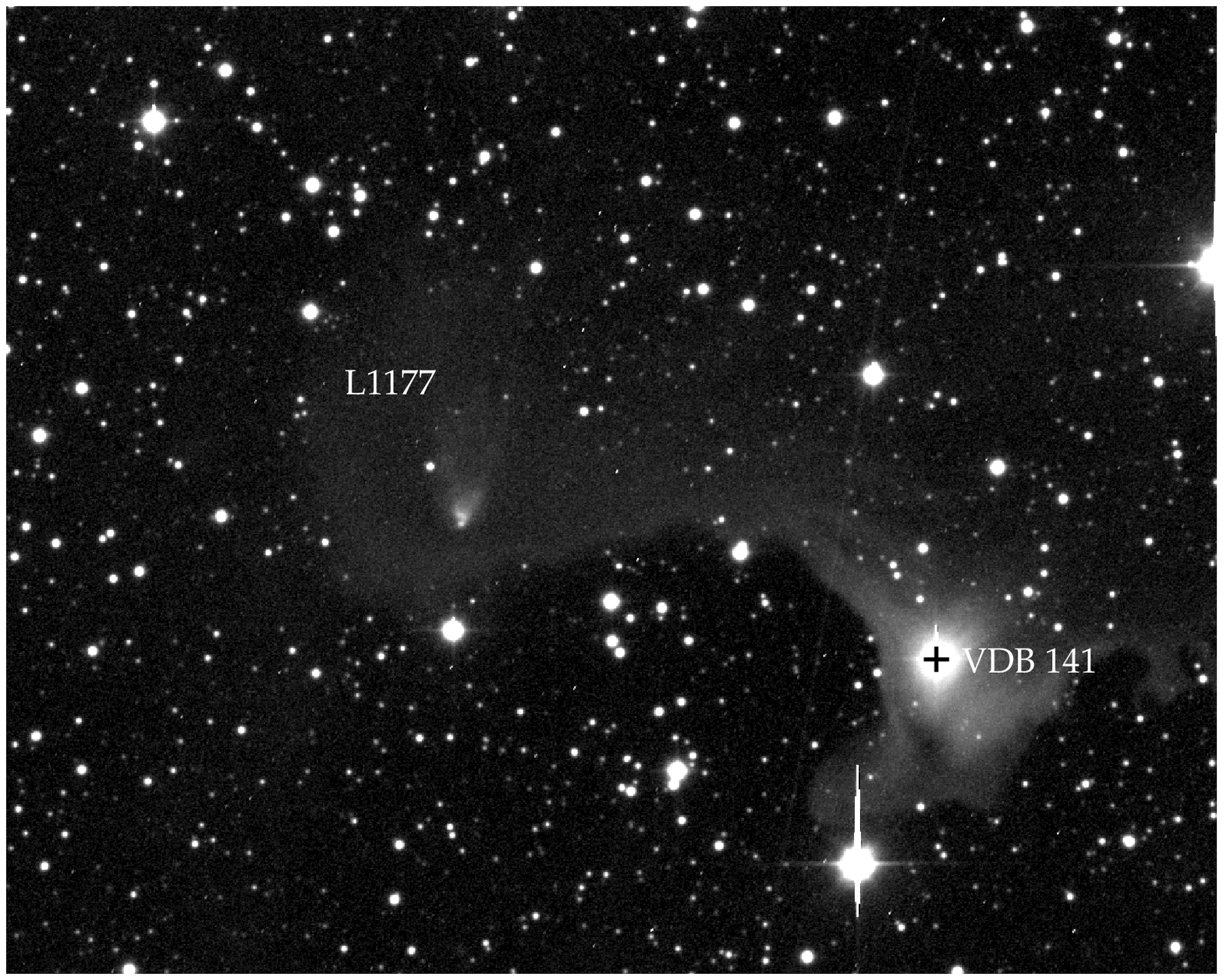}
\caption{An I-band image of a $17\arcmin \times13\farcm5$ area of the dark cloud 
L\,1177 (CB 230), and the reflection nebula VDB\,141, obtained with
the Schmidt telescope of the Konkoly Observatory. L\,1177 is obviously
connected with the cloud illuminated by VDB~141, though the presence
of the star within the cloud is accidental.}
\label{Fig_VDB141}
\end{figure}

\clearpage

\clearpage

\begin{figure}
\centering
\includegraphics[width=12cm]{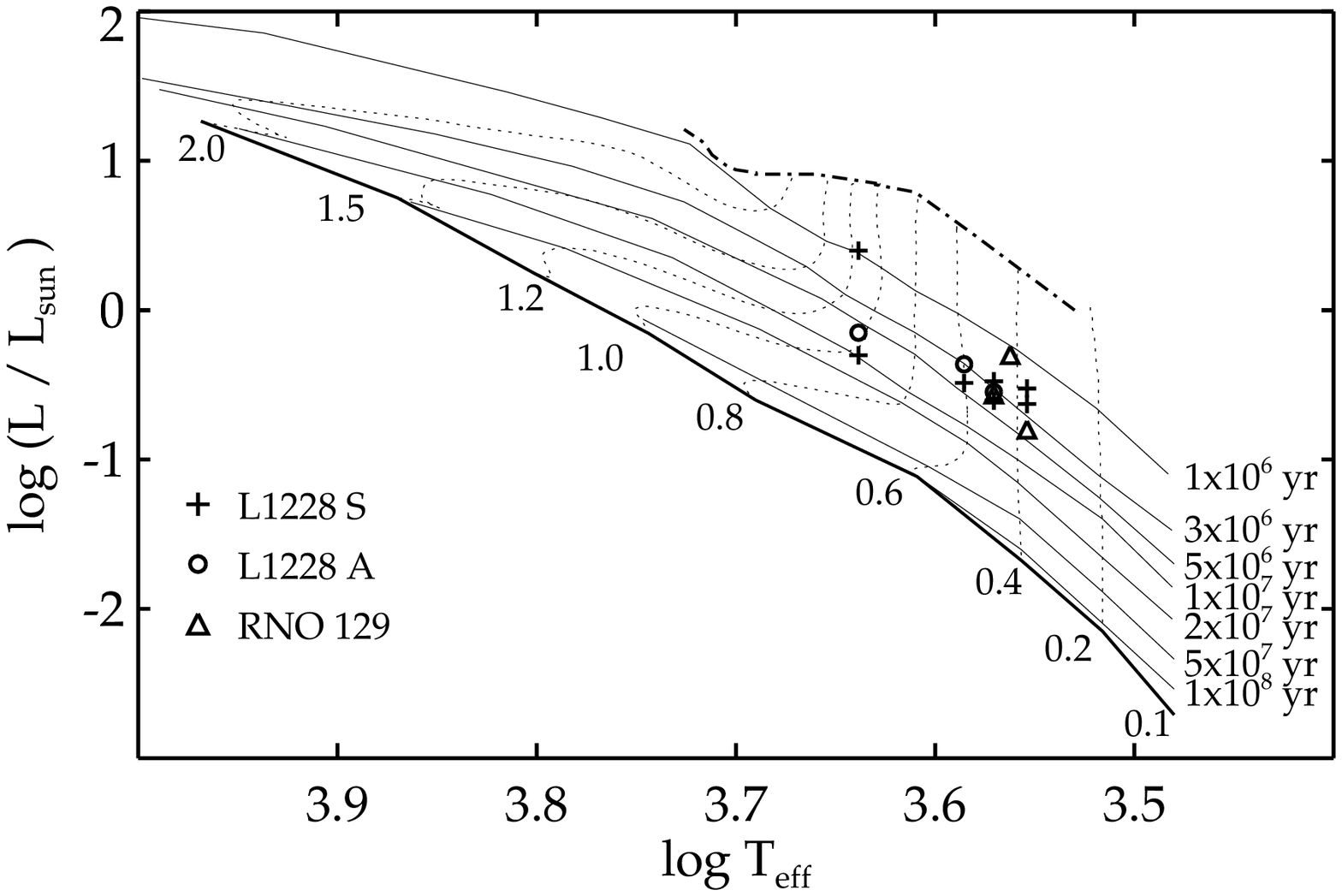}
\caption{Hertzsprung--Russell diagram of the PMS stars of L\,1228, assuming a 
distance of 200~pc. Thin solid lines
indicate the isochrones as labelled, and dotted lines show the
evolutionary tracks for the masses (in solar masses) indicated at the
lower end of the tracks according to \citet{PS99} model. The
dash-dotted line corresponds to the birthline and thick solid line
indicates the zero age main sequence (ZAMS). Plusses indicate the
members of the aggregate L\,1228\,S
\citep{Padgett}, circles are the stars near the core L\,1228\,A \citep{Bally95}, and 
triangles show the three stars embedded in the reflection nebula RNO\,129 \citep{MM04}.}
\label{Fig_hrd1}
\end{figure}


\begin{figure}
\centering
\includegraphics[width=12cm]{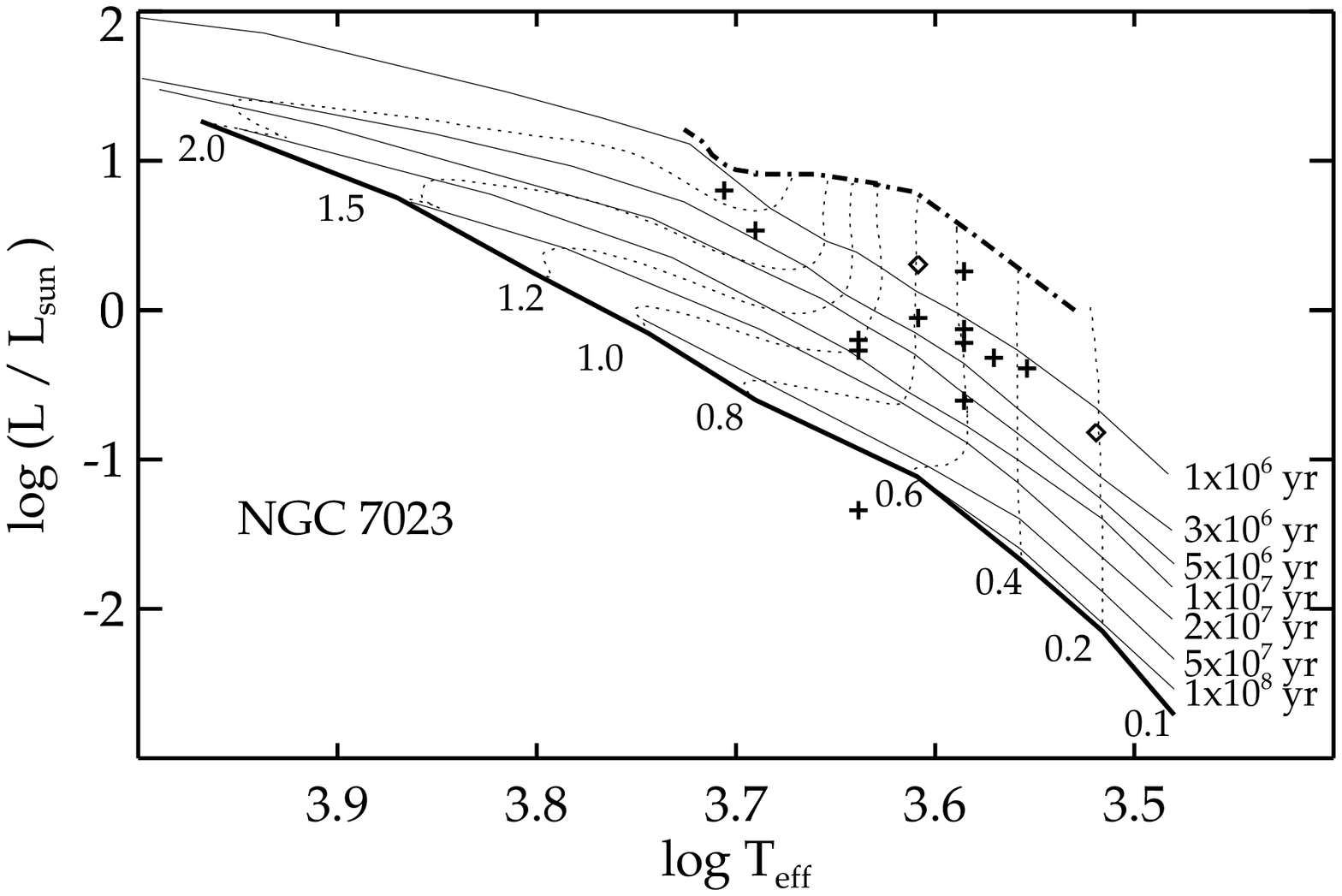}
\caption{Same as  Fig.~\ref{Fig_hrd1}, for the PMS stars of NGC\,7023, assuming a 
distance of 330~pc. Diamonds show the positions of both components of
the visual binary Lk\ha428. The star below the ZAMS is
NGC\,7023~RS\,10.}
\label{Fig_hrd2}
\end{figure}

\clearpage

\begin{figure}
\centering
\includegraphics[width=12cm]{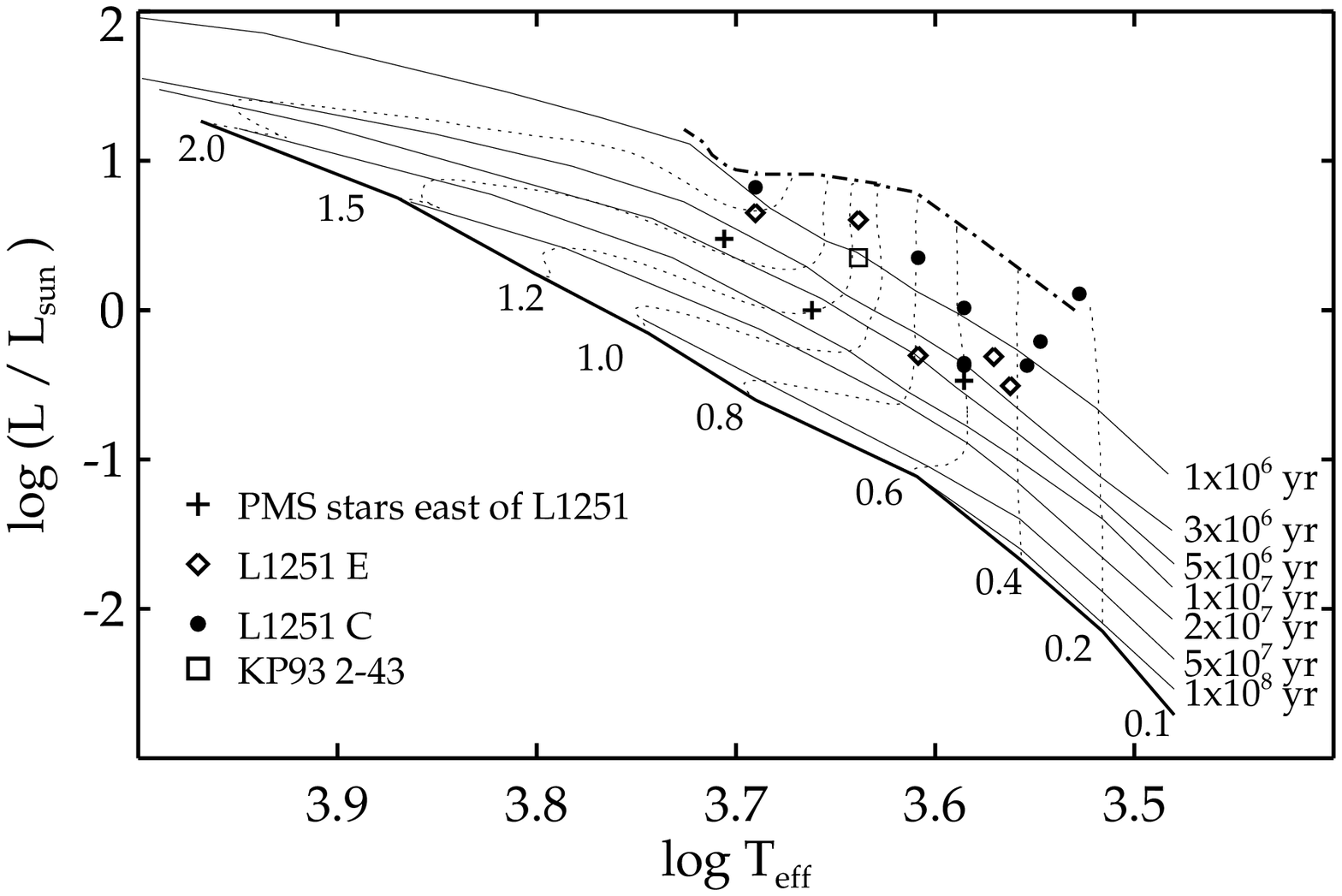}
\caption{Same as  Fig.~\ref{Fig_hrd1}, for the PMS stars of L\,1251, assuming a 
distance of 320~pc. 
Plusses show the three PMS stars on the high-longitude side of the
cloud, diamonds are the stars associated with the C$^{18}$O core
L\,1251\,E \citep{SMN94,Lee06}, and dots are the stars associated with
the core L\,1251\,C \citep{SMN94}. Open square marks the WTTS [KP93]
2-43. It can be seen that the PMS stars on the high-longitude side 
of the cloud are older than those projected inside the molecular 
cloud, indicating that star formation propagates from the higher towards the 
lower Galactic longitudes.}
\label{Fig_hrd3}
\end{figure}


\begin{figure}
\centering
\includegraphics[width=12cm]{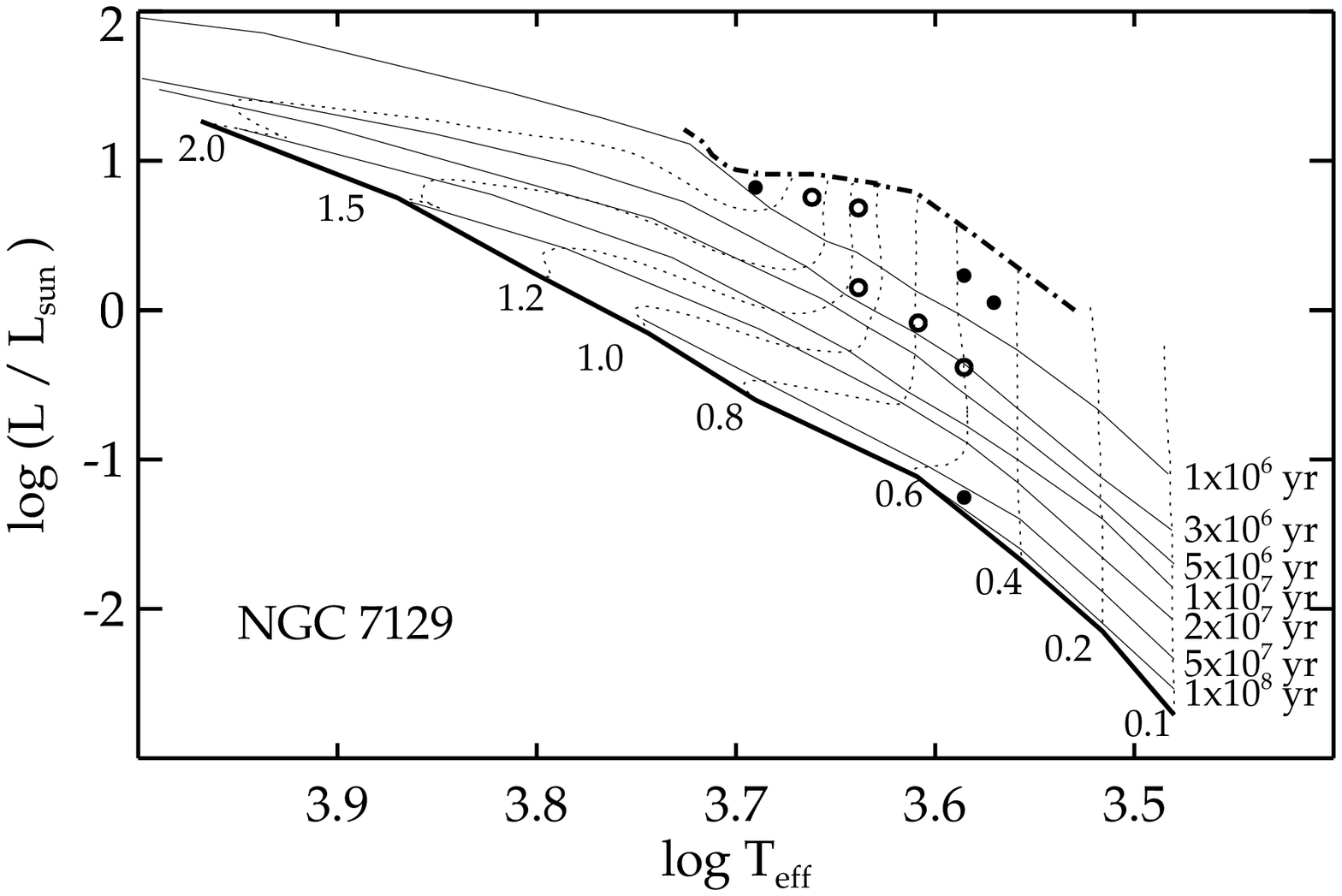}
\caption{Same as  Fig.~\ref{Fig_hrd1}, for the PMS stars of NGC\,7129, assuming a 
distance of 800~pc. Filled circles show the stars associated with the
central cluster, and open symbols indicate those found in the
high-latitude clump of L\,1181.  The stars close to the birthline
suggest that NGC\,7129 is probably not farther than the adopted
distance. The star near the ZAMS is GGD~33a, probably a Class~I
source.}
\label{Fig_hrd4}
\end{figure}

\clearpage

\begin{figure}
\centering
\includegraphics[width=12cm]{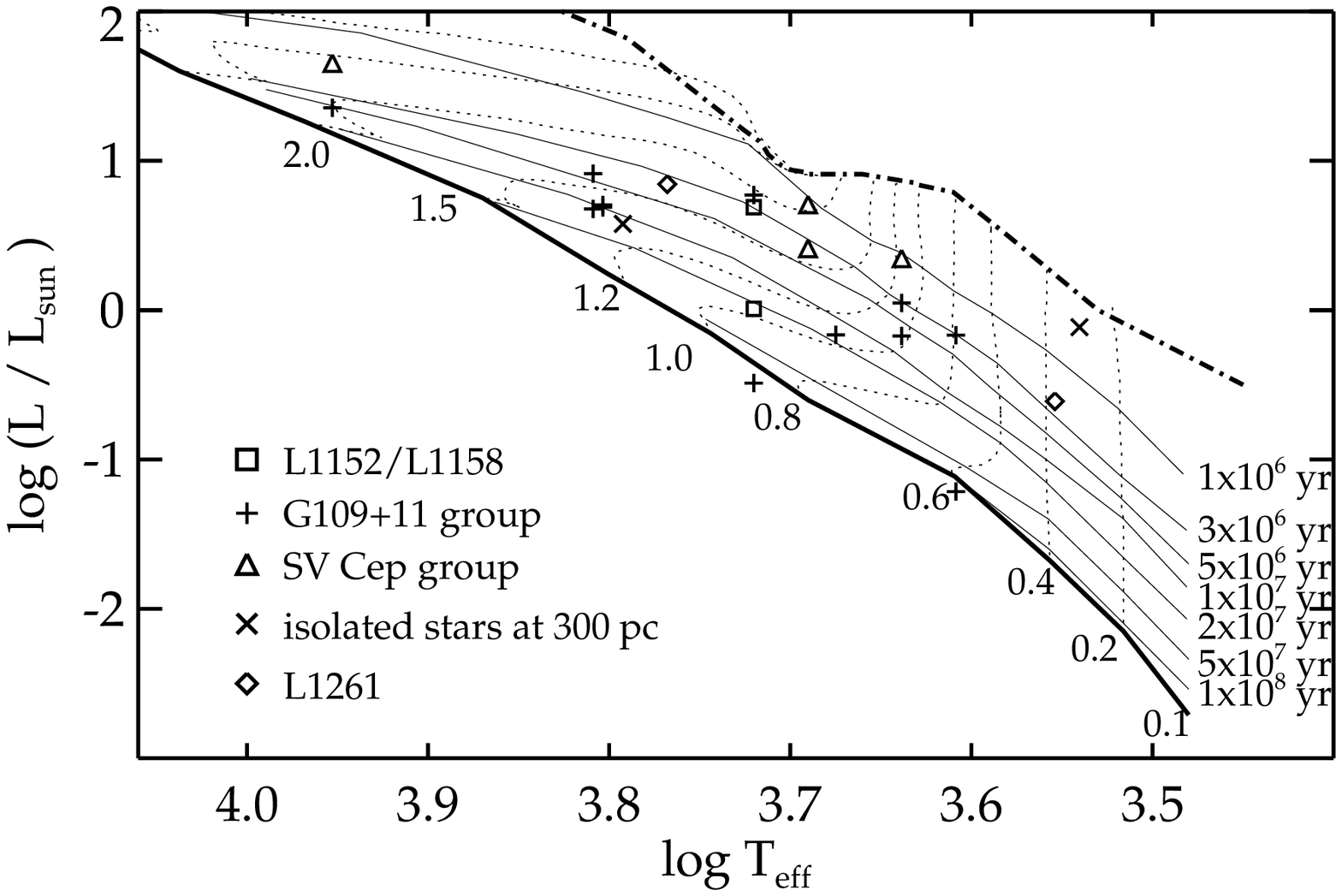}
\caption{HRD of the PMS stars associated with L\,1152/L\,1158, L\,1219, 
SV~Cep, L\,1261/L\,1262, and the PMS stars dispersed over the Cepheus
flare. The stars below the ZAMS are [K98c]\,Em*~58 and [K98c]\,Em*~119\,S, 
probably edge-on disk systems. IRAS~F22219+7901 is not plotted.}
\label{Fig_hrd5}
\end{figure}


\begin{figure}
\centering
\includegraphics[width=8cm]{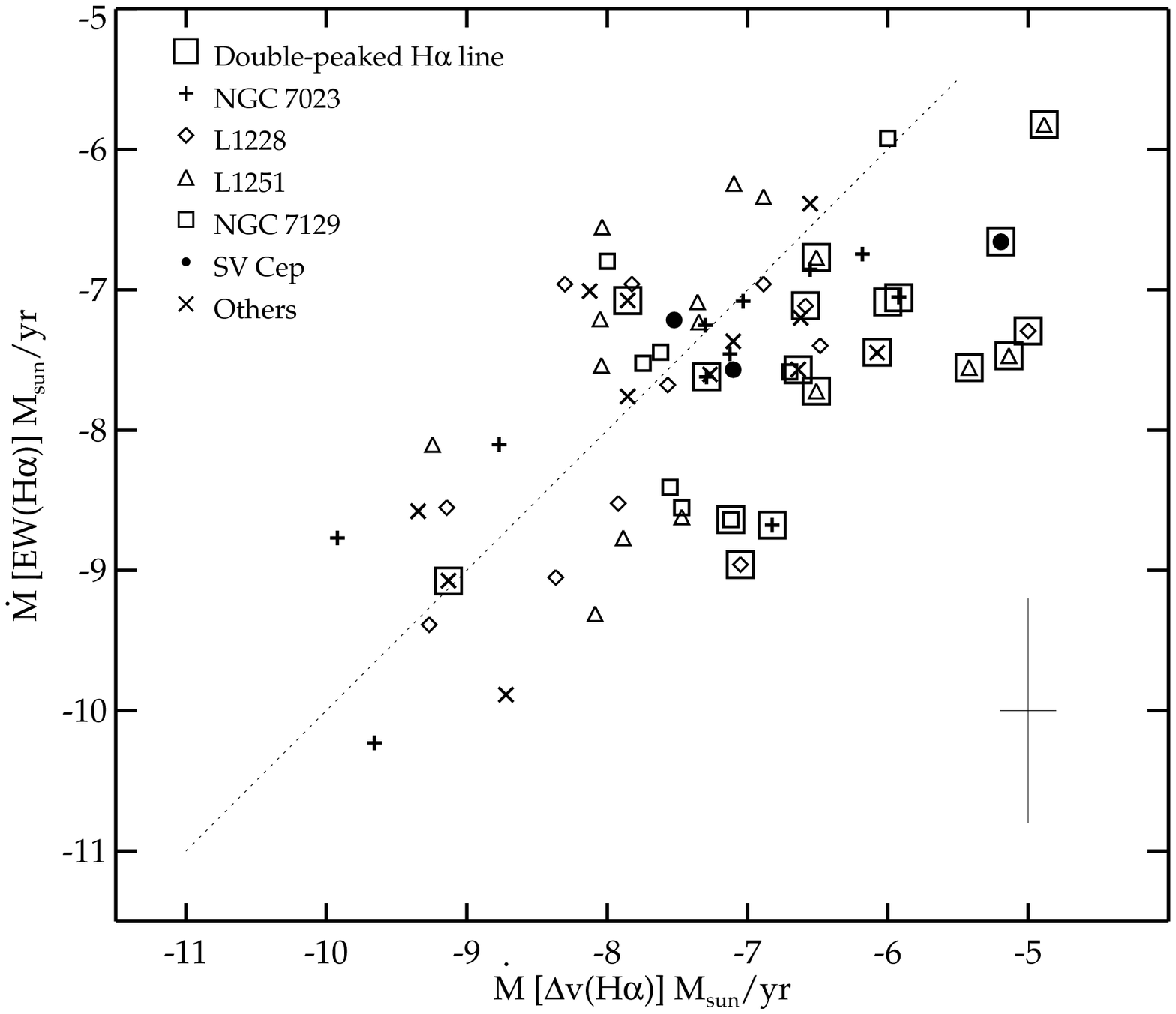}
\caption{Accretion rates derived from the H$\alpha$ luminosities,
plotted against those derived from the 10\% width of the H$\alpha$
emission line. Different symbols indicate the stars belonging to
different clouds. Symbols framed by large squares mark the stars
exhibiting double-peaked H$\alpha$ line. The formal estimated error of
the points is shown in the lower right corner.}
\label{Fig_mdot1}
\end{figure}

\clearpage

\begin{figure}
\centering
\includegraphics[width=8cm]{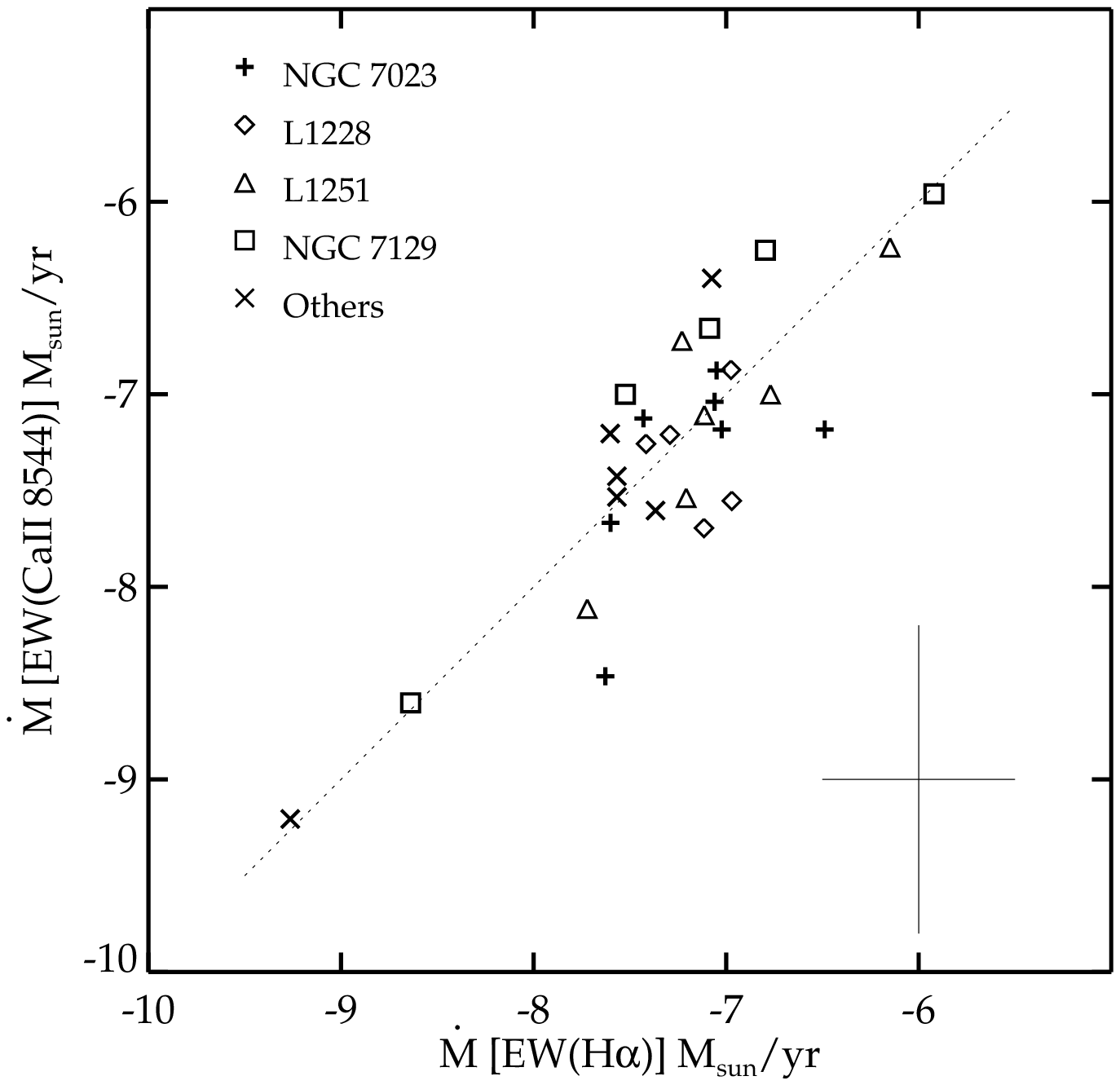}
\caption{Accretion rates derived from the \ion{Ca}{2} luminosities,
plotted against those derived from the H$\alpha$ luminosities. The
errorbar is shown in the lower right corner.}
\label{Fig_haca}
\end{figure}


\begin{figure}
\centering
\includegraphics[width=8cm]{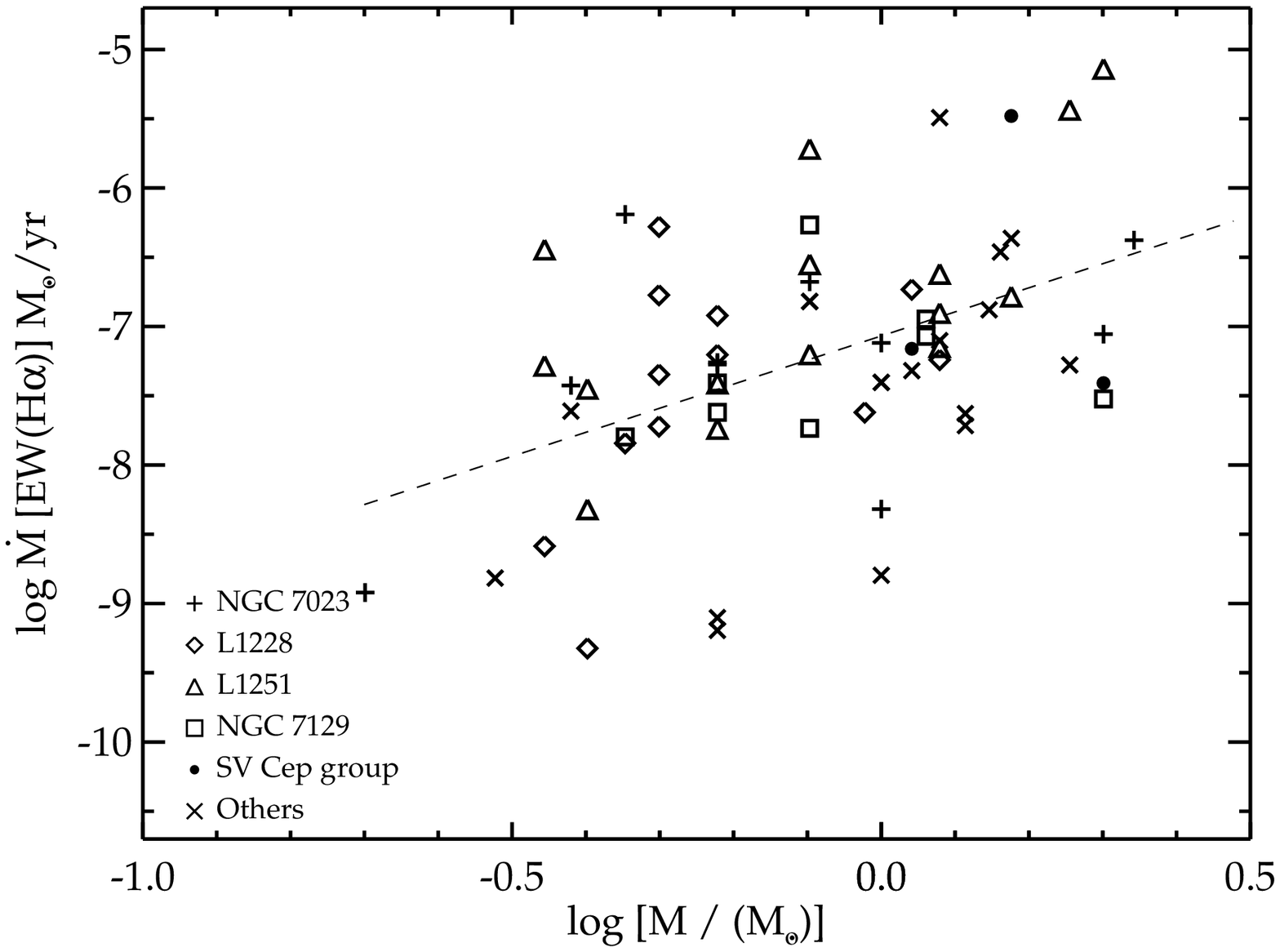}
\caption{Accretion rates derived from the H$\alpha$ lines plotted
against the stellar masses determined from the HRD. The dashed line,
fitted to the data corresponds to $\log \dot{M} = (1.74\pm0.40) \times
\log M - 7.07\pm0.10$.}
\label{Fig_mdotm}
\end{figure}

\clearpage

\begin{figure*}
\centering{\includegraphics[width=14cm]{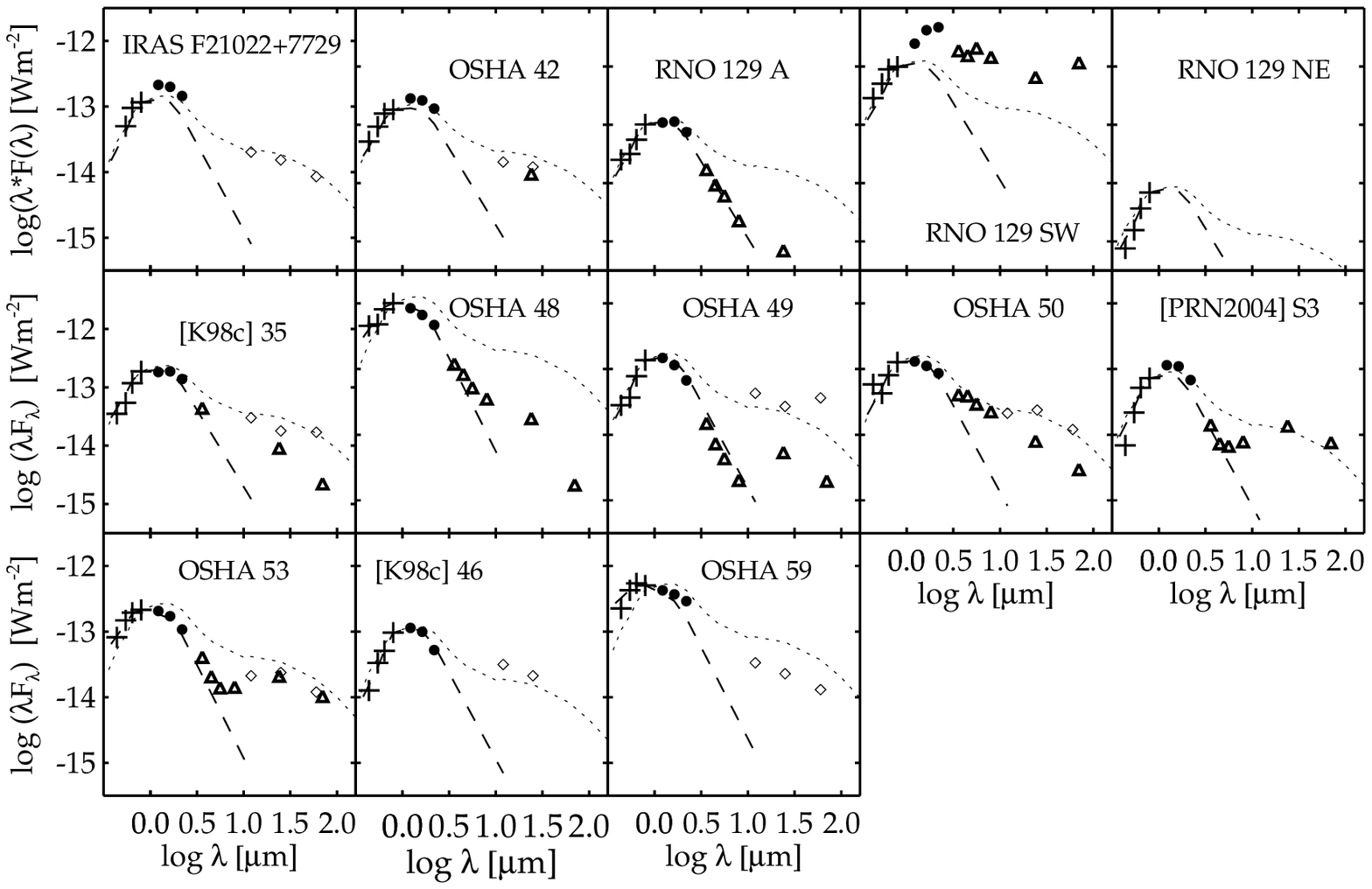}}
\caption{Spectral energy distributions, corrected for interstellar
extinction, of the PMS stars in the region of L\,1228, based on our
optical photometry (plusses), 2MASS (dots), \textit{Spitzer} IRAC and
MIPS (triangles), as well as \textit{IRAS} (thin diamonds)
data. Dashed lines show the photospheric SEDs.  Their values were
determined from the dereddened $I_{C}$ magnitudes, and from the colour
indices corresponding to the spectral types. For comparison, dotted
lines show the median SED of the T~Tauri stars of the Taurus star
forming region \citep{DAlessio}.}
\label{Fig_sed1}
\end{figure*}

\clearpage

\begin{figure*}
\centering{\includegraphics[width=14cm]{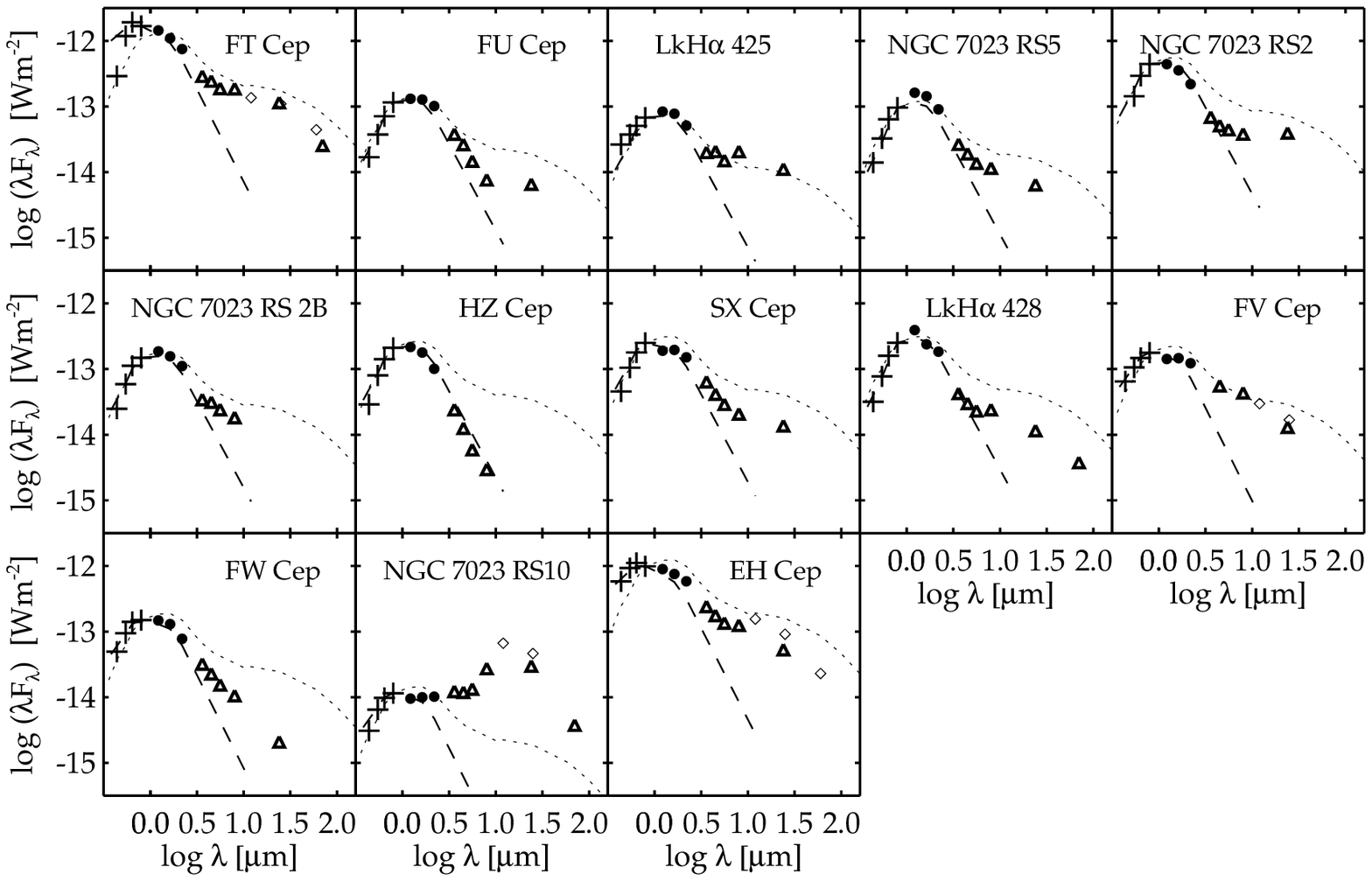}} \caption{Same as
Fig.~\ref{Fig_sed1}, but for the PMS stars of NGC\,7023.}
\label{Fig_sed2} \end{figure*}

\clearpage

\begin{figure*}
\centering{\includegraphics[width=14cm]{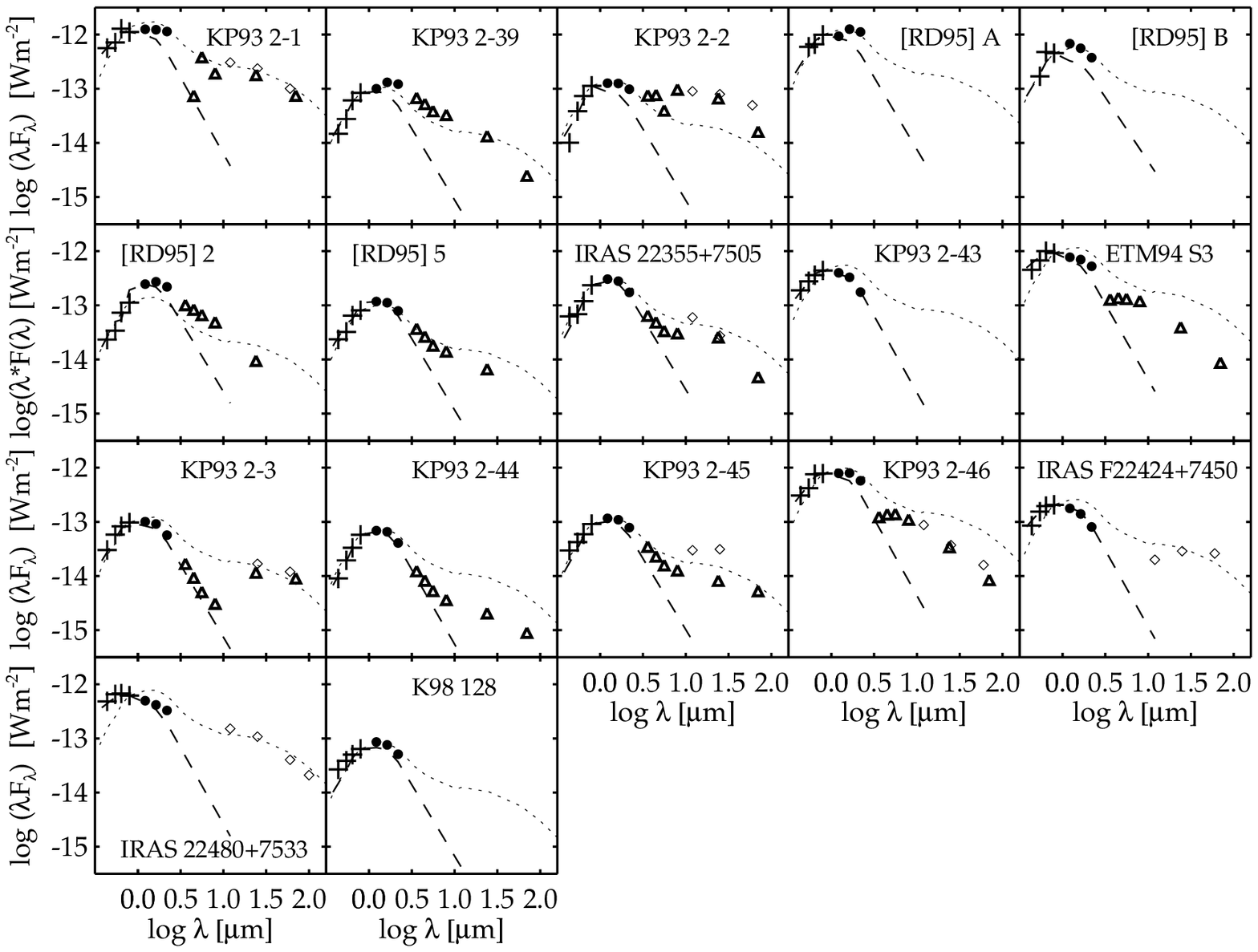}}
\caption{Same as Fig.~\ref{Fig_sed1}, but for the PMS stars of L\,1251.}
\label{Fig_sed3}
\end{figure*}

\clearpage

\begin{figure*}
\centering{\includegraphics[width=14cm]{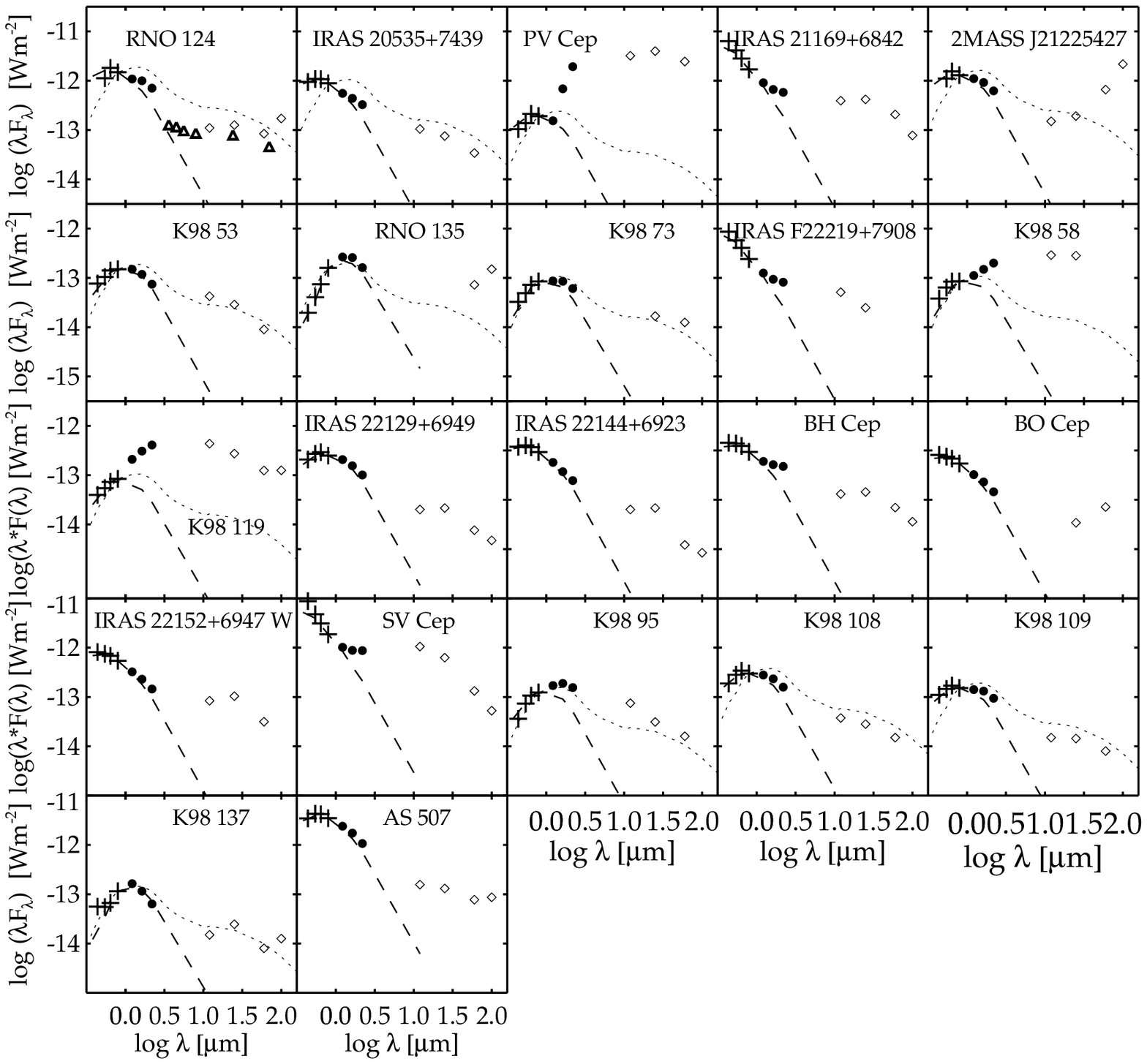}}
\caption{Same as Fig.~\ref{Fig_sed1}, but for the stars associated
with L\,1152/L\,1158, L\,1177, L\,1219, SV~Cep, L\,1261/L\,1262, and
the PMS stars dispersed over the cloudy region.}
\label{Fig_sed4}
\end{figure*}

\clearpage

\begin{figure*}
\centering{\includegraphics[width=14cm]{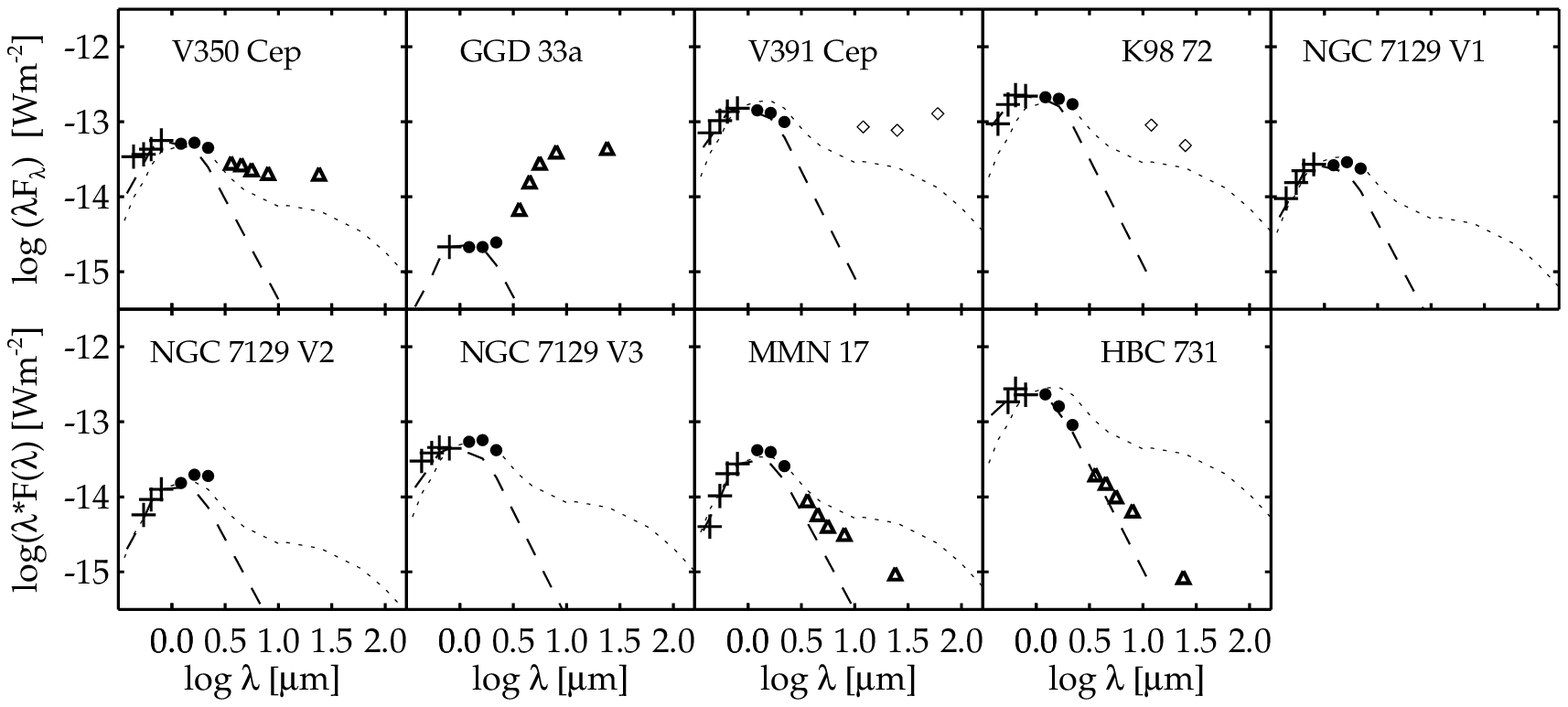}}
\vskip -2.0cm
\caption{Same as Fig.~\ref{Fig_sed1}, but for the observed PMS stars of NGC\,7129.}
\label{Fig_sed5}
\end{figure*}


\begin{figure*} 
\centerline{\includegraphics[width=5cm]{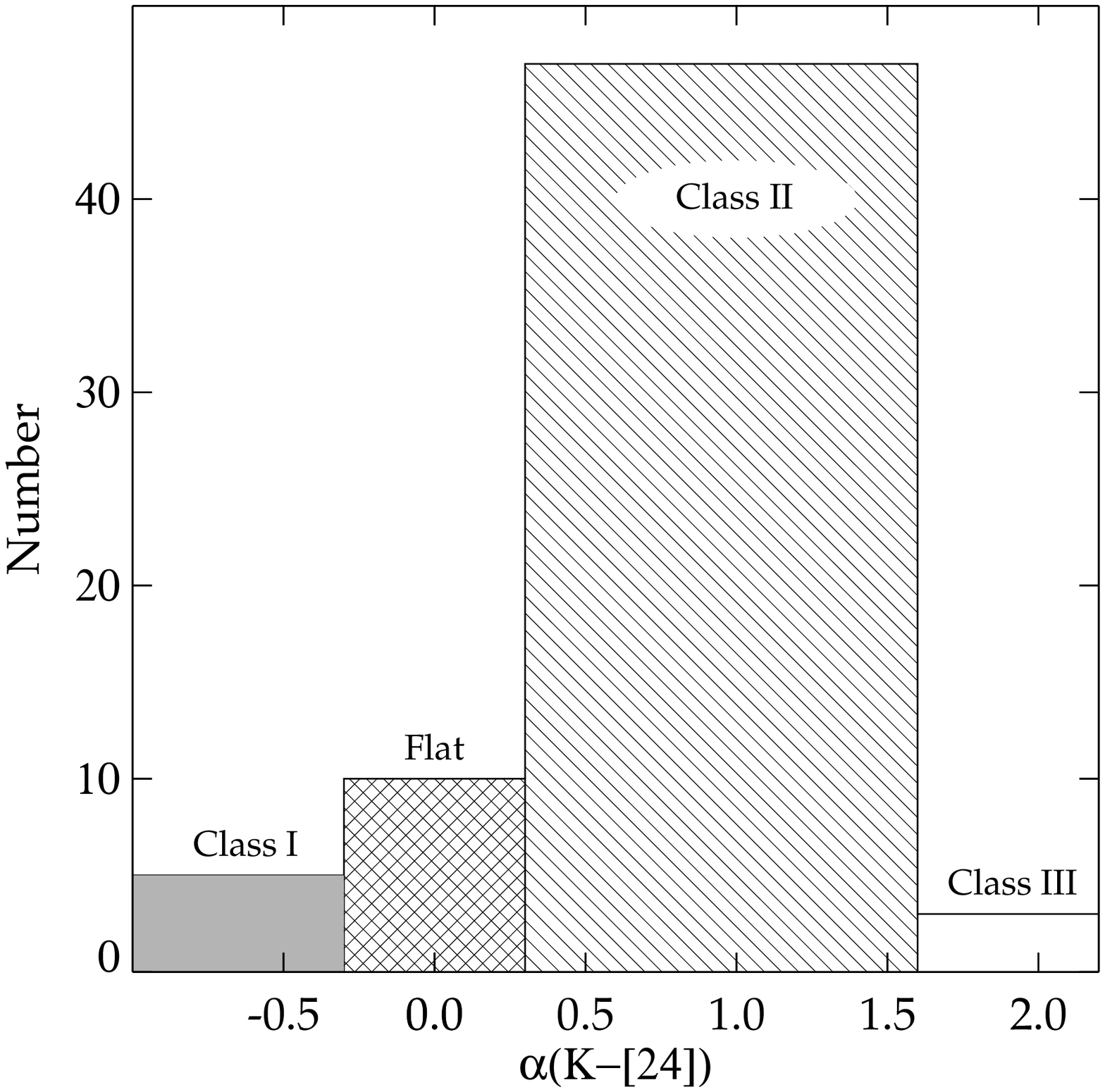}\hskip5mm
\includegraphics[width=5cm]{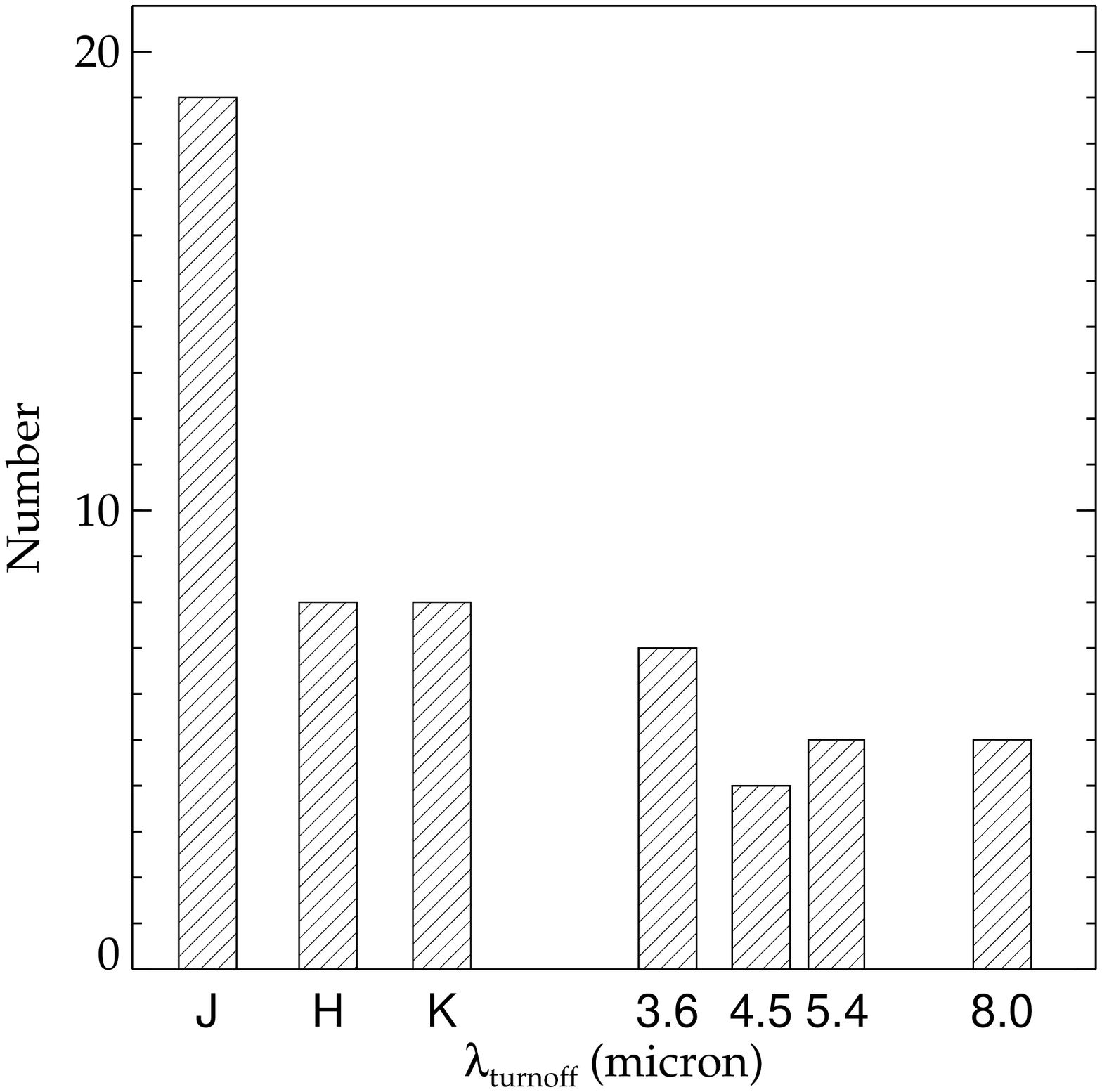}}
\caption{ {\it Left}: distribution of the SED classes defined by the 
slope $\alpha_{24-K}$;
{\it Right}: Number of stars whose observed SED has the last photospheric value
in the photometric band indicated in the horizontal axis.}
\label{Fig_dhist}
\end{figure*}

\clearpage

\begin{figure*} 
\centerline{\includegraphics[width=5cm]{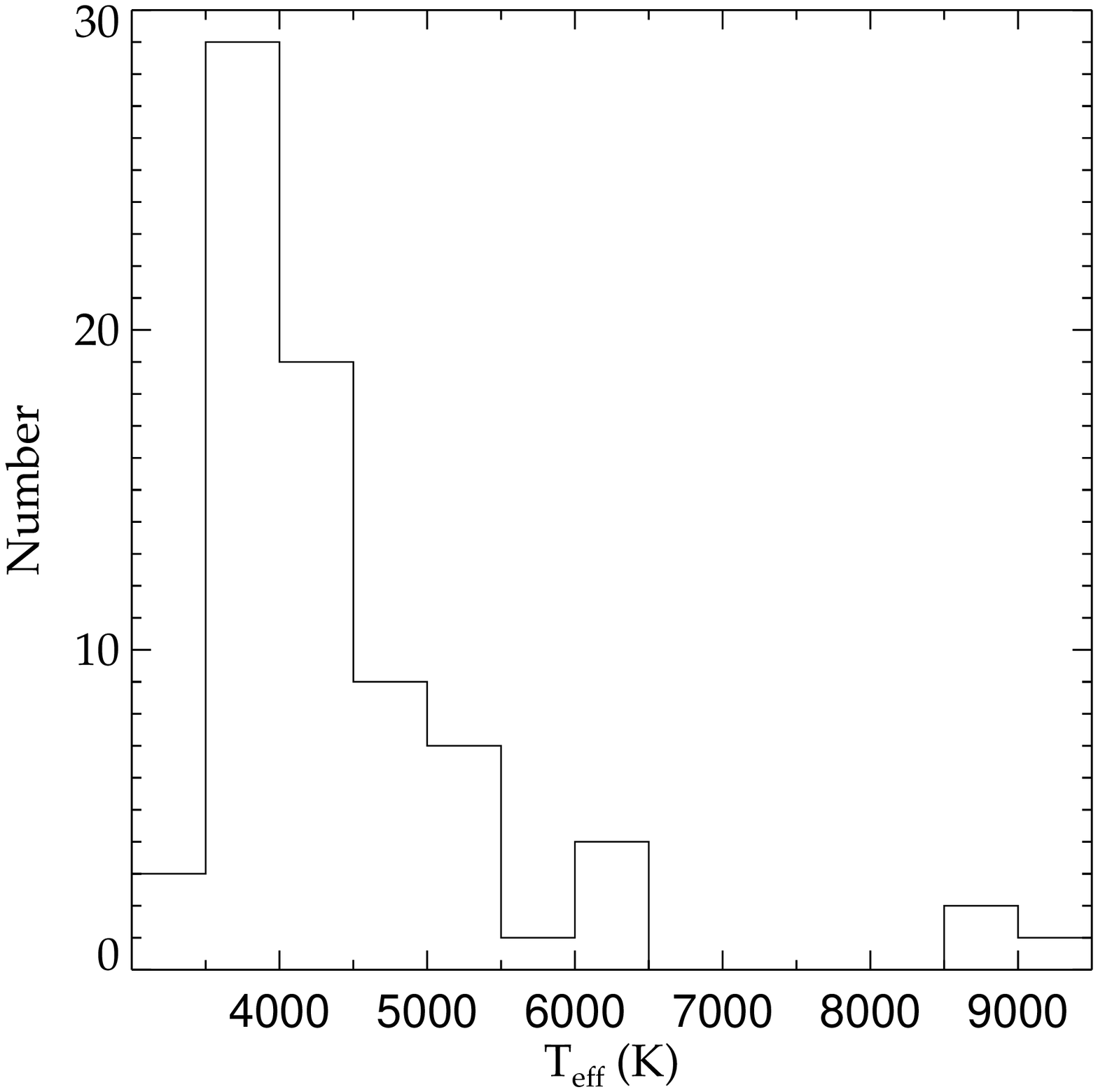}
\includegraphics[width=5cm]{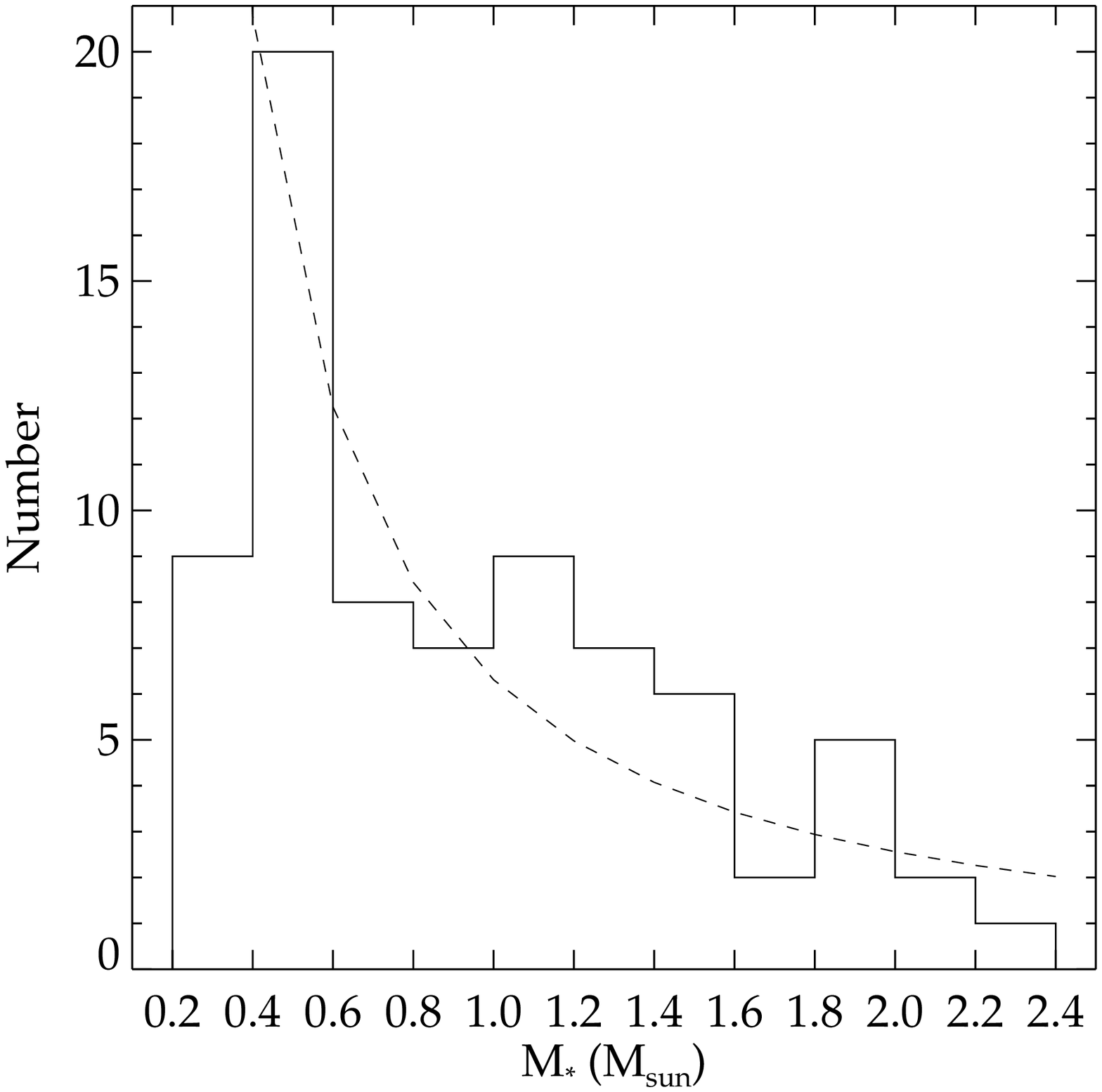}
\includegraphics[width=5cm]{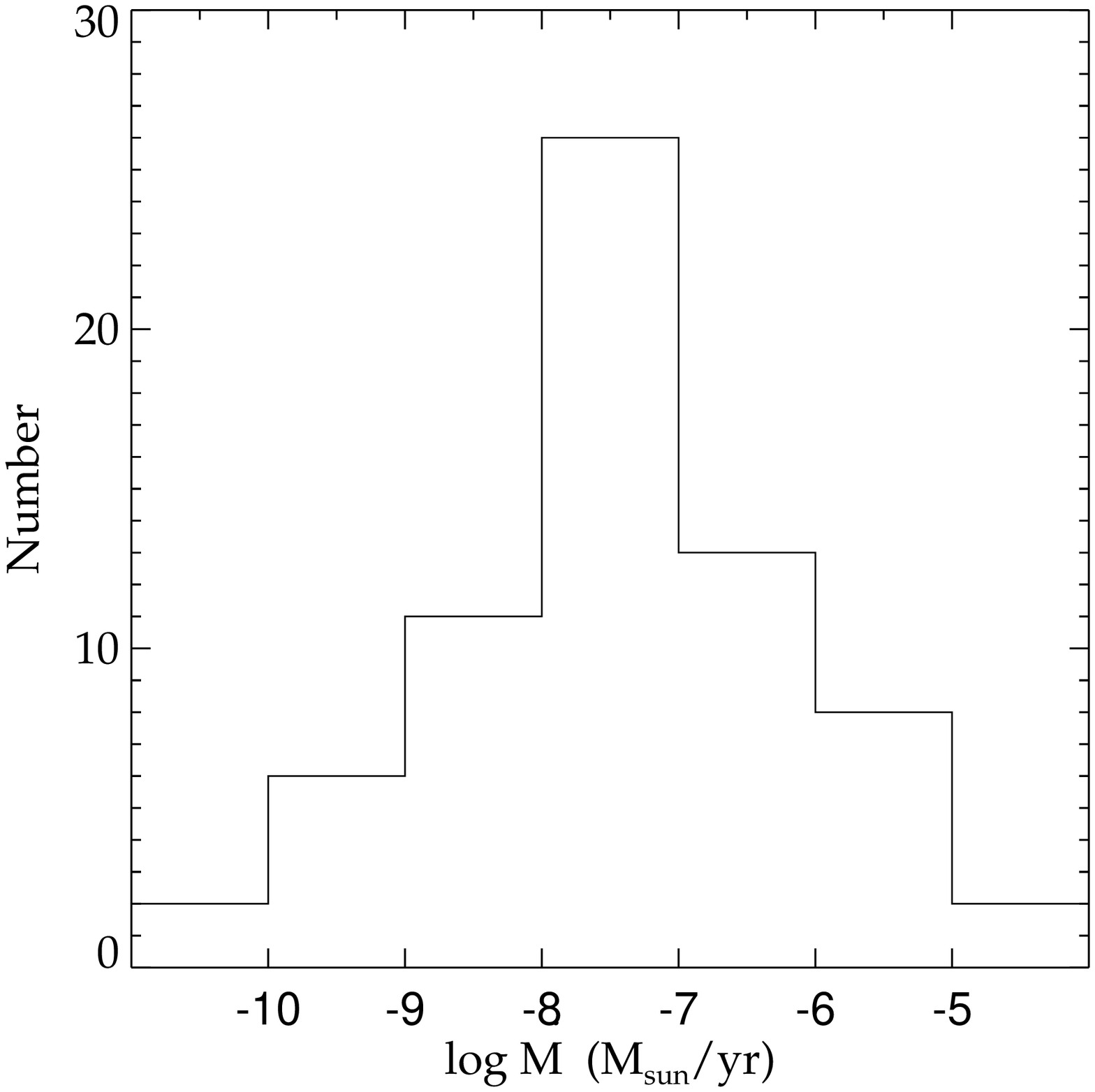}}
\caption{Frequency distribution of effective temperatures, masses, and accretion rates.
The shape of the Salpeter IMF is indicated by dashed line in the
middle panel.}
\label{Fig_hist}
\end{figure*}

\clearpage



\begin{deluxetable}{lrcclll}
\tabletypesize{\scriptsize}
\tablecolumns{7}
\tablewidth{0cm}
\rotate
\tablecaption{List of the observed PMS stars and associated clouds\label{TabList0}}
\tablehead{
\colhead{Names}  &  \colhead{IRAS}  &  2MASS  & Spitzer Id.$^*$ & \colhead{Membership} 
 &  \colhead{Cand. Type}   & \colhead{Ref.} } 
\startdata 
RNO\,124, HBC\,695, $[$K98c$]$\,Em*\,6        & 20359+6745  & 20361986+6756316 & 3 & L\,1152 & CTTS & 3  \\  
PV Cep, HBC\,696, RNO\,125, $[$K98c$]$\,Em*\,9 & 20453+6746 & 20455394+6757386 & 135 & L\,1155  & CTTS & 3 \\
GSC 04472-00143                   & 20535+7439  & 20530638+7450348 &\nodata &\nodata & IRAS  & 8 \\   
OSHA\,42, $[$K98c$]$\,Em*\,30                 & F20598+7728 & 20584668+7740256 & 123 & L\,1228 & H$\alpha$ & 5,8 \\
RNO\,129 A                        &             & 20590373+7823088 &\nodata & L\,1228 & red, nebulous & 18 \\
RNO\,129 SW, Bern\,48, $[$K98c$]$\,Em*\,32, OSHA\,44 & 21004+7811 & 20591409+7823040 &\nodata & L\,1228 & IRAS & 5,18 \\
RNO\,129 NE\tablenotemark{a}                       & 21004+7811 & 20591409+7823040 &\nodata & L\,1228 & IRAS & 5,18 \\
FT Cep, NGC\,7023 RS\,1, $[$K98c$]$\,Em*\,26  & 20587+6802  & 20592284+6814437 & 15 & NGC\,7023 & variable, \ha & 3 \\
$[$K98c$]$\,Em*\,35, $[$PRN2004$]$ S5                   & F21016+7651 & 21005285+7703149 & 124 & L\,1228 & H$\alpha$, MIR &  8,12   \\
\nodata                           & F21022+7729 & 21011339+7741091 &\nodata & L\,1228 & IRAS  & 8 \\
NGC\,7023 RS\,2,  NGC\,7023 S\,J   &             & 21012637+6810385 & 27 & NGC\,7023 & variable & 1,4  \\
NGC\,7023 RS\,2\,B\tablenotemark{b}  &             & 21012706+6810381 & 137 & NGC\,7023 & variable & 1  \\
OSHA\,48\,A, $[$PRN2004$]$ S6               &             & 21012919+7702373 & 30 & L\,1228 & H$\alpha$  & 5,12   \\
OSHA\,48\,B                      &             & 21012919+7702373 &\nodata & L\,1228 & H$\alpha$  & 5,12   \\
OSHA\,49, $[$K98c$]$\,Em*\,40, $[$PRN2004$]$ S7        & F21023+7650 & 21013097+7701536 & 32 & L\,1228 & H$\alpha$ & 5,8,12  \\
OSHA\,50, $[$K98c$]$\,Em*\,41, $[$PRN2004$]$ S8        &             & 21013267+7701176 & 33 & L\,1228 & H$\alpha$ & 5,8,12   \\
PW Cep, Lk\ha 425, NGC\,7023 RS\,3, NGC\,7023 S\,E  &  & 21013590+6808219 & 104 & NGC\,7023 & variable, \ha & 1,4 \\
SX Cep, NGC\,7023 S\,L  &      & 21013754+6811309 &\nodata & NGC\,7023 & variable & 1,4  \\ 
NGC\,7023 RS\,5, NGC\,7023 S\,N &             & 21014250+6812572 & 38 & NGC\,7023 & variable & 1,4 \\
HZ Cep, RS S\,3, NGC\,7023 S\,D  &             & 21014358+6809361 &\nodata & NGC\,7023 & variable &  1,4 \\     
FU Cep, LkH$\alpha$ 427, HBC\,304, NGC\,7023 S\,C  & 21009+6758  & 21014672+6808453 &\nodata & NGC\,7023 & CTTS &  2,4 \\
$[$PRN2004$]$ S3                            &             & 21014960+7705479 & 42 & L\,1228 & MIR excess & 12 \\
OSHA\,53, $[$K98c$]$\,Em*\,43, $[$PRN2004$]$ S9        & F21028+7645 & 21020488+7657184 & 43 & L\,1228 & H$\alpha$ & 5,8,12  \\
FV Cep, LkH$\alpha$ 275, HBC\,306, $[$K98c$]$\,Em*\,38 & F21017+6813  & 21022039+6825240 & 46 &  NGC\,7023 & CTTS & 3  \\
LkH$\alpha$ 428 N, NGC\,7023 RS\,8  &           & 21022829+6803285 & 125 & NGC\,7023 & H$\alpha$  & 1  \\
LkH$\alpha$ 428 S\tablenotemark{b}  &           & 21022829+6803285 & 52 & NGC\,7023 & H$\alpha$  & 1  \\
FW Cep, NGC\,7023 RS\,9           &             & 21023299+6807290 & 55 & NGC\,7023 & variable & 1 \\
NGC\,7023 RS\,10                  & 21023+6754  & 21025943+6806322 & 57 & NGC\,7023 & variable, IRAS & 1   \\   
$[$K98c$]$\,Em*\,46                           & F21037+7614 & 21030242+7626538 &\nodata & L\,1228 & H$\alpha$, IRAS & 8  \\
EH Cep, LkH$\alpha$ 276, HBC\,307, $[$K98c$]$\,Em*\,42 & 21027+6747 &  21032435+6759066 & 60 & NGC\,7023 & CTTS & 3 \\	 
OSHA\,59, $[$K98c$]$\,Em*\,49                 & F21066+7710 & 21055189+7722189 &\nodata & L\,1228 & H$\alpha$  & 5,8 \\
$[$K98c$]$\,Em*\,53                           &             & 21153595+6940477 &\nodata &\nodata & H$\alpha$ & 8 \\
BD $+68\degr1118$, GSC\,04461-01336& 21169+6842 & 21173917+6855097 &\nodata  &L\,1177? & IRAS & 8 \\
$[$K98c$]$\,Em*\,58                           & F21202+6835 & 21205785+6848183 &\nodata &\nodata &  H$\alpha$, IRAS & 8 \\
2MASS 21225427+6921345         &             & 21225427+6921345  &\nodata&\nodata & K-band excess & 17 \\
RNO\,135, $[$K98c$]$\,Em*\,61                 & 21326+7608  & 21323108+7621567 &\nodata & L\,1241? & H$\alpha$, IRAS & 8 \\
NGC\,7129 S V1                    &             & 21401174+6630198 &\nodata & NGC\,7129 & variable & 9 \\
NGC\,7129 S V2                    &             & 21402277+6636312 &\nodata & NGC\,7129 & variable & 9 \\
V391 Cep, $[$K98c$]$\,Em*\,71                 & F21394+6621 & 21402755+6635214 &\nodata & NGC\,7129 & CTTS & 8,19 \\
NGC\,7129 S V3                    &             & 21403852+6635017 &\nodata & NGC\,7129 & variable & 9  \\
$[$K98c$]$\,Em*\,72                           & F21404+6608 & 21413315+6622204 &\nodata & NGC\,7129 & H$\alpha$  & 8 \\	  
HBC\,731, $[$SVS76$]$ NGC 7129 6  &             & 21425961+6604338 & NGC 7129 30 & NGC\,7129 & CTTS & 2,3  \\  
HBC\,732, V350\,Cep, $[$MMN2004b$]$\,13  &             & 21430000+6611279 & NGC 7129 31 & NGC\,7129 & CTTS & 3,10,11  \\
GGD 33A                           &             & 21430320+6611150 & NGC 7129 5 & NGC\,7129 & HH, H$\alpha$ & 10 \\
$[$MMN2004b$]$\,17                 &             & 21431161+6609114 & NGC 7129 43 & NGC\,7129 & H$\alpha$ & 11  \\     
$[$K98c$]$\,Em*\,73                           &             & 21443229+7008130 &\nodata & \nodata & H$\alpha$  & 8 \\
BH Cep, HBC\,734, $[$K98c$]$\,Em*\,83         & 22006+6930  & 22014287+6944364 &\nodata & L\,1219? & HAeBe & 3 \\ 
$[$K98c$]$\,Em*\,95                           &             & 22131219+7332585 &\nodata & L\,1235? & H$\alpha$ & 8 \\
GSC 04467-00835                   & 22129+6949  & 22140608+7005043 &\nodata & L\,1219 &  IRAS  & 8 \\ 
TYC2 4467-324-1                   & 22144+6923  & 22154189+6938566 &\nodata & L\,1219 & IRAS & 8 \\
TYC 4467-836-1                    & 22152+6947  & 22163111+7002393 &\nodata & L\,1219 &  IRAS  & 8 \\
BO Cep, HBC\,735, $[$K98c$]$\,Em*\,100        & F22156+6948 & 22165406+7003450 &\nodata & L\,1219? &  HAeBe & 3  \\ 
$[$K98c$]$\,Em*\,109                          &             & 22190169+7346072 &\nodata & L\,1235? & H$\alpha$  & 8 \\
$[$K98c$]$\,Em*\,108                          &             & 22190203+7319252 &\nodata & L\,1235? & H$\alpha$  & 8 \\
SV Cep, HBC\,736, $[$K98c$]$\,Em*\,113        & 22205+7325  & 22213319+7340270 &\nodata & \nodata  &  HAeBe & 3 \\
TYC\,4608-2063-1                  & 22219+7908  & 22220233+7923279 &\nodata &\nodata &  IRAS & 8 \\
$[$K98c$]$\,Em*\,119 N                        & 22256+7102  & 22265660+7118011 &\nodata & \nodata& H$\alpha$  & 8 \\ 
$[$K98c$]$\,Em*\,119 S                        & 22256+7102  & 22265660+7118011 &\nodata & \nodata& H$\alpha$  & 8 \\
$[$KP93$]$ 2-1, XMMU J223412.2 +751809    & 22331+7502  & 22341189+7518101 & 142 & L\,1251 & H$\alpha$  & 6,13 \\
$[$KP93$]$ 39, XMMU J223516.6 +751848   &             & 22351668+7518471 & 76 & L\,1251 & H$\alpha$  & 6,13 \\
$[$RD95$]$~B, VLA 10?              &  22343+7501 & 22352442+7517037 & 143? & L\,1251 & IR, radio? & 14,15,16  \\
$[$RD95$]$~A, VLA 7, VLA B         &  22343+7501 & 22352497+7517113 & 143? & L\,1251 & IR, radio & 14,15,16 \\
$[$RD95$]$~2         &             & 22352542+7517562 & 77 & L\,1251 & K-band excess & 16 \\
$[$RD95$]$~5         &             & 22352722+7518019 & 79 & L\,1251 & K-band excess & 16 \\
$[$KP93$]$ 2-2, XMMU J223605.8 +751831    & 22350+7502  & 22360559+7518325 & 80 & L\,1251 & H$\alpha$  & 6,13 \\
$[$KP93$]$ 3-10                         & 22355+7505  & 22363545+7521352 & 81 & L\,1251 & IRAS  & 6 \\
$[$KP93$]$ 2-43, XMMU J223727.7 +751525   &             & 22372780+7515256 & & L\,1251 & H$\alpha$ & 6,13 \\
$[$KP93$]$ 2-3, XMMU J223750.1 +750408    &             & 22374953+7504065 & 83 & L\,1251 &  H$\alpha$  & 6,13 \\
$[$ETM94$]$ IRAS 22376+7455 Star 3, & & 22381872+7511538 & 86 & L\,1251 &  H$\alpha$  & 6,10  \\
~~~~~~XMMU J223818.8 +751154 \\
$[$KP93$]$ 2-44, $[$ETM94$]$IRAS 22376+7456 Star 1,  &  & 22384249+7511455 & 107 & L\,1251 &  H$\alpha$  & 6,7,13 \\
~~~~~~XMMU J223842.5+751146 & \\
$[$KP93$]$ 2-45, XMMU J223927.3 +751029   &            & 22392717+7510284 & 96 & L\,1251 &  H$\alpha$  & 6,13 \\
$[$KP93$]$ 2-46, XMMU J223942.9 +750644   & 22385+7457 & 22394030+7513216 & 97 & L\,1251 &  H$\alpha$ & 6,13 \\
GSC 04601-03483                   & F22424+7450 & 22433926+7506302 &\nodata & L\,1251 & IRAS  & 8 \\
$[$K98c$]$\,Em*\,128                          &             & 22490470+7513145 &\nodata & L\,1251 &  H$\alpha$ & 8 \\ 
TYC 4601-1543-1                   & 22480+7533  & 22491626+7549438 &\nodata & L\,1251 &  IRAS & 8 \\ 
$[$K98c$]$\,Em*\,137                          &             & 23143398+7350022 &\nodata & L\,1261 & H$\alpha$ & 8 \\        
AS\,507, HBC\,741, V395 Cep, $[$K98c$]$\,Em*\,140 & 23189+7357  & 23205208+7414071 &\nodata & L\,1261 & CTTS & 3 \\ 
\enddata 
\tablerefs{(1) \citet{Rosino}; (2) \citet{SVS76}; (3) \citet{HBC}; 
(4) \citet{Sellgren};  (5) \citet{OS90}; (6) \citet{KP93}; (7) \citet{ETM94}; (8) \citet{Kun98};
(9) \citet{Semkov};  (10) \citet{MEFG94};  (11) \citet{MMN04}; (12) \citet{Padgett};
(13) \citet{Simon}; (14) \citet{Beltran01}; (15) \citet{Meehan}; (16) \citet{RD95}; (17) \citet{Cutri};
 (18) \citet{MM04}; (19) \citet{Semkov93}.}
\tablenotetext{*}{This column shows the identifier of the stars observed by Spitzer in 
\citet{Guter09} (NGC 7129), and in \citet{Kirk09} (other stars).}
\tablenotetext{a}{Binarity of the star embedded in RNO\,129 was reported by \citet{MMN04}. Our observations
have shown that both components are PMS stars.}
\tablenotetext{b}{Binarity of NGC\,7023\,RS2 and Lk\ha 428 was reported by \citet{Rosino}. Our observations
have shown that both components of these doubles are PMS stars.}
\end{deluxetable} 


\clearpage


\clearpage 

\begin{deluxetable}{lllll}
\tabletypesize{\scriptsize}
\tablecolumns{5}
\tablecaption{List of the observed stars rejected as PMS stars\label{Tab_rej}}
\tablehead{
\colhead{Name}  &  \colhead{2MASS} &  \colhead{Instrument} & \colhead{Comment} & \colhead{Ref.}} 
\startdata 
$[$K98c$]$\,Em*\,2 &  20200929+7351431  & CAFOS+R100 & dMe \\
$[$K98c$]$\,Em*\,3 &  20220638+7301104 & CAFOS+R100 & dMe \\
$[$K98c$]$\,Em*\,4 & 20233729+6807137 & CAFOS+R100 & planetary nebula \\
$[$K98c$]$\,Em*\,7 & 20422058+7229495  & FAST & no \ha\ emission \\
F20432+7457, 4C 74.26  & 20423730+7508024 & CAFOS+R100 & quasar & 1 \\
OSHA 1 & 20433097+7903247 & CAFOS+R100 & no \ha\ emission  \\
OSHA 2 & 20441641+7659130 & CAFOS+R100 &  no \ha\ emission \\
OSHA 3 & 20445359+7542255 & CAFOS+R100 & no \ha\ emission  \\  
\enddata 
\tablecomments{Table \ref{Tab_rej} is published in its entirety in the 
electronic edition of the {\it Astrophysical Journal}.  A portion is 
shown here for guidance regarding its form and content.}
\tablerefs{(1) \citet{Veron}}
\end{deluxetable} 

\clearpage 

\clearpage


\begin{deluxetable}{lcccl} 
\tabletypesize{\scriptsize}
\rotate
\tablecolumns{5} 
\tablewidth{0pc} 
\tablecaption{Summary of spectroscopic observationa\label{Tab_sp0}} 
\tablehead{ 
\colhead{Id.} &  \colhead{Instrument} &  \colhead{Sp.T.}   & \colhead{H$\alpha$ shape\tablenotemark{a}}  & 
\colhead{Other emission lines} } 
\startdata 
RNO\,124          & CAFOS+R100 &  K0       & I     &  $[$\ion{O}{1}$]$\,6300, 6363, \ion{He}{1}\,5875, 6678, \ion{O}{1}\,7773, 8446,
                                            \ion{Ca}{2} triplet, Pa\,11,12,14  \\
PV Cep            & CAFOS+R100 &   cont.   & IIIB  &  $[$\ion{O}{1}$]$\,6300, 6363, \ion{Fe}{2}\,6456, 6516, \ion{He}{1}\,5875, 
                                           6678, $[$\ion{S}{2}$]$\,6716, 6731, $[$\ion{Fe}{2}$]$\,7155 \\
                                          &&&& \ion{K}{1}\,7664, 7698, \ion{O}{1}\,7773, 8446, \ion{Ca}{2} triplet, Pa\,11,12,13,14,19   \\ 
IRAS 20535+7439   & CAFOS+R100, FAST & F8  & IVR   &  \\ 
OSHA\,42         & CAFOS+R100 &   M0       & I     &  $[$\ion{O}{1}$]$\,6300, 6363, \ion{He}{1}\,6678, $[$\ion{S}{2}$]$\,6716, 6731 \\
                                          &&&& \ion{O}{1}\,7773, 8446, \ion{Ca}{2} triplet \\ 
RNO\,129\,A       & CAFOS+R100 &   M1      & I     &   $[$\ion{O}{1}$]$\,6300, 6363 \\ 
RNO\,129\,SW    & CAFOS+R100, FAST & M1.5  & IIIB  &  $[$\ion{O}{1}$]$\,6300, 6363, $[$\ion{S}{2}$]$\,6716, 6731  \\ 
                                          &&&& $[$\ion{Fe}{2}$]$\,7155, \ion{Ca}{2} triplet \\
RNO\,129\,NE      & CAFOS+R100, FAST & M2  & IIIB  & \ion{Ca}{2} H\&K, Balmer ser.,  \ion{Fe}{2}\,4351, \ion{Fe}{1}\,4922, 5016 \\
                                        &&&& $[$\ion{O}{1}$]$\,6300, 6363, $[$\ion{S}{2}$]$\,6716, 6731, $[$\ion{Fe}{2}$]$\,7155 \\ 
FT Cep           & CAFOS+R100 &   K1       & I     &  $[$\ion{O}{1}$]$\,6300, \ion{O}{1}\,7773, 8446, \ion{Ca}{2} triplet \\ 
$[$K98c$]$\,Em*\,35           & CAFOS+R100 &   M2      & I     &   \\ 
F21022+7729       & CAFOS+R100 &   M1      & IIB   &  $[$\ion{O}{1}$]$\,6300, \ion{O}{1}\,7773, 8446, \ion{Ca}{2} triplet, Pa\,11,12,14,19 \\
NGC\,7023 RS\,2    & CAFOS+R100 & M0IV     & IIB   &  $[$\ion{O}{1}$]$\,6300  \\
NGC\,7023 RS\,2\,B & CAFOS+R100 & M0     & I     &  \\ 
OSHA\,48         & CAFOS+R100, FAST & K5   & I    &  \\
OSHA\,48 B        & CAFOS+R100 & cont.     & I    &   \\ 
OSHA\,49          & CAFOS+R100, FAST & M1  & I    & \\ 
OSHA\,50          & CAFOS+R100, FAST & M0  & I    & \ion{Ca}{2} H\&K, Balmer ser., \ion{Fe}{1}\,4922, 5016, $[$\ion{O}{1}$]$\,6300 \\ 
                                          &&&& 6363, \ion{He}{1}\,5875, 6678, \ion{Ca}{2} triplet \\ 
LkH$\alpha$ 425   & CAFOS+R100 &    M0     & I     & $[$\ion{O}{1}$]$\,6300, 6363, \ion{He}{1}\,6678, $[$\ion{S}{2}$]$ 6716, 6731, \ion{O}{1}\,8446 \\
                                          &&&&  \ion{Ca}{2} triplet  \\
SX Cep            & CAFOS+R100 &   K7      & abs.  &  \\ 
NGC\,7023 RS\,5   & CAFOS+R100 &   M2      & I     & $[$\ion{O}{1}$]$\,6363, \ion{O}{1}\,8446, \ion{Ca}{2} triplet \\ 
HZ Cep            & CAFOS+R100 &   M0      & I     &  \\  
FU Cep            & CAFOS+R100 &   M1      & IIIB  &  \ion{He}{1}\,6678, \ion{O}{1}\,8446, \ion{Ca}{2} triplet  \\ 
$[$PRN2004$]$ S3  & CAFOS+R100&   M1	 & I  &  \\
OSHA\,53         & CAFOS+R100, FAST & K5 &  I	 &  H$\beta$ \\
FV Cep            & CAFOS+R100 &   K5      & I    & \ion{He}{1}\,6678 \\ 
LkH$\alpha$ 428 N & CAFOS+R100 &   K7      & I    & \ion{O}{1}\,8446, \ion{Ca}{2} triplet  \\ 
LkH$\alpha$ 428 S & CAFOS+R100 & M4.5IV    & I    & \\
FW Cep           & CAFOS+R100 &   K5	 &  I	 &    \\
NGC\,7023 RS\,10 & CAFOS+R100 &   K4	 &  I	 &   \\ 
$[$K98c$]$\,Em*\,46          & CAFOS+R100 &   M2	 &  I	 &   \\ 
EH Cep           & CAFOS+R100 &   K2	 &  I	 & \ion{Ca}{2} triplet \\ 
OSHA\,59         & CAFOS+R100, FAST & K5  &  I    &   \\ 
$[$K98c$]$\,Em*\,53          & CAFOS+R100, FAST & K5  &  I    & \ion{Ca}{2} H\&K, Balmer ser., \ion{He}{1}\,5875, 6678, \ion{O}{1}\,7773, 8446, 
                                           \ion{Ca}{2} triplet \\ 
BD +68\degr1118       & FAST &   A2           &  I    & \\ 
$[$K98c$]$\,Em*\,58          & CAFOS+R100, FAST & K7  &  IIIB & $[$\ion{O}{1}$]$\,6300, 6363, $[$\ion{S}{2}$]$\,6716, 6731, \ion{Ca}{2} triplet \\ 
2MASS 21225427+6921345 & CAFOS+R100 &  K3 & I    &   $[$\ion{S}{2}$]$\,6717 \\
RNO\,135          & CAFOS+R100 &   M3IV   &  I    &    \\
NGC\,7129 S V1    & CAFOS+R100 &   K7     &  I    &  \ion{He}{1}\,6678 \\
NGC\,7129 S V2    & CAFOS+R100 &   M0     &  IIB  &    \\ 
V391 Cep, $[$K98c$]$\,Em*\,71 & CAFOS+R100 &   K5     &  IVB  &  $[$\ion{O}{1}$]$\,6300, \ion{O}{1}\,7773, 8446, 6431, \ion{Fe}{2} 6456, \ion{He}{1}\,6678 \\
                                          &&&& \ion{Ca}{2} triplet, \ion{Fe}{1} 8387, 8514, Pa\,11,12,14,19 \\ 
NGC\,7129 S V3    & CAFOS+R100 &   K5     &  I    &   \\ 
$[$K98c$]$\,Em*\,72           & CAFOS+R100 &   K4     &  IIIB & $[$\ion{O}{1}$]$\,6300, \ion{O}{1}\,7773, 8446, \ion{Ca}{2} triplet \\ 
HBC 731             & CAFOS+G100 &   K2     &  I    & \\
V350\,Cep         & CAFOS+R100 &   M0     &  I    &   $[$\ion{O}{1}$]$\,6300, \ion{He}{1}\,6678, $[$\ion{S}{2}$]$\,6716, 6731, \ion{Fe}{2}\,7711 \\
                                          &&&& \ion{O}{1}\,7773, 8446, \ion{Fe}{1} 8327, \ion{Ca}{2} triplet \\ 
GGD 33A           & CAFOS+R100 & M0    &  IIB  & $[$\ion{O}{1}$]$\,6300, 6363, $[$\ion{S}{2}$]$ 6716, 6731, \ion{Fe}{1}\,7282, 
                                            \ion{O}{1}\,8446, \ion{Ca}{2} triplet \\
$[$MMN2004b$]$\,17 & CAFOS+R100 &   M1     &  I    &   \\ 
$[$K98c$]$\,Em*\,73           & CAFOS+R100 &   K7     &  I    &   \ion{He}{1}\,6678, \ion{O}{1}\,8446, \ion{Ca}{2} triplet \\ 
BH Cep            & CAFOS+G100 &    F5    &  IIB  &   \\
$[$K98c$]$\,Em*\,95           & CAFOS+R100, FAST & K5 &  IIB  &  \ion{O}{1}\,7773, 8446, \ion{Ca}{2} triplet \\
IRAS 22129+6949   & CAFOS+G100, FAST & K0 &  I    &  \\ 
IRAS 22144+6923   & CAFOS+R100   &  F6    & IIR & \ion{O}{1}\,8446, \ion{Ca}{2} triplet \\ 
IRAS 22152+6947W  & CAFOS+G100, FAST  & F5 &  abs. &  \\
BO Cep            & CAFOS+G100 &    F5    & IIR   &   \\
$[$K98c$]$\,Em*\,109         & CAFOS+R100, FAST  & K2 &  IIR  &  Balmer ser., \ion{He}{1}\,5875, 6678, $[$\ion{O}{1}$]$\,6300, 
                                          $[$\ion{S}{2}$]$\,6731, \ion{O}{1}\,7773, 8446, \ion{Ca}{2} triplet \\ 
$[$K98c$]$\,Em*\,108         & CAFOS+R100, FAST  & K2 &  I    &   \\
SV Cep            & CAFOS+G100 &    A2    &   I   &  \\ 
IRAS F22219+7908  & CAFOS+R100, FAST & A1 &  I    &  \ion{O}{1}\,7773, 8446, \ion{Ca}{2} triplet   \\
$[$K98c$]$\,Em*\,119\,N        & CAFOS+R100, FAST  & K5 &  I    &   \ion{He}{1}\,6678, \ion{O}{1}\,8446, $[$\ion{S}{2}$]$ 6731, \ion{Ca}{2} triplet \\
$[$K98c$]$\,Em*\,119\,S        & CAFOS+R100, FAST &  K0 &  I    &  \ion{Ca}{2} H\&K, H$\beta$, \ion{Fe}{1}\,4922, \ion{Na}{1} D, $[$\ion{O}{1}$]$\,6300, \ion{O}{1}\,8446\\
                                          &&&&  \ion{He}{1}\,5875, 6678, 7065, $[$\ion{S}{2}$]$\,6716, 6731,
					  $[$\ion{Fe}{2}$]$\,7155 \\ 
$[$KP93$]$ 2-1         & CAFOS+G100 &   K2      &  IIB  & H$\beta$, \ion{Fe}{2}\,4923, $[$\ion{O}{1}$]$\,6300, 6364, \ion{He}{1}\,5875, 6678,
 $[$\ion{S}{2}$]$\,6717, 6731 \\ 
$[$KP93$]$ 2-39           & CAFOS+G100 &   M0     &  I    &  \\ 
$[$RD95$]$\,B   & CAFOS+R100 &   M0      &  I    &  $[$\ion{O}{1}$]$\,6300, \ion{Ca}{2} triplet  \\ 
$[$RD95$]$\,A   & CAFOS+R100 &   K7      &  I    &  $[$\ion{O}{1}$]$\,6300, 6363, \ion{Ca}{2} triplet \\
$[$RD95$]$ 2 & CAFOS+R100 & M2.5IV   &  I &  \ion{Ca}{2} triplet  \\
$[$RD95$]$ 5 &CAFOS+R100& M0IV       &  I   & $[$\ion{O}{1}$]$\,6300 \\
$[$KP93$]$ 2-2           & CAFOS+R100 &    M2    &  I    &  $[$\ion{O}{1}$]$\,6300, \ion{He}{1}\,6678, $[$\ion{S}{2}$]$\,6716, 6731,
 \ion{O}{1}\,7773, 8446 \\
IRAS 22355+7505   & CAFOS+G200 &    M4    &  I    &  \\ 
$[$KP93$]$ 2-43          & FAST &   K5           &  I    & \\ 
$[$KP93$]$ 2-3           & CAFOS+G100 &   K7     &  I    &  \\
$[$ETM94$]$ Star 3      & CAFOS+R100 &   K2     &  IIB  &  \ion{O}{1}\,8446, \ion{Ca}{2} triplet  \\
$[$KP93$]$ 2-44          & CAFOS+G100 &     M1.5 &  I    & H$\beta$, $[$\ion{O}{1}$]$\,6300, 6363, $[$NII$]$\,6584, $[$\ion{S}{2}$]$\,6716, 6731 \\ 
$[$KP93$]$ 2-45          & CAFOS+G100 &    M1    &  I    & H$\beta$, \ion{He}{1}\,5875, 6678  \\
$[$KP93$]$ 2-46          & CAFOS+G100 &    K5    &  I    & \\
IRAS F22424+7450   & CAFOS+R100, FAST & K4 &  I &   \\
$[$K98c$]$\,Em*\,128          & CAFOS+R100 &  M0      &  IIIB &   \ion{He}{1}\,6678, \ion{O}{1}\,7773, \ion{Ca}{2} triplet, Pa\,11,12,14 \\
IRAS 22480+7533   & CAFOS+R100 &  K1      &  IIB  &  \ion{O}{1}\,7773, 8446, \ion{Ca}{2} triplet   \\
$[$K98c$]$\,Em*\,137          & CAFOS+R100 &  M2      &  I    &  \ion{He}{1}\,6678  \\ 
AS\,507           & CAFOS+G100 &    G2    &  I    &  \\ 
\enddata 
\tablenotetext{a}{The type of the \ha\
emission line is indicated using the scheme introduced by
\citet*{RPL}, i.e. type I lines are symmetric, 
IIR and IIB are double-peaked with a secondary peak higher than half
of the primary located on its red and the blue side,
respectively. Type III lines are double-peaked with a secondary peak
lower than half of the primary, and IVB and IVR denotes the P Cygni
and reversed P Cygni type profiles, respectively.}
\end{deluxetable} 


\clearpage


\begin{deluxetable}{lccccccc} 
\tabletypesize{\scriptsize}
\tablecolumns{8} 
\tablewidth{0pc} 
\tablecaption{Spectral characteristics of the observed stars\label{Tab_sp1}} 
\tablehead{ 
\colhead{Id.} & \multicolumn{2}{c}{H$\alpha$ line} & \multicolumn{3}{c}{EW(CaII triplet)} & \colhead{EW(LiI)} & 
\colhead{PMS Class} \\
\cline{2-3} \cline{4-6} \\
  & \colhead{EW} & $\Delta$v(10\%)   & \colhead{8499} & \colhead{8545} & \colhead{8663} & \\
& \colhead{(\AA)} & (km\,s$^{-1}$) &  \colhead{(\AA)} & \colhead{(\AA)} & \colhead{(\AA)} & \colhead{(\AA)}} 
\startdata 
RNO\,124          & $-$52.2 & 615   & $-$9.3  & $-$8.7  & $-$8.3 & 0.30  & CTTS \\
PV Cep            & $-$71.2 & 481   & $-$76.5  & $-$71.6  & $-$68.0 & \nodata & CTTS  \\ 
IRAS 20535+7439   & \phn$-$3.8  & 399   &\nodata  &\nodata  &\nodata  & 0.32 & CTTS	\\ 
OSHA\,42          & $-$82.3 & 516   & $-$4.9  & $-$2.9  & $-$2.3 & \nodata & CTTS    \\ 
RNO\,129\,A       & \phn$-$6.0 & 430   &\nodata &\nodata &\nodata & \nodata  & WTTS   \\ 
RNO\,129\,SW      & $-$36.8 & 658   & $-$8.4  & $-$8.3  & $-$4.6 & \nodata  & CTTS  \\
RNO\,129\,NE      & \phn$-$5.0  & 622   & $-$1.2  & $-$0.8  & $-$0.8 & \nodata & CTTS     \\ 
FT Cep            & \phn$-$9.6  & 573   & $-$2.8  & $-$2.1  & $-$2.2 & 0.36 & CTTS  \\ 
$[$K98c$]$\,Em*\,35           & \phn$-$7.4  & 426   &\nodata &\nodata &\nodata & 0.48  & CTTS  \\ 
F21022+7729       & $-$91.0 & 793   & $-$10.8  & $-$13.3  & $-$12.9 & \nodata  & CTTS	\\
NGC\,7023 RS\,2    & $-$59.8 & 543   & $-$0.6  & $-$0.6  & $-$0.9 & \nodata   & CTTS   \\
NGC\,7023 RS\,2\,B& $-$63.9 & 488   &\nodata &\nodata &\nodata  & \nodata   & CTTS  \\ 
OSHA\,48          & $-$12.6,$-$2.3 & 399   &\nodata &\nodata &\nodata & 0.48  & CTTS	 \\
OSHA\,48 B        & $-$79.6 & 551   &\nodata &\nodata &\nodata  & \nodata  & CTTS   \\ 
OSHA\,49    & $-$7.3 & 283 &\nodata &\nodata &\nodata & 0.60  & CTTS  \\ 
OSHA\,50 & $-$65.0, $-$78.8 & 487   & $-$11.0  & $-$11.0  & $-$10.8 & \nodata  & CTTS	\\ 
LkH$\alpha$ 425   & $-$106.0& 563   & $-$21.5  & $-$22.3  & $-$18.5 & \nodata  & CTTS	\\
SX Cep            & ~~2.15 &\nodata & +1.2  & +2.3  & +0.9 & 0.61 & WTTS   \\ 
NGC\,7023 RS\,5   & $-$28.8 & 544   &\nodata  & \nodata & \nodata & 0.50  & CTTS   \\ 
HZ Cep            & \phn$-$3.9  & 239   &\nodata &\nodata &\nodata & 0.46 & WTTS  \\ 
FU Cep            & $-$64.5 & 691   & $-$3.5  & $-$3.4  & $-$3.1 & \nodata  & CTTS   \\ 
$[$PRN2004$]$ S3  & \phn$-$9.0  & 526   &\nodata &\nodata &\nodata & \nodata   & CTTS   \\
OSHA\,53 & $-$20.5, $-$28.4 & 515  &\nodata &\nodata &\nodata & \nodata  & CTTS   \\
FV Cep            & $-$11.5 & 378    &\nodata &\nodata &\nodata & 0.55  & CTTS  \\ 
LkH$\alpha$ 428 N & $-$56.0 & 624    & $-$2.1  & $-$2.6  & $-$1.4 & 0.63  & CTTS   \\ 
LkH$\alpha$ 428 S & \phn$-$7.5  & 361    &\nodata  &\nodata  &\nodata  & 0.50  & CTTS   \\
FW Cep            & \phn$-$4.0  & 596    &\nodata &\nodata &\nodata & 0.56  & CTTS   \\
NGC\,7023 RS\,10  & \phn$-$8.1  & 272    & +0.5  & +1.6  & +1.0 & \nodata	 & CTTS  \\ 
$[$K98c$]$\,Em*\,46           & \phn$-$5.3  & 321    &\nodata &\nodata &\nodata & \nodata  & CTTS  \\ 
EH Cep            & $-$13.1 & 665   & $-$4.0  & $-$2.8  & $-$2.9 & 0.32 & CTTS  \\ 
OSHA\,59 & $-$27.0, $-$13.7 & 632  & $-$4.1  & $-$4.3  & $-$3.5 & 0.38 & CTTS    \\ 
$[$K98c$]$\,Em*\,53           & $-$58.0 & 565   & \nodata  &\nodata &\nodata & 0.55 & CTTS  \\ 
BD +68\degr1118   & $-$24.1 & \nodata	&\nodata &\nodata &\nodata & \nodata & HAe  \\ 
$[$K98c$]$\,Em*\,58           & $-$38.3 & 727     & $-$3.5  & $-$2.7  & $-$2.1 & \nodata& CTTS \\ 
2MASS J\,21225427 & \phn$-$7.3  & 385  &\nodata  &\nodata  &\nodata  & 0.80 & CTTS  \\
RNO\,135          & \phn$-$5.0  & 310  & $-$0.4  & +0.9  & +1.0 & 0.27 & WTTS   \\
NGC\,7129 S V1    & $-$19.4 & 620   &\nodata &\nodata &\nodata & \nodata  & CTTS   \\
NGC\,7129 S V2    & $-$45.0 & 636   & $-$2.6  & $-$2.1  & $-$3.7 & \nodata & CTTS   \\ 
V391 Cep          & $-$62.0 & 455   & $-$29.8  & $-$36.3  & $-$31.4 & \nodata & CTTS   \\ 
NGC\,7129 S V3    & $-$76.5 & 684   &\nodata &\nodata &\nodata & 0.86 & CTTS 	 \\ 
$[$K98c$]$\,Em*\,72           & $-$34.0 & 687    & $-$12.0  & $-$11.1  & $-$8.6 & 0.86  & CTTS   \\ 
HBC 731             & $-$36.5 & 509   &\nodata  & \nodata &\nodata &\nodata & CTTS \\
V350\,Cep         & $-$70.0 & 495   & $-$15.9  & $-$19.7  & $-$17.1 & \nodata & CTTS  \\ 
GGD 33A           & $-$93.8 & 764   & $-$57.1  & $-$64.9  & $-$80.8 & \nodata & CTTS 	  \\
$[$MMN2004b$]$\,17 & $-$11.0 & 515   &\nodata  &\nodata &\nodata & 0.37  & CTTS   \\ 
$[$K98c$]$\,Em*\,73           & $-$87.0 & 618   & $-$0.8  & $-$0.5  &  & 0.42  & CTTS   \\ 
BH Cep            & \phn$-$2.0  & \nodata  &  &&& \nodata & HAe \\
$[$K98c$]$\,Em*\,95           & $-$23.8 & 540   & $-$2.0  & $-$2.6  & $-$1.9 & 0.42  & CTTS  \\
IRAS 22129+6949   & \phn$-$9.4, $-$9.1  & 446  &\nodata &\nodata &\nodata & 0.25  & CTTS   \\ 
IRAS 22144+6923   & \phn$-$7.8  & 676 & $-$2.5  & 0.9 & 0.9 & 0.22 & CTTS \\ 
IRAS 22152+6947   & \phn2.4 &  &\nodata &\nodata &\nodata & 0.20 & WTTS? \\
BO Cep            & \phn$-$6.8  & 712 & \nodata  &\nodata &\nodata & \nodata & HAe  \\
$[$K98c$]$\,Em*\,109          & $-$52.9 & 770   &\nodata &\nodata &\nodata & 0.54 & CTTS   \\ 
$[$K98c$]$\,Em*\,108          & $-$17.4 & 520   &\nodata &\nodata &\nodata & 0.50 & CTTS 	 \\
SV Cep            & \phn$-$8.0  & \nodata & \nodata  &\nodata &\nodata & 0.27 & HAe \\ 
IRAS F22219+7908  & $-$31.0, 5.4 & 472 & $-$1.9  & $-$1.7 &$-$1.3 & \nodata & HAe?   \\
$[$K98c$]$\,Em*\,119\,N         & $-$61.8 & 500   & $-$19.2  & $-$19.2  & $-$13.6 & 0.44  & CTTS   \\
$[$K98c$]$\,Em*\,119\,S         & $-$37.4 & 547    &\nodata &\nodata &\nodata & 0.25 & CTTS   \\ 
$[$KP93$]$ 2-1              & $-$81.5 & 803   &\nodata &\nodata &\nodata & \nodata  & CTTS  \\ 
$[$KP93$]$ 2-39             & \phn$-$6.0  & 662   &\nodata &\nodata &\nodata &  0.37  & CTTS   \\ 
$[$RD95$]$ B    & $-$27.4 & 461   & $-$1.6  & $-$1.2  & $-$1.5 & \nodata & CTTS  \\ 
$[$RD95$]$ A    & $-$24.6 & 537    & $-$1.3 & $-$1.1 & $-$0.6 & \nodata  & CTTS   \\
$[$RD95$]$ 2  & $-$46.6 & 465  & $-$3.8 & $-$3.15 & $-$2.42 & & CTTS \\
$[$RD95$]$ 5  & $-$19.8 & 324   &  & & & 0.58 & CTTS \\
$[$KP93$]$ 2-2              & $-$69.5 & 538   & $-$14.7  & $-$15.3  & $-$12.6 & \nodata	& CTTS \\
IRAS 22355+7505   & $-$30.8 & 620    &\nodata &\nodata &\nodata & \nodata  & CTTS  \\ 
$[$KP93$]$ 2-43             & \phn$-$3.2  & 316    &\nodata &\nodata &\nodata & \nodata & WTTS   \\ 
$[$KP93$]$ 2-3              & $-$23.6 & 746    &\nodata &\nodata &\nodata &  0.36  & CTTS  \\
$[$ETM94$]$ Star 3          & \phn$-$7.1  & 776   &$-$2.0 & $-$0.6 & $-$0.9 & 0.26 & CTTS   \\
$[$KP93$]$ 2-44             & $-$12.7 & 479    &\nodata &\nodata &\nodata & 0.44 & CTTS  \\ 
$[$KP93$]$ 2-45             & $-$33.2 & 432    &\nodata &\nodata &\nodata & 0.32 & CTTS   \\
$[$KP93$]$ 2-46             & $-$57.8 & 566    &\nodata &\nodata &\nodata & 0.55 & CTTS  \\
IRAS F22424+7450  & \phn$-$0.4  & \nodata  &\nodata &\nodata &\nodata & 0.57 & WTTS  \\
$[$K98c$]$\,Em*\,128          & $-$83.5 & 630     & $-$3.5  & $-$3.5  & $-$3.1  & \nodata& CTTS  \\
IRAS 22480+7533   & $-$25.7 & 673   & $-$2.8  & $-$3.2  & $-$2.9 & \nodata & CTTS  \\
$[$K98c$]$\,Em*\,137          & $-$44.3 & 484     &\nodata &\nodata &\nodata & \nodata & CTTS \\ 
AS\,507           & \phn$-$9.4  & 625     &\nodata &\nodata &\nodata & 0.36 & CTTS  \\ 
\enddata 
\end{deluxetable} 


\clearpage


\def\ha{\relax \ifmmode {\rm H}\alpha\else H$\alpha$ \fi}

\begin{deluxetable}{lccccl} 
\tabletypesize{\scriptsize}
\tablecolumns{6} 
\tablewidth{0pc} 
\tablecaption{Photometric data of the observed stars. Numbers in parentheses indicate the 
mean absolute deviation of the derived magnitudes. \label{Tab_phot}} 
\tablehead{ 
\colhead{Id.}   &   \colhead{B}   & \colhead{V} & \colhead{R$_C$}   & \colhead{I$_C$}  & 
\colhead{Comment}}
\startdata 
RNO 124              &  \nodata    &    17.81\,(0.25) & 16.02\,(0.12)  & 14.34\,(0.12)  & var. $BVR_\mathrm{C}I_\mathrm{C}$ \\  
PV Cep               & 19.27\,(0.85)  & 17.46\,(0.70) & 15.98\,(0.50)  & 14.60\,(0.40) & var. $BVR_\mathrm{C}I_\mathrm{C}$ \\ 
IRAS 20535+7439      & 13.44\,(0.05)  & 12.37\,(0.03) & 11.76\,(0.03)  & 11.11\,(0.03) &  \\ 
OSHA\,42             & 18.36\,(0.30)  & 16.74\,(0.30) & 15.38\,(0.15)  & 14.06\,(0.20) & var. $BVR_\mathrm{C}I_\mathrm{C}$ \\ 
RNO\,129\,A	     & 20.76\,(0.05)  & 18.76\,(0.03) & 16.83\,(0.04)  & 14.89\,(0.02) &  \\   
RNO\,129\,SW         & 17.43\,(0.07)  & 15.68\,(0.07) & 14.17\,(0.07)  & 12.75\,(0.07) & \\  
RNO\,129\,NE	     & \nodata        & 19.60\,($\cdots$) & 17.86\,($\cdots$)  & 15.85\,($\cdots$) & 2.1\arcsec, 61\fdg3  \\ 
FT Cep               & 19.37\,(1.10)  & 16.41\,(0.83) & 14.79\,(0.80)  & 13.30\,(0.95) & var. $BVR_\mathrm{C}I_\mathrm{C}$ \\ 
FT Cep B             & 18.59\,(0.07)  & 16.97\,(0.01) & 15.90\,(0.01)  & 14.97\,(0.02) & 5.4\arcsec, 27\degr  \\
$[$K98c$]$\,Em*\,35  & 17.04\,(0.07)  & 15.64\,(0.11) & 14.04\,(0.25)  & 12.72\,(0.25) & var. V$R_\mathrm{C}I_\mathrm{C}$ \\ 
F21022+7729          & 20.30\,(0.72)  & 19.89\,(0.65) & 18.09\,(0.80)  & 16.22\,(0.45) & var. $BVR_\mathrm{C}I_\mathrm{C}$ \\   
NGC\,7023 RS\,2      & 18.61\,(0.05)  & 16.50\,(0.03) & 14.90\,(0.03)  & 13.25\,(0.07) & \\ 
NGC\,7023 RS\,2\,B   & 20.37\,(0.30)  & 18.26\,(0.42) & 16.63\,(0.65)  & 14.98\,(0.73) & var. $BVR_\mathrm{C}I_\mathrm{C}$  \\ 
OSHA\,48	     & 15.64\,(0.12)  & 14.31\,(0.05) & 12.75\,(0.04)  & 11.51\,(0.22) & var. $I_\mathrm{C}$ \\ 
OSHA\,48 B	     & 16.88\,(0.26)  & 15.37\,(0.30) & 13.85\,(0.15)  & 12.66\,(0.12) &  1.5\arcsec, 86\degr\\
OSHA\,49	     & 17.97\,(0.06)  & 16.58\,(0.05) & 14.89\,(0.05)  & 13.38\,(0.08) &  \\ 
OSHA\,50	     & 18.07\,(0.05)  & 17.11\,(0.05) & 15.38\,(0.05)  & 13.86\,(0.05) &   \\ 
OSHA\,50 B	     & 19.43\,($\cdots$)  & 18.44\,($\cdots$) & 16.599\,(0.11)  & 14.702\,(0.14) & 1.4\arcsec, 81\degr \\
LkH$\alpha$ 425      & 17.00\,(0.05)  & 15.95\,(0.02) & 15.04\,(0.02)  & 13.93\,(0.02) &  \\
SX Cep               & 16.21\,(0.16)  & 14.67\,(0.30) & 13.53\,(0.36)  & 12.40\,(0.38) & var. $BVR_\mathrm{C}I_\mathrm{C}$ \\ 
NGC\,7023 RS\,5      & 19.09\,(0.10)  & 17.29\,(0.10) & 15.84\,(0.51)  & 14.35\,(0.60) & var. $BVR_\mathrm{C}I_\mathrm{C}$ \\ 
NGC\,7023 RS\,5 B    & \nodata  & \nodata & 19.08\,($\cdots$)  & 17.501\,($\cdots$) &  1.5\arcsec, 158\degr  \\
HZ Cep               & 17.00\,(0.10)  & 15.22\,(0.06) & 14.01\,(0.10)  & 12.76\,(0.35) & var. $I_\mathrm{C}$ \\ 
FU Cep               & 17.70\,(1.72)  & 16.13\,(0.67) & 14.83\,(0.51)  & 13.47\,(0.03) & var. $BVR_\mathrm{C}$ \\ 
PRN S3               & 21.86\,(0.20)  & 18.94\,(0.21) & 16.73\,(0.34)  & 15.11\,(0.05) & var. $BVR_\mathrm{C}I_\mathrm{C}$ \\   
OSHA\,53             & 15.88\,(0.12)  & 14.37\,(0.16) & 13.35\,(0.14)  & 12.47\,(0.15) & var. $BVR_\mathrm{C}I_\mathrm{C}$ \\   
FV Cep               & 15.58\,(0.55)  & 14.45\,(0.40) & 13.56\,(0.29)  & 12.63\,(0.43) & var. $BVR_\mathrm{C}I_\mathrm{C}$ \\ 
LkH$\alpha$ 428 N    & 18.84\,(0.30)  & 16.89\,(0.06) & 15.32\,(0.01)  & 13.68\,(0.01) &  \\ 
LkH$\alpha$ 428 S    & 19.91\,(0.30)  & 17.73\,(0.02) & 15.86\,(0.01)  & 13.95\,(0.05) &  2.1\arcsec, 166\degr \\
FW Cep               & 16.17\,(0.10)  & 14.83\,(0.01) & 13.83\,(0.03)  & 12.98\,(0.03) &  \\ 
NGC\,7023 RS\,10     & 16.92\,(0.07)  & 15.81\,(0.05) & 15.04\,(0.05)  & 14.46\,(0.06) &  \\ 
$[$K98c$]$\,Em*\,46  & 17.11\,(0.05)  & 15.39\,(0.04) & 14.34\,(0.02)  & 12.99\,(0.02) &  \\
$[$K98c$]$\,Em*\,46B & \nodata  & \nodata & 17.50\,(0.10) & 15.22\,(0.08)  & 1.0\arcsec, 270\degr \\  
EH Cep               & 13.28\,(0.07)  & 12.16\,(0.02) & 11.42\,(0.02)  & 10.81\,(0.02) &  \\ 
OSHA\,59             & 15.47\,(0.07)  & 13.99\,(0.05) & 13.06\,(0.02)  & 12.21\,(0.02) &  \\   
$[$K98c$]$\,Em*\,53  & 15.83\,(0.10)  & 14.83\,(0.12) & 13.92\,(0.12)  & 13.06\,(0.14) & var. $BVR_\mathrm{C}I_\mathrm{C}$ \\   
BD +68\degr1118      & 10.08\,(0.04)  & \phn9.94\,(0.04) &  \phn9.811\,(0.03)  &  \phn9.750\,(0.03) &  \\ 
$[$K98c$]$\,Em*\,58  & 18.79\,(0.15)  & 17.62\,(0.20) & 16.77\,(0.12)  & 16.03\,(0.25) & var. $BVR_\mathrm{C}I_\mathrm{C}$ \\ 
2MASS 21225427+6921345 & \nodata& 19.27\,(0.25) & 17.78\,(0.12)  & 16.23\,(0.10) &  \\
RNO\,135	     &18.67\,(0.03)  & 16.72\,(0.01) & 15.13\,(0.01)  & 13.35\,(0.01) &   \\ 
NGC\,7129 S V1       &17.37\,(0.35)  & 16.18\,(0.36) & 15.20\,(0.40)  & 14.34\,(0.50) & var. $BVR_\mathrm{C}I_\mathrm{C}$ \\ 
NGC\,7129 S V2       &19.32\,(0.50)  & 17.72\,(0.45) & 16.53\,(0.42)  & 15.45\,(0.36) & var. $BVR_\mathrm{C}I_\mathrm{C}$ \\ 
V391 Cep 	     &15.24\,(0.06)  & 14.16\,(0.04) & 13.27\,(0.04)  & 12.50\,(0.03) &  \\ 
NGC\,7129 S V3       &19.55\,(0.45)  & 17.78\,(0.42) & 16.45\,(0.50)  & 15.34\,(0.55) & var. $BVR_\mathrm{C}I_\mathrm{C}$ \\
NGC\,7129 S V3\,B    &\nodata & \nodata &\nodata & 16.77\,($\cdots$) &  3.2\arcsec, 165\degr \\
$[$K98c$]$\,Em*\,72  & 17.22\,(0.06)  & 15.67\,(0.03) & 14.61\,(0.04)  & 13.57\,(0.03) &  \\
HBC 731             & \nodata &   19.34\,(0.15) & 17.69\,(0.12)  & 16.08\,(0.23) & var. $BVR_\mathrm{C}I_\mathrm{C}$\\
V350\,Cep           & 17.24\,(0.10)  & 16.20\,(0.10) & 15.23\,(0.06)  & 14.11\,(0.07) &  \\ 
GGD 33A             & \nodata &\nodata &  \nodata      & 18.49\,($\cdots$) &  \\ 
$[$MMN2004b$]$\,17  & 18.06\,(0.10)  & 16.31\,(0.04) & 15.23\,(0.36)  & 14.14\,(0.01) & var. $R_\mathrm{C}$ \\ 
$[$K98c$]$\,Em*\,73 & 15.58\,(0.05)  & 14.59\,(0.04) & 13.66\,(0.03)  & 12.90\,(0.03) &  \\ 
BH Cep              & 11.54\,(0.05)  & 11.01\,(0.03) & 10.62\,(0.03)  & 10.32\,(0.03) &  \\ 
$[$K98c$]$\,Em*\,95 & 15.25\,(0.10)  & 13.98\,(0.15) & 13.11\,(0.12)  & 12.39\,(0.07) & var. $VR_\mathrm{C}$ \\
$[$K98c$]$\,Em*\,95\,B	     & 18.379\,($\cdots$)  & 17.54\,($\cdots$) & 16.14\,($\cdots$)  & 14.502\,($\cdots$) & 1.8\arcsec, 83\degr  \\
IRAS 22129+6949      & 12.52\,(0.01)  & 11.63\,(0.04) & 11.01\,(0.02)  & 10.57\,(0.02) & \\ 
IRAS 22144+6923      & 12.40\,(0.04)  & 11.60\,(0.02) & 11.07\,(0.02)  & 10.63\,(0.02) &  \\ 
IRAS 22152+6947\,W   & 11.74\,(0.03)  & 11.40\,(0.02) & 11.15\,(0.03)  & 10.84\,(0.03)  & \\
BO Cep               & 11.98\,(0.08)  & 11.55\,(0.07) & 11.16\,(0.06)  & 10.84\,(0.03) &  \\ 
$[$K98c$]$\,Em*\,109 & 16.30\,(0.01)  & 14.96\,(0.02) & 13.94\,(0.02)  & 13.18\,(0.01) &  \\ 
$[$K98c$]$\,Em*\,108 & 14.47\,(0.04)  & 13.27\,(0.03) & 12.43\,(0.02)  & 11.87\,(0.02) &  \\ 
SV Cep               & 10.70\,(0.10)  & 10.52\,(0.12) & 10.28\,(0.06)  & 10.08\,(0.05) &  \\ 
IRAS 22219+7908      & 12.65\,(0.06)  & 12.39\,(0.04) & 12.16\,(0.02)  & 12.05\,(0.02) &  \\
$[$K98c$]$\,Em*\,119\,N & 15.67\,(0.08)  & 14.81\,(0.08) & 14.00\,(0.06)  & 13.19\,(0.07) &  \\   
$[$K98c$]$\,Em*\,119\,S & 18.41\,(0.07)  & 17.27\,(0.07) & 16.22\,(0.11)  & 15.26\,(0.01) &  4\arcsec, 184\degr \\
$[$KP93$]$ 2-1 & 17.12\,(0.75)  & 15.42\,(0.42) & 13.80\,(1.03)  & 12.53\,(0.07)  &var. $BVR_\mathrm{C}$ \\ 
$[$KP93$]$ 39   & 17.34\,(0.25)  & 15.89\,(0.46) & 14.38\,(0.43)  & 13.31\,(0.12)  & var. $BVR_\mathrm{C}I_\mathrm{C}$\\
$[$RD95$]$ B	     & \nodata   & 22.99\,($\cdots$)  & 20.22\,(0.25)  & 17.76\,(0.30)  & var. $BVR_\mathrm{C}I_\mathrm{C}$ \\ 
$[$RD95$]$ A	     & \nodata   & 20.94\,($\cdots$)  & 19.12\,(0.76)  & 16.09\,(0.60)  & var. $R_\mathrm{C}I_\mathrm{C}$ \\ 
2MASS 22352542+7517562 & 18.93\,(0.85) & 17.20\,(0.90) & 15.59\,(0.75) &  14.15\,(0.55) & var. $R_\mathrm{C}I_\mathrm{C}$ \\
2MASS 22352722+7518019 & 19.40\,(0.60) & 17.66\,(0.65) & 15.83\,(0.40) & 14.45\,(0.45) & var. $BVR_\mathrm{C}I_\mathrm{C}$ \\
$[$KP93$]$ 2-2             & 19.66\,(1.60)  & 16.97\,(1.85) & 15.29\,(0.45)  & 13.84\,(1.10) & var. $BVR_\mathrm{C}I_\mathrm{C}$\\ 
IRAS 22355+7505      & 21.26\,(0.60)  & 19.82\,(0.45) & 17.48\,(0.34)  & 15.09\,(0.50)  & var. $BVR_\mathrm{C}I_\mathrm{C}$\\ 
$[$KP93$]$ 2-43            & 19.18\,(0.09)  & 16.87\,(0.08) & 15.17\,(0.05)  & 13.57\,(0.05)  & \\ 
$[$KP93$]$ 2-43B           & 19.37\,(0.12)  & 18.71\,(0.04) & 16.71\,(0.05)  & 14.86\,(0.04) & 5.2\arcsec, 245\fdg5 \\
$[$KP93$]$ 2-3             & 16.50\,(0.06)  & 15.04\,(0.08) & 14.01\,(0.03)  & 13.13\,(0.03) & \\ 
$[$ETM94$]$ IRAS 22376+7455 Star 3 & 19.80\,(0.25)  & 17.44\,(0.06) & 15.75\,(0.06)  & 13.983\,(0.05) &  \\ 
$[$KP93$]$ 2-44            & 17.65\,(0.10)  & 16.10\,(0.07) & 14.90\,(0.03)  & 13.62\,(0.01)  & \\ 
$[$KP93$]$ 2-45            & 17.06\,(0.12)  & 15.79\,(0.02) & 14.70\,(0.34)  & 13.44\,(0.45)  & var. $R_\mathrm{C}I_\mathrm{C}$\\ 
$[$KP93$]$ 2-46            & 19.05\,(0.22)  & 17.01\,(0.02) & 15.24\,(0.02)  & 13.56\,(0.03)  & \\ 
IRAS F22424+7450     & 14.37\,(0.07)  & 13.22\,(0.07) & 12.48\,(0.04)  & 11.86\,(0.02) &  \\ 
$[$K98c$]$\,Em*\,128             & 16.87\,(0.12)  & 15.66\,(0.07) & 14.70\,(0.08)  & 13.69\,(0.04) &  \\ 
IRAS 22480+7533      & 12.95\,(0.05)  & 12.00\,(0.03) & 11.41\,(0.02)  & 10.87\,(0.02) &  \\ 
$[$K98c$]$\,Em*\,137             & 17.66\,(0.11)  & 16.48\,(0.06) & 15.321\,(0.03)  & 13.76\,(0.03) &  \\ 
AS\,507    & 11.17\,(0.06)  & 10.22\,(0.06) &  9.63\,(0.18)  &  9.16\,(0.05) &  \\ 
AS 507 B             & 16.93\,(0.07)  & 16.08\,(0.04) & 14.69\,(0.03)  & 12.71\,(0.03) & 9\arcsec, 156\degr  \\
\enddata  
\end{deluxetable} 

\clearpage

\clearpage

\begin{table}
\begin{center}
\caption{Adopted cloud distances.}
\label{Tab_dist}
{\scriptsize
\begin{tabular}{lcl}
\hline
\noalign{\smallskip}
Cloud & D (pc) & References \\
\noalign{\smallskip}
\hline
L\,1147 / L\,1158 & 325 & \citet{SCKM} \\
NGC 7023 & 282 & \citet{Zdana09} \\
L\,1228 &  200 & \citet{Kun98} \\
L\,1251 & 320 & average from \citet{Kun98} and \citet{Balazs04} \\
L\,1219 & 400 & \citet{Racine} \\
L\,1261 / L\,1262 & 180 & \citet{Kun98} \\
NGC 7129 & 800 & \citet{ABK00} \\
\noalign{\smallskip}
\hline
\end{tabular}}
\end{center}
\end{table}

\clearpage


\begin{deluxetable}{lcccccccc} 
\tabletypesize{\scriptsize}
\tablecolumns{9} 
\tablewidth{0pt} 
\tablecaption{Parameters of the observed stars, derived from our optical data.\label{Table_res}} 
\tablehead{ 
\colhead{Star} & \colhead{A$_{\mathrm V}$ (mag)} & \colhead{L$_\mathrm{bol}$(L$_{\sun}$)} 
& \colhead{$T_\mathrm{eff}$ (K)} 
& \colhead{$M_{*}$ (M$_{\sun}$)} & \colhead{Age (Myr)} & 
\multicolumn{3}{c}{$\log \dot{M}_{d}$ (M$_{\sun}$\,yr$^{-1}$)} \\
\cline{7-9} \\
&   &  &  &  &  & \colhead{$\Delta$ V(H$\alpha$)} & \colhead{EW(H$\alpha$)} & \colhead{EW(8542)} } 
\startdata 
\cutinhead{L1228}
F21022  & 5.90 &  0.34 & 3720 &  0.5\phn  &  2.8 &  $-$4.98 & $-$7.29  & $-$7.21   \\
OSHA 42 & 3.42 &  0.49  & 3850 &  0.6\phn  &  2.8 &  $-$6.90 & $-$6.97 & $-$7.55   \\
$[$K98c$]$\,Em* 35  & 0.92 & 0.34 & 3580 &  0.35 &  2.0 &  $-$8.36  & $-$8.30  &\nodata    \\
OSHA 48 & 2.89  & 2.83  & 4350  & 1.2\phn  & 1.0 &  $-$8.60  & $-$6.96 & \nodata    \\
OSHA 49 & 2.26 &  0.38 &  3720 &  0.5\phn &  3.0 &  $-$9.11 & $-$8.60 & \nodata \\
OSHA 50 & 3.05  & 0.28  & 3850 &  0.60 &  4.5  & $-$7.82  & $-$6.97 & $-$6.87   \\
$[$PRN2004$]$ S3  & 4.04 &  0.21 &  3720 &  0.5\phn   & 4.0 &  $-$7.94  & $-$8.53 & \nodata  \\
OSHA 53 & 1.55 &  0.57 &  4350 &  0.95   & 10.0  & $-$7.56 & $-$7.67 & \nodata  \\
OSHA 59 & 1.74 &  0.67 &  4350 &  1.0\phn  &  6.0  &  $-$6.49 & $-$7.39 & \nodata  \\
RNO 129 SW& 4.20  & 0.42  & 3720  & 0.5\phn   & 3.0 & $-$6.58 & $-$7.12 & $-$6.69  \\
RNO 129 NE& 4.40 &  0.42  & 3580  & 0.35  & 3.0  & $-$6.24 & $-$7.41 & $-$7.26  \\
RNO 129\,A& 4.27 &  0.22  & 3650  & 0.42   & 1.2   & $-$12.47 & $-$11.95 & \nodata  \\
$[$K98c$]$\,Em* 46  & 0.00  & 0.16  & 3580  & 0.35 & 2.5  & $-$9.27 & $-$9.39 & \nodata  \\ 
\cutinhead{NGC 7023}
FT Cep      & 5.85 & 6.33 &  5080  &  2.1\phn  &  1.2 &  $-$7.03 & $-$7.08 & $-$6.90  \\	 
FU Cep      & 1.82 & 0.48 &  3720  &  0.48 &  1.6  &  $-$5.94  & $-$7.05 & $-$7.21  \\	
Lk\ha 425   & 1.47 & 0.25 & 3850   & 0.6\phn  & 5.0  & $-$7.12 & $-$7.46 & $-$7.15  \\
NGC\,7023 RS2 & 3.65 & 1.82 &  3850  & 0.6\phn  & 0.5  & $-$7.31 & $-$7.26 & $-$8.21  \\		
NGC\,7023 RS2B & 4.41 & 0.60 & 3850 & 0.6\phn  & 2.0 &  $-$7.81 & $-$7.19  &\nodata  \\
NGC\,7023 RS5 & 2.83 & 0.41  & 3580  & 0.38  & 1.3  & $-$7.30 & $-$7.63 & $-$7.69  \\       
HZ Cep      & 1.74  & 0.75  & 3850  & 0.6\phn  & 1.5  & \nodata & \nodata & \nodata  \\   
SX Cep      & 1.46 & 0.89 &  4060  & 0.8\phn  & 2.1  & \nodata  & \nodata  & \nodata  \\
Lk\ha 428\,N& 4.47 & 2.02  & 4060  & 0.8\phn  & 0.8  & $-$6.56 & $-$6.87 & $-$7.02  \\	   
Lk\ha 428\,S& 0.68 & 0.15  & 3305 & 0.2\phn  & 1.5  & $-$8.93 & $-$8.92 & \nodata \\ 
FV Cep      & 1.27 &  0.63  & 4350   & 1.0\phn   & 7.0  & $-$8.78 & $-$8.10 & \nodata  \\  
FW Cep      & 1.52 &  0.54 &  4350  & 1.0\phn  & 8.5 & $-$6.82 & $-$8.68 & \nodata  \\	
NGC\,7023 RS\,10 & 0.00 &  0.07 &  4590 & \nodata & \nodata & $-$9.66 & $-$10.22\phn & \nodata  \\ 
EH Cep      & 1.39 & 3.41 & 4900  & 1.8\phn  & 2.9  & $-$6.18 & $-$6.76 & $-$6.84 \\
\cutinhead{L1251}
$[$KP93$]$ 2-1    & 4.67  & 6.61 &  4900 &  2.1\phn   & 0.9 &  $-$4.88 & $-$5.83 & \nodata  \\
$[$KP93$]$ 2-39   & 1.25 &  0.44 &  3850 & 0.6\phn   & 3.0  & $-$7.46 & $-$8.63 & \nodata \\
$[$KP93$]$ 2-2    & 2.01 &  0.43 & 3580 & 0.38 & 1.1  & $-$7.35 & $-$7.23 & $-$6.78  \\
$[$KP93$]$ 3-10   & 6.16 &  1.28 & 3370 & 0.2\phn  & 0.1  & $-$5.48 & $-$6.24 &\nodata  \\
$[$RD95$]$ A      & 8.88 & 2.24   & 4060 & 0.8\phn  & 0.8  & $-$7.36 & $-$7.09 & $-$7.22  \\
$[$RD95$]$ B      & 9.88 & 1.03   & 3850 & 0.6\phn  & 0.9  & $-$8.05 & $-$7.11 & $-$7.53  \\
$[$RD95$]$ 2  & 3.19 &  0.62 &  3525 &  0.35 & 0.8  & $-$8.01 & $-$6.15 & $-$6.24  \\
$[$RD95$]$ 5  & 3.19 &  0.43 &  3580 &  0.6\phn  &  3.0 &  $-$9.24 & $-$7.11  & \nodata \\
$[$KP93$]$ 2-43    & 4.74 & 2.25  & 4350  & 1.2\phn   & 1.0 & $-$9.31 & $-$9.31 &\nodata \\
$[$ETM94$]$ Star 3 & 6.47 &  4.49 &  4900 &  2.0\phn &  1.8 &  $-$5.14  & $-$7.47 &\nodata \\
$[$KP93$]$ 2-3     & 1.21 & 0.50 & 4060 & 0.8\phn  & 4.0 & $-$5.42 & $-$7.91 &\nodata  \\ 
$[$KP93$]$ 2-44    & 1.08 & 0.31 & 3650 & 0.41 & 2.5 & $-$7.87  & $-$8.76 &\nodata  \\
$[$KP93$]$ 2-45    & 1.62 & 0.49 & 3720 &  0.5\phn  &  1.8 &  $-$8.31 & $-$7.54 &\nodata  \\
$[$KP93$]$ 2-46    & 5.47 &  4.02 &  4350 & 1.2\phn  &  0.5 &  $-$7.09 & $-$6.24 &\nodata  \\ 
IRAS 22424         & 0.44 &  1.00 &  4590 &  1.2\phn  &  7.5 &  \nodata  & $-$9.29 & \nodata \\ 
IRAS 22480         & 0.79 & 3.00  & 5080 &  1.5\phn  &  4.0 &  $-$6.51 & $-$6.77 & $-$6.99  \\  
$[$K98c$]$\,Em* 128 & 1.39 & 0.34  & 3850 &  0.6\phn & 4.0  & $-$6.50 & $-$7.73 & $-$8.11  \\ 
\cutinhead{G\,109+11 association}
BD+68\degr1118 & 0.99  & 21.40  & 8970 &  2.0  &  7.0    & \nodata & \nodata & \nodata \\ 
$[$K98c$]$\,Em*\,53        & 1.59   & 1.65  & 4350  & 1.2   & 1.5  &  $-$7.10 & $-$7.37 & $-$7.60  \\ 
$[$K98c$]$\,Em*\,58        & 1.34 & 0.10   & 4060  & 0.6  & \nodata & $-$5.60 & $-$9.26 & $-$9.21  \\ 
2MASS 212254 & 5.66  & 0.87  & 4730 &  1.0 &  20: & $-$8.72 &  $-$8.88 & \nodata  \\
$[$K98c$]$\,Em*\,73        & 0.57 & 0.69   & 4060   &  0.8  & 3.0  & $-$6.61 & $-$7.20 & $-$8.47  \\
BH Cep     & 0.62  & 8.20  & 6440 & 1.5  & 6.0  & \nodata  &  $-$7.85 &\nodata \\
BO Cep     & 0.49 & 4.75   & 6440 & \phn1.35  & 15.0\phn   & \nodata  &  $-$7.59 &  \nodata  \\
IRAS 22129 & 0.72 &  5.90  & 5250  & 1.8   & 3.0  & $-$8.18 & $-$7.01 & \nodata \\
IRAS 22144 & 0.21 &  4.88  & 6360  & 1.5   & 5.0  & $-$6.08 & $-$7.44 & \nodata  \\
IRAS 22152 & 0.00 &  5.07  & 6440  & 1.3  & 10.0\phn  & \nodata & \nodata & \nodata  \\
$[$K98c$]$\,Em*\,119N  & 0.90  & 0.69  & 4350  &  1.0  & 5.0  & $-$7.70 &  $-$7.60 & $-$7.20  \\
$[$K98c$]$\,Em*\,119S  & 2.89  & 0.33  & 5250  & \phn0.85  &\nodata & $-$7.27 &  $-$8.48 & \nodata  \\
\cutinhead{SV Cep group}
$[$K98c$]$\,Em* 95     & 0.40 & 2.22  & 4350 & 1.1  & 3.0  & $-$7.33 & $-$7.57 & $-$7.53  \\	    
$[$K98c$]$\,Em* 108    & 1.15 & 3.01  & 4900 & 2.0  & 1.5  & $-$7.52 & $-$7.22 &\nodata \\    
$[$K98c$]$\,Em* 109    & 3.12   & 2.59  & 4900 & 1.5  & 3.5  & $-$5.19 & $-$6.66 &\nodata \\    
SV Cep     & 1.48  & 44.67  & 8970  & 2.4   & 4.0  & \nodata  & $-$6.63 & \nodata  \\ 
\cutinhead{NGC 7129}				        	    
V350 Cep       & 1.20 &  1.27 &  3850  & 0.58 & 0.5 & $-$7.74 & $-$6.99 & $-$6.99 \\
GGD 33a        & 2.84 &  0.06  & 3850 &  0.4\phn &  \nodata  &  $-$5.52 & $-$9.13 & $-$8.29  \\
$[$MMN2004b$]$ 17 &  1.00 &  1.06 & 3720 &  0.45  & 0.5 & $-$7.56 & $-$8.41 & \nodata  \\ 
HBC 731        & 7.71 & 6.53 & 4900 & 2.0\phn  & 1.0  & $-$7.62 & $-$7.45 & \nodata  \\ 
V391 Cep       & 0.96 &  4.76  & 4350  & 1.15  & 0.2  & $-$8.10 & $-$6.80 & $-$6.25  \\
NGC\,7129 S V1 & 0.92  & 0.81  & 4060  & 0.8\phn  & 2.5  & $-$7.47 & $-$8.55 & \nodata \\
NGC\,7129 S V2 & 1.38  & 0.41  & 3850  & 0.6\phn   & 3.0 &  $-$7.12 & $-$8.63 & $-$8.61 \\
NGC\,7129 S V3& 3.50 & 1.40  & 4350  & 1.15 & 2.5  & $-$6.69 & $-$7.59 & \nodata \\
$[$K98c$]$\,Em* 72         & 3.05 &  5.70 &  4590  & 1.6\phn & 0.5 &  $-$5.97 & $-$5.91 & $-$5.96  \\
\cutinhead{Small groups and dispersed stars}
RNO 124    & 6.53 &   0.92  & 5250 &  1.4\phn  & 3.0 &  $-$6.64 & $-$7.57 & $-$7.43  \\
PV Cep     & 4.87 & 0.90  & 5250 & 1.1\phn  & \nodata   & $-$7.87 & $-$7.08 & $-$6.40  \\
IRAS F20535    & 1.76  &  4.04 &  6200 &  1.35  &  9.0 &  $-$8.60 & $-$7.76 & \nodata \\ 
RNO\,135       & 2.49  &  0.34 &  3470 &  0.30  & 1.0 &   $-$9.36 &  $-$8.58 &\nodata  \\ 
IRAS F22219 & 1.17  &  5.33  & 9230 &  2.0\phn  & \nodata  &$-$7.94 & $-$7.67 &\nodata  \\
$[$K98c$]$\,Em* 137 & 2.59 &  0.20 &  3580  & 0.38  & 5.0  & $-$7.84 & $-$7.45 &\nodata  \\ 
AS\,507      & 1.07 &  5.77 &  5860 &  1.55 &  3.1  & $-$6.55 & $-$6.70 &\nodata  \\ 
\enddata 			         
\flushleft{
{\footnotesize Column 1: Name of the star; Col. 2: The visual extinction A$_{\mathrm V}$, derived from the 
color excesses; Col.~3: the bolometric luminosity of the star, derived from the
optical data; Col. 3: $T_\mathrm{eff}$ derived from the spectral classification; 
Col.~4: Mass of the star derived from the HRD; Col.~5: Age of the star, derived from the HRD; Cols.~6--9: 
Logarithm of accretion rates from the disk in M$_{\sun}$ yr$^{-1}$, derived from the 10\% width of the \ha line
\citep{Natta04}, from the luminosity of \ha emission line, and  
the CaII\,$\lambda$8542 line \citep{Dahm08}, respectively.}}

\end{deluxetable}


\clearpage

\begin{deluxetable}{lcccccccc} 
\tabletypesize{\scriptsize}
\tablecolumns{9} 
\tablewidth{0pc} 
\tablecaption{Median properties of observed stars for individual regions \label{Tab_med}} 
\tablehead{ 
\colhead{Region} & \colhead{A$_\mathrm{V}$}  & \colhead{T$_\mathrm{eff}$} &
 \colhead{M$_{*}$} &  \colhead{Age} & \colhead{$\log(\dot{M}^d_\mathrm{acc})$}& $\alpha_{24-K}$ 
 & $\lambda_\mathrm{turnoff}$ & N \\
 & \colhead{(mag)} & \colhead{(K)} & \colhead{(M$_{\sun}$)} & \colhead{(Myr)} 
 & \colhead{(M$_{\sun}\,yr^{-1}$)} & & \micron } 
\startdata
NGC\,7023     & 1.6 & 3850 & 0.6 & 1.6 & $-$7.3 & 0.98 & 3.6 & 14 \\
L\,1228       & 3.1 & 3720 & 0.5 & 3.0 &\nodata & 0.90 & 3.6 & 13 \\
\phn L\,1228\,S  & 2.3 & 3720 & 0.5 & 3.0 & $-$8.3 & 0.90 & 5.8 & 7  \\
L\,1251       & 3.2 & 3850 & 0.6 & 1.8 & $-$7.1 & 0.93 & 2.2 & 17 \\
\phn L\,1251\,C  & 4.7 & 3850 & 0.6 & 0.9 & $-$7.1 & 0.93 & 1.2 & 8 \\
\phn L\,1251\,E  & 1.6 & 4060 & 0.8 & 1.8 & $-$7.5 & 0.94 & 5.8 & 5 \\
G\,109+11 ass.& 1.0 & 5250 & 1.5 & 6.0 &\nodata & 0.28 & \nodata & 12 \\
SV Cep group  & 1.3 & 4900 & 1.7 & 3.0 &\nodata & 0.68 & \nodata  &  4 \\
NGC\,7129     & 1.4 & 4060 & 0.8 & 1.0 & $-$7.5 & 0.29 & 1.6 & 9 \\ 
\enddata
\flushleft{
{\small The quantities listed in Cols. 2--6 are determined form the present spectroscopic 
and photometric observations; Columns 7 and 8 show the median SED slope $\alpha_{24-K}$,
and $\lambda_\mathrm{turnoff}$, respectively. Column~9 shows the number of group members 
studied.}}
\end{deluxetable}

\clearpage

\end{document}